\documentclass[paper]{JHEP3}

\usepackage{makeidx}         
\usepackage{graphicx}        

\usepackage{multicol}        
\usepackage[bottom]{footmisc}

\def\a{\alpha}
\def\b{\beta}
\def\c{\gamma}
\def\d{\delta}
\def\e{\epsilon}

\def\l{\lambda}
\def\m{\mu}
\def\n{\nu}
\def\o{\theta}
\def\p{\pi}
\def\r{\rho}
\def\s{\sigma}

\def\C{\Gamma}
\def\D{\Delta}
\def\L{\Lambda}

\def\pl{\partial}

\def\rta{\rightarrow}

\def\la{\langle}
\def\ra{\rangle}
\def\DD{{\cal D}}

\def\OO{{\cal O}}

\def\BB{{\cal B}}

\def\RR{{\cal R}}

\def\Dslash{\,{\raise.15ex\hbox{/}\mkern-12mu D}}

\def\be{\begin{equation}}
\def\ee{\end{equation}}

\def\SIMQ{\mathrel{\mathop \sim_{Q^2\rightarrow\infty}}} 
\def\SIMEQz{\mathrel{\mathop \simeq_{z \sim 1}}}
\def\SIMEQz0{\mathrel{\mathop \simeq_{z \sim 0}}}

\def\hD{\stackrel{\leftrightarrow}{D}}

\title{The $U(1)_A$ Anomaly and QCD Phenomenology\thanks{To appear
in the volume {\it String Theory and Fundamental Interactions}, published
in honour of Gabriele Veneziano on his 65th birthday, eds.~M.~Gasperini
and J.~Maharana, Lecture Notes in Physics, Springer, Berlin/Heidelberg
2007.}}

\author{G.M. Shore\\
        Department of Physics\\
        University of Wales, Swansea\\
        Swansea SA2 8PP, U.K.\\
        E-mail: \email{g.m.shore@swansea.ac.uk}}

\abstract{
The role of the $U(1)_A$ anomaly in QCD phenomenology is reviewed,
focusing on the relation between quark dynamics and gluon topology.
Topics covered include a generalisation of the Witten-Veneziano
formula for the mass of the $\eta'$, the determination of pseudoscalar
meson decay constants, radiative pseudoscalar decays and the $U(1)_A$
Goldberger-Treiman relation. Sum rules are derived for the proton and 
photon structure functions $g_1^p$ and $g_1^\c$ measured in polarised 
deep-inelastic scattering. The first moment sum rule for $g_1^p$ (the
`proton spin' problem) is confronted with new data from COMPASS and HERMES
on the deuteron structure function and shown to be quantitatively explained 
in terms of topological charge screening. Proposals for experiments on 
semi-inclusive DIS and polarised two-photon physics at future $ep$ and 
high-luminosity $e^+ e^-$ colliders are discussed. }

\begin{document}

\section{Introduction}

The $U(1)_A$ anomaly has played an important historical role in 
establishing QCD as the theory of the strong interactions. 
The description of radiative decays of the pseudoscalar mesons 
in the framework of a gauge theory requires the existence of the 
electromagnetic axial anomaly and determines the number of colours
to be $N_c = 3$. The compatibility of the symmetries of QCD with the
absence of a ninth light pseudoscalar meson -- the so-called 
`$U(1)_A$ problem' -- in turn depends on the contribution of the colour
gauge fields to the anomaly. More recently, it has become clear how the 
anomaly-mediated link between quark dynamics and gluon topology 
(the non-perturbative dynamics of topologically non-trivial gluon 
configurations) is the key to understanding a range of phenomena in 
polarised QCD phenomenology, most notably the `proton spin' sum rule 
for the first moment of the structure function $g_1^p$.

In this paper, based on original research performed in a long-standing 
collaboration with Gabriele Veneziano, we review the role of the $U(1)_A$
anomaly in describing a wide variety of phenomena in QCD, ranging from the 
low-energy dynamics of the pseudoscalar mesons to sum rules in polarised 
deep-inelastic scattering. The aim is to show how these experiments reveal 
subtle aspects of quantum field theory, in particular topological gluon 
dynamics, which go beyond simple current algebra or parton model 
interpretations.

We begin in section 2 with a brief review of the essential theoretical 
toolkit: anomalous chiral Ward identities, Zumino transforms, the 
renormalisation group, and the range of expansion schemes associated with 
large $N_c$, notably the OZI approximation. Then, in section 3, we build on
Veneziano's seminal 1979 paper \cite{Veneziano:1979ec} to describe 
how the pseudoscalar mesons saturate the Ward identities in a way compatible
with both the renormalisation group and large-$N_c$ constraints and
derive a generalisation of the famous Witten-Veneziano mass formula for the 
$\eta'$ which incorporates, but goes beyond, the original large-$N_c$
derivation \cite{Shore:1999tw,Shore:2006mm}.

In section 4, we turn to QCD phenomenology and describe how this intuition 
on the resolution of the $U(1)_A$ problem allows a quantitative description
of low-energy pseudoscalar meson physics, especially radiative decays,
the determination of the pseudoscalar decay constants, and meson-nucleon 
couplings. We review the $U(1)_A$ extension of the Goldberger-Treiman formula 
first proposed by Veneziano \cite{Veneziano:1989ei}
as the key to understanding the `proton spin' problem and test an
important hypothesis on the origin of OZI violations and their relation to 
the renormalisation group. Low-energy $\eta$ and $\eta'$ physics is
currently an active experimental field and we explain the importance
of an accurate determination of the couplings $g_{\eta NN}$ and 
$g_{\eta' NN}$ in elucidating the role of gluon topology in QCD.  

All of these low-energy phenomena have counterparts in high-energy,
polarised deep-inelastic scattering. 
This enables us to formulate a new sum rule for the first moment of the 
polarised photon structure function $g_1^\c$ (section 6). The
dependence of this sum rule on the invariant momentum of the off-shell
target photon measures the form factors of the 3-current AVV Green
function and encodes a wealth of information about the realisation of
chiral symmetry in QCD, while its asymptotic limit reflects both the
electromagnetic and colour $U(1)_A$ anomalies. We show how this sum
rule, which we first proposed in 1992 \cite{Narison:1992fd,Shore:1992pm},
may soon be tested if the forthcoming generation of high-luminosity
$e^+ e^-$ colliders, currently conceived as $B$ factories, are run
with polarised beams \cite{Shore:2004cb}.

The most striking application of these ideas is, however, to the
famous `proton spin' problem, which originated with the observation of
the violation of the Ellis-Jaffe sum rule for the first moment of the
polarised proton structure function $g_1^p$ by the EMC collaboration
at CERN in 1988. This experiment, and its successors at SLAC, DESY
(HERMES) and CERN (SMC, COMPASS) determined the axial charge $a^0$ of
the proton. In the simple valence-quark parton model, this can be
identified with the quark spin and its observed suppression led to an
intense experimental and theoretical search over two decades for the
origin of the proton spin. In fact, as Veneziano was the first to
understand \cite{Veneziano:1989ei}, $a^0$ does not measure spin in QCD
itself and its suppression is related to OZI violations induced by the 
$U(1)_A$ anomaly.

In a series of papers, summarised in section 5, we have shown how
$a^0$ decouples from the real angular momentum sum rule for the proton
(the form factors for this sum rule are given by generalised parton
distributions (GPDs) which can be extracted from less inclusive
measurements such as deeply-virtual Compton scattering) and is instead
related to the gluon topological susceptibility 
\cite{Shore:1990zu,Shore:1991dv}. The experimentally observed
suppression is a manifestation of topological charge screening
in the QCD vacuum. 
In a 1994 paper with Narison \cite{Narison:1994hv},
using QCD spectral sum rule methods, we were able to compute the slope
of the topological susceptibility and give a quantitative prediction
for $a^0$. Our prediction, $a^0 = 0.33$, has within the last few
months been spectacularly confirmed by the latest data on the deuteron
structure function from the COMPASS and HERMES collaborations.

Hopefully, this impressive new evidence for topological charge screening 
will provide fresh impetus to experimental `spin' physics - first,
to verify the real angular momentum sum rule by measuring the relevant
GPDs, and second, to pursue the programme of target-fragmentation
studies in semi-inclusive DIS at polarised $ep$ colliders which we
have proposed as a further test of our understanding of the $g_1^p$
sum rule \cite{Shore:1997tq}.

\vskip0.5cm
This review has been prepared in celebration of the 65th birthday of
Gabriele Veneziano. I first met Gabriele when I came to Geneva
as a CERN fellow in 1981. In fact, our first interaction was across 
a tennis court, in a regular Friday doubles match with Daniele Amati 
and Toine Van Proeyen. I like to think that in those days 
I could show Gabriele a thing or two about tennis -- 
physics, of course, was a different matter. 
It has been my privilege through these ensuing 25 years to 
collaborate with one of the most brilliant and innovative physicists 
of our generation. But it has also been fun. As all his collaborators 
will testify, his good humour, generosity to younger colleagues, 
and enthusiasm in thinking out solutions to the deepest and most
fundamental problems in particle physics and cosmology make working with
Gabriele not only intellectually rewarding but hugely enjoyable.

In his contribution to the `Okubofest' in 1990 \cite{Veneziano:Okubo}, 
Gabriele concluded an account of the relevance of the OZI rule to $g_1^p$ 
by hoping that he had `made Professor Okubo happy'. In turn, I hope that
this review will make Gabriele happy:~ happy to recall how his original
ideas on the $U(1)_A$ problem have grown into a quantitative
description of anomalous QCD phenomenology, and happy at the prospect
of new discoveries from a rich programme of experimental physics at 
future polarised colliders. 
It is my pleasure to join all the contributors to this volume in
wishing him a happy birthday.

\section{The $U(1)_A$ anomaly and the topological susceptibility}

We begin by reviewing some essential features of the $U(1)_A$ anomaly,
chiral Ward identities and the renormalisation group,
placing particular emphasis on the role of the gluon topological 
susceptibility. As we shall see, the anomaly provides the vital link 
between quark dynamics and gluon topology which is essential in
understanding a range of phenomena in polarised QCD phenomenology.

\subsection{Anomalous chiral Ward identities}
\label{sec 2.1}

An anomaly arises when a symmetry which is present in the classical limit 
cannot be consistently imposed in a quantum field theory. The original
example of an anomaly, and one which continues to have far-reaching
implications for the phenomenology of QCD, is the famous Adler-Bell-Jackiw
axial anomaly \cite{Adler:1969, BellJackiw:1969, AdlerBardeen:1969}, 
which was first understood in its present form in 1969.
In fact, calculations exhibiting what we now recognise as the anomaly
had already been performed much earlier by Steinberger in his analysis
of meson decays \cite{Steinberger:1949} and by Schwinger
\cite{Schwinger:1951}.

Anomalies manifest themselves in a number of ways. The original derivations
of the axial anomaly involved the impossibility of simultaneously
imposing conservation of both vector and axial currents due to 
regularisation issues in the AVV triangle diagram in QED. More generally,
they arise as anomalous contributions to the commutation relations in 
current algebra. A modern viewpoint, due to Fujikawa \cite{Fujikawa:1979},
sees anomalies as due to the non-invariance of the fermionic measure in 
the path integral under transformations corresponding to a symmetry of 
the classical Lagrangian. In this approach, the result of a chiral 
transformation $q \rta e^{i\a^a T^a \c_5}q$ on the quark fields in 
the QCD generating functional $W[V_{\m 5}^a,V_\m^a,\o,S_5^a,S^a]$ defined
as\footnote{Our notation follows that of 
ref.\cite{Shore:2006mm}.
The currents and pseudoscalar fields $J_{\m5}^a$, $Q$, $\phi_5^a$ 
together with the scalar $\phi^a$ are defined by
\begin{eqnarray*}
&J_{\m5}^a = \bar q \c_\m \c_5 T^a q ~~~~~~~~
&J_\m^a = \bar q \c_\m T^a q ~~~~~~~~
Q = {\a_s\over8\pi} {\rm tr} G_{\m\n} \tilde G^{\m\n} 
\nonumber\\
&\phi_5^a = \bar q \c_5 T^a q ~~~~~~~~~
&\phi^a = \bar q T^a q ~~~~~~~~~~~~
\end{eqnarray*}
where $G_{\m\n}$ is the field strength for the gluon field.
Here, $T^i = {1\over2}\l^i$ are flavour $SU(n_f)$ generators,
and we include the singlet $U(1)_A$ generator 
$T^0 = {\bf 1}/\sqrt{2n_f}$ 
and let the index $a = 0, i$. With this normalisation, ${\rm tr} T^a T^b
= {1\over2}\d^{ab}$ for all the generators $T^a$.
This accounts for the rather unconventional factor $\sqrt{2n_f}$ in the 
anomaly equation but has the advantage of giving a consistent normalisation
to the full set of decay constants including the flavour singlets
$f^{0\eta'}$ and $f^{0\eta}$.

We will only need to consider fields where $i$ corresponds to a generator
in the Cartan sub-algebra, so that $a = 3, 8, 0$ for $n_f = 3$ quark
flavours. We define $d$-symbols by $\{T^a,T^b\} = d_{abc} T^c$. 
For $n_f = 3$, the explicit values are $d_{000} =
d_{033} = d_{088} =  d_{330} = d_{880} = \sqrt{2/3}, ~d_{338} = d_{383} =
-d_{888} = \sqrt{1/3}$.}

\begin{equation}
e^{iW} = \int \DD A \DD\bar q \DD q ~\exp \biggl[i\int dx\bigl(
{\cal L}_{\rm QCD} + V^{\m a}_5 J_{\m5}^a + 
V^{\m a} J_\m^a + \o Q + S_5^a \phi_5^a
+ S^a \phi^a\bigr)\biggr]
\label{eq:ba}
\end{equation}
is
\begin{equation}
\int \DD A \DD\bar q \DD q~\Bigl[\pl^\m J_{\m5}^a - \sqrt{2n_f} \d^{a0} Q
-d_{abc}m^b\phi_5^c - \d \Bigl(\int d^4x {\cal L}_{\rm QCD}\Bigr)\Bigr]~
\exp \biggl[ \ldots \biggr] = 0
\label{eq:bb}
\end{equation}
The terms in the square bracket are simply those arising from
Noether's theorem, including soft breaking by the quark masses, 
with the addition of the anomaly involving the gluon topological 
charge density $Q$.
Re-expressing the chiral variation of the elementary fields in terms
of a variation with respect to the sources $V_{\m5}^a, V_\m^a, \theta, 
S_5^a, S^a$ then gives the functional form of the anomalous chiral Ward 
identities:
\begin{eqnarray}
&\pl_\m W_{V_{\m5}^a} &- \sqrt{2n_f} \d_{a0} W_{\o}-
d_{abc} m^b W_{S_5^c} 
\nonumber\\
&{}&+ f_{abc} V_\m^b W_{V_{\m 5}^c}
+ f_{abc} V_{\m 5}^b W_{V_\m^c}
+ d_{abc} S^b W_{S_5^c}
- d_{abc} S_5^b W_{S^c}
= 0
\label{eq:bc}
\end{eqnarray}
where we have abbreviated functional derivatives as suffices.
This is the key to all the results derived in this section. 
It makes precise the familiar statement of the anomaly as
\begin{equation}
\pl^\m J_{\m5}^a - \sqrt{2n_f} Q \d_{a0} - d_{abc}m^b \phi_5^c ~\sim~0
\label{eq:bd}
\end{equation}

The chiral Ward identities for two and higher-point Green functions
are found by taking functional derivatives of eq.(\ref{eq:bc}) with respect
to the sources. The complete set of identities for two-point functions is
given in our review \cite{Shore:1998dm}. 
As an example, we find\footnote{We use the following $SU(3)$ notation 
for the quark masses and condensates: 
$$
\left(\matrix{m_u &0 &0 \cr
0 &m_d &0 \cr
0 &0 &m_s \cr}\right)
= \sum_{a=0,3,8} m^a T^a
$$
and 
$$
\left(\matrix{ \langle \bar u u\rangle &0 &0 \cr 0 &\langle \bar d d\rangle
&0 \cr
0 &0 &\langle \bar s s\rangle \cr}\right) = 
2 \sum_{0,3,8} \langle \phi^a \rangle T^a
$$
where $\langle \phi^a\rangle$ is the VEV of $\phi^a = \bar{q}~ T^a q$.
It is also convenient to use the compact notation 
$$
M_{ab} = d_{acb} m^c
~~~~~~~~~~~~~~~~~~
\Phi_{ab} = d_{abc} \langle \phi^c\rangle 
$$
} 
\begin{equation}
\pl_\m W_{V_{\m5}^a S_5^b} - \sqrt{2n_f}\d_{a0}W_{\o S_5^b} - 
M_{ac}W_{S_5^c S_5^b} - \Phi_{ab} ~=~ 0
\label{eq:be}
\end{equation}
which in more familiar notation reads
\begin{equation}
\pl^\m \langle 0| T~ J_{\m 5}^a~ \phi_5^b |0\rangle - \sqrt{2n_f}\d_{a0}
\langle 0| T~ Q~\phi_5^b|0\rangle - d_{adc}m^d \langle 0|
T~\phi_5^c ~ \phi_5^b |0\rangle - d_{abc}\langle \phi^c\rangle = 0
\label{eq:bf}
\end{equation}

The anomaly breaks the original $U(n_f)_L \times U(n_f)_R$ chiral symmetry 
to $SU(n_f)_L \times SU(n_f)_R \times U(1)_V / Z_{n_f}^V$ 
and the quark condensate spontaneously breaks this further to the 
coset $SU(n_f)_L \times SU(n_f) / SU(n_f)_V$.
Goldstone's theorem follows immediately. In the chiral limit, 
there are $(n_f^2 - 1)$ massless Nambu-Goldstone bosons, 
which acquire masses of order $\sqrt{m}$ for non-zero quark mass. 
There is no flavour singlet Nambu-Goldstone boson since the corresponding 
current is anomalous. 

The zero-momentum Ward identities are especially important here,
since they control the low-energy dynamics. With the assumption 
that there are no exactly massless particles coupling to the currents, 
we find 
\begin{eqnarray}
&\sqrt{2n_f} \d_{a0} W_{\o\o} + M_{ac} W_{S_5^c \o} = 0 
\nonumber\\ 
&\sqrt{2n_f} \d_{a0} W_{\o S_5^b} + M_{ac} W_{S_5^c S_5^b} + \Phi_{ab} = 0 
\label{eq:bg}
\end{eqnarray}

Another key element of our analysis will be the chiral Ward identities for 
the effective action $\C[V_{\m 5}^a,V_\m^a, Q, \phi_5^a, \phi^a]$, defined
as the generating functional for vertices which are 1PI with respect to the
set of fields $Q, \phi_5^a$ and $\phi^a$ but {\it not} the currents 
$J_{\m 5}^a$, $J_\m^a$. This is achieved using the partial Legendre 
transform (or Zumino transform): 
\begin{equation}
\C[V_{\m5}^a, V_\m^a, Q, \phi_5^a, \phi^a]~=~ W[V_{\m5}^a, V_\m^a, \o,
S_5^a, S^a] - \int dx~\Bigl(\o Q + S_5^a \phi_5^a + S^a \phi^a \Bigr) 
\label{eq:bh}
\end{equation}
The chiral Ward identities for $\C$ are 
\begin{eqnarray}
&\pl_\m \C_{V_{\m5}^a} &- \sqrt{2n_f} \d_{a0} Q - d_{abc}m^b \phi_5^c 
\nonumber\\
&{}&+ f_{abc} V_{\m}^b \C_{V_{\m5}^c}
+ f_{abc} V_{\m5}^b \C_{V_{\m}^c}
- d_{abc} \phi_5^c \C_{\phi^b}  
+ d_{abc} \phi^c \C_{\phi_5^b} = 0
\label{eq:bi}
\end{eqnarray}
Again, the zero-momentum identities for the two-point vertices
play an important role:
\begin{eqnarray}
&\Phi_{ac}\C_{\phi_5^c Q} - \sqrt{2n_f} \d_{a0} = 0 
\nonumber\\
&\Phi_{ac} \C_{\phi_5^c \phi_5^b} - M_{ab} = 0 
\label{eq:bj}
\end{eqnarray}
These will be used in section 3 to construct an effective action
which captures the low-energy dynamics of QCD in the pseudoscalar sector.

\subsection{Topological susceptibility}

The connection with topology arises through the identification of
the gluon operator $Q$ in the anomaly with a topological charge density.
$Q$ is a total divergence:
\begin{equation}
Q ~=~ {\a_s\over8\pi}~{\rm tr}~G_{\m\n} \tilde G^{\m\n} ~=~ \pl^\m K_\m 
\label{eq:bk}
\end{equation}
where $K_\m$ is the Chern-Simons current,
\begin{equation}
K_\m = {\a_s\over4\pi}\e_{\m\n\r\s}{\rm tr}\bigl(A^\n G^{\r\s} - 
{1\over3}gA^\n [A^\r,A^\s]\bigr)
\label{eq:bl}
\end{equation}
Nevertheless, the integral over (Euclidean) spacetime of $Q$ need 
not vanish. In fact, for gauge field configurations such as instantons 
which become pure gauge at infinity,
\begin{equation}
\int d^4 x~Q ~=~ n \in {\bf Z}
\label{eq:bm}
\end{equation}
where the integer $n$ is the topological winding number, an element of the 
homotopy group $\pi_3(SU(N_c))$. 

The form of the anomaly is then understood as follows.
Under a chiral transformation, the fermion measure in the path integral
transforms as (for one flavour)
\begin{equation}
\DD \bar q \DD q ~\rta~ e^{-2i\a\int dx \varphi_i^\dagger \c_5 \varphi_i}~
\DD \bar q \DD q ~=~ \exp^{-2i\a (n_+ - n_-)}~\DD\bar q \DD q
\label{eq:bn}
\end{equation}
where $\varphi_i$ is a basis of eigenfunctions of the Dirac operator
$\Dslash$ in the background gauge field. The non-zero eigenvalues
are chirality paired, so the Jacobian only depends on the difference
$(n_+ - n_-)$ of the positive and negative chirality zero modes of
$\Dslash$. Finally, the index theorem relates the anomaly to the topological
charge density:
\begin{equation}
{\rm ind}\Dslash ~=~ n_+ - n_- ~=~ \int d^4 x ~Q
\label{eq:bo} 
\end{equation}

The topological susceptibility $\chi(p^2)$ is defined as the two-point 
Green function of $Q$, viz.
\begin{equation}
\chi(p^2) = i\int dx~e^{ipx}\langle 0|T~Q(x)~Q(0)|0\rangle
\label{eq:bp}
\end{equation}
We are primarily concerned with the zero-momentum limit $\chi(0) =
W_{\o\o}(0)$. Combining eqs.(\ref{eq:bg}) gives the crucial Ward 
identity satisfied by $\chi(0)$:
\begin{equation}
2n_f \chi(0) = M_{0a} W_{S_5^a S_5^b} M_{b0} + (M\Phi)_{00}
\label{eq:bq}
\end{equation}
that is,
\begin{equation}
n_f^2 \int dx~\langle 0|T~Q(x)~Q(0)|0\rangle ~=~
\int dx~m^a m^b\langle 0|T~\phi_5^a(x)~\phi_5^b(0)|0\rangle 
~+~ m^a \langle \phi^a\rangle
\label{eq:br}
\end{equation}
Determining exactly how this is satisfied in QCD is at the heart of the
Witten-Veneziano approach to the $U(1)_A$ problem 
\cite{Witten:1979vv,Veneziano:1979ec}.

The zero-momentum Ward identities allow us to write a precise form for
the topological susceptibility in QCD in terms of just one unknown
dynamical constant \cite{DiVecchia:1980ve}.
To derive this, recall that the matrix of two-point 
vertices is simply the inverse of the two-point Green function matrix, 
so in the pseudoscalar sector we have the following inversion formula:
\begin{eqnarray}
&\C_{QQ} = - \Bigl(W_{\o\o} - W_{\o S_5^a} (W_{S_5 S_5})_{ab}^{-1} 
W_{S_5^b \o} \Bigr)^{-1} 
\label{eq:bs}
\end{eqnarray}
Using the identities (\ref{eq:bg}) and (\ref{eq:bq}), this implies that
at zero momentum
\begin{equation}
\C_{QQ}^{-1} = - \chi \Bigl(1 - 2n_f \chi (M\Phi)_{00}^{-1} \Bigr)^{-1}
\label{eq:bt}
\end{equation}
and inverting this relation gives
\begin{equation}
\chi = - \C_{QQ}^{-1} \Bigl(1 - 2n_f \C_{QQ}^{-1}
(M\Phi)_{00}^{-1} \Bigr)^{-1}
\label{eq:bu}
\end{equation}
Finally, substituting for $(M\Phi)_{00}^{-1}$ using the definitions above, 
we find the following important identity which determines the quark
mass dependence of the topological susceptibility in QCD:
\begin{equation}
\chi(0) = -A \biggl(1 - A \sum_q {1\over m_q \la \bar q q\ra}\biggr)^{-1} 
\label{eq:bv}
\end{equation}
where we identify the non-perturbative coefficient $A$ as $\C_{QQ}^{-1}$.

Notice immediately how this expression exposes the well-known result 
that $\chi(0)$ vanishes if any quark mass is set to zero. In section 3, 
we will see how it also clarifies the role of the $1/N_c$
expansion in the $U(1)_A$ problem.

\subsection{Renormalisation group}

The conserved current corresponding to a non-anomalous symmetry is not
renormalised and has vanishing anomalous dimension. However, an anomalous
current such as the flavour singlet axial current $J_{\m5}^0$ is
renormalised. The composite operator renormalisation and mixing in
the $J_{\m5}^0, Q$ sector is as follows \cite{Espriu:1982bw}:
\begin{eqnarray}
&J_{\m5R}^0 = Z J_{\m5B}^0  ~~~~~~~~~~
\nonumber\\
&Q_R = Q_B - {1\over\sqrt{2n_f}}(1-Z) \pl^\m J_{\m5B}^0 
\label{eq:bw}
\end{eqnarray}
Notice the form of the mixing of the operator $Q$ with $\pl^\m J_{\m5}^0$ 
under renormalisation. This ensures that the combination 
$\bigl(\pl^\m J_{\m5}^0 - \sqrt{2n_f} Q\bigr)$ occurring in the $U(1)_A$ 
anomaly equation is RG invariant. The chiral Ward identities therefore take 
precisely the same form expressed in terms of the bare or renormalised 
operators, making precise the notion of `non-renormalisation of the 
anomaly'. We may therefore interpret the above Ward identities, which
were derived in terms of the bare operators, as identities for the
renormalised composite operators (and omit the suffix ${}_R$ for notational 
simplicity).

The renormalisation group equation (RGE) for the generating functional 
$W[V_{\m5}^a, V_\m^a, \o, S_5^a, S^a]$ 
follows immediately from the definitions 
(\ref{eq:bw}) of the renormalised composite operators.
Including also a standard multiplicative renormalisation $Z_\phi = 
Z_m^{-1}$ for the pseudoscalar and scalar operators $\phi_5^a$ and 
$\phi^a$ and denoting the anomalous dimensions corresponding to
$Z$ and $Z_\phi$ by $\c$ and $\c_\phi$ respectively, we find\footnote{
The notation $+\ldots$ refers to additional terms which are required 
to produce the contact term contributions to the RGEs for $n$-point
Green functions and vertices of composite operators. These are discussed 
fully in refs.\cite{Shore:1990wp,Shore:1991dv,Shore:1991pn},
but will be omitted here for simplicity. 
They vanish at zero-momentum.}
\begin{equation}
\DD W = \c\Bigl(V_{\m 5}^0 - 
{1\over\sqrt{2n_f}}\pl_\m \theta\Bigr)W_{V_{\m5}^0}
+ \c_\phi\Bigl(S_5^a W_{S_5^a} + S^a W_{S^a}\Bigr) + \ldots
\label{eq:bx}
\end{equation}
where $\DD = \Bigl(\m{\pl\over\pl\m} + \b{\pl\over\pl g} - \c_m\sum_q
m_q{\pl\over\pl m_q}\Bigr)\Big|_{V,\theta,S_5,S}$.

The RGEs for Green functions are found by functional differentiation
of eq.(\ref{eq:bx}) and can be simplified using the Ward identities. 
For example, for $W_{\o\o}$ we find
\begin{equation}
\DD W_{\o\o} ~=~ 2\c W_{\o\o} + 2\c {1\over\sqrt{2n_f}}M_{0b}W_{\o S_5^b}
+ \ldots
\label{eq:by}
\end{equation}
At zero momentum, we can then use the first identity in eq.(\ref{eq:bg})
to prove that the topological susceptibility $\chi(0)$ is RG invariant,
\begin{equation} 
\DD \chi(0) ~=~ 0
\label{eq:bz}
\end{equation}
which is consistent with its explicit expression (\ref{eq:bv}).

A similar RGE holds for the effective action 
$\C[V_{\m 5}^a,V_\m^a,Q,\phi_5^a,\phi_5]$, which allows the scaling
behaviour of the proper vertices involving $Q$ and $\phi_5^a$ to be 
determined \cite{Shore:1990wp,Shore:1991dv,Shore:1991pn}. This reads
\begin{equation}
\DD\C = \c\Bigl(V_{\m5}^0 - {1\over\sqrt{2n_f}}\C_Q\pl_\m\Bigr)
\C_{V_{\m5}^0} - \c_\phi\Bigl(\phi_5^a \C_{\phi_5^a} + 
\phi^a \C_{\phi^a}\Bigr) +\ldots
\label{eq:bzz}
\end{equation}
An immediate consequence is that $\DD \C_{QQ} = 0$ at zero momentum, 
which ensures the compatibility of (\ref{eq:bv}) with the RG invariance 
of $\chi(0)$.

\subsection{$1/N_c$, the topological expansion and OZI}

The final theoretical input into our analysis of the $U(1)_A$ problem
and phenomenological implications of the anomaly concerns the range of
dynamical approximation schemes associated with the large-$N_c$ limit.
At various points we will refer either to the original large-$N_c$  
expansion of 't Hooft \cite{Hooft:1974}, the topological expansion 
introduced by Veneziano \cite{Veneziano:TE} and the OZI limit 
\cite{Okubo,Zweig,Iizuka}.
A very clear summary of the distinction between them is given in 
Veneziano's `Okubofest' review \cite{Veneziano:Okubo}, which we
follow here.

\begin{figure}
\centering
\includegraphics[height=3.3cm]{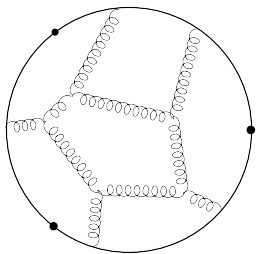} 
\caption{A typical Feynman diagram allowed in the large-$N_c$ limit.
The dots on the quark loop represent external sources.}
\label{fig:largeN}     
\end{figure}

In terms of Feynman diagrams, the leading order in the large $N_c$,
fixed $n_f$ ('t Hooft) limit is the most restrictive of these 
approximations, including only planar diagrams with sources on a single
quark line and no further quark loops (Fig.~\ref{fig:largeN}). 

\begin{figure}
\centering
\includegraphics[height=3.3cm]{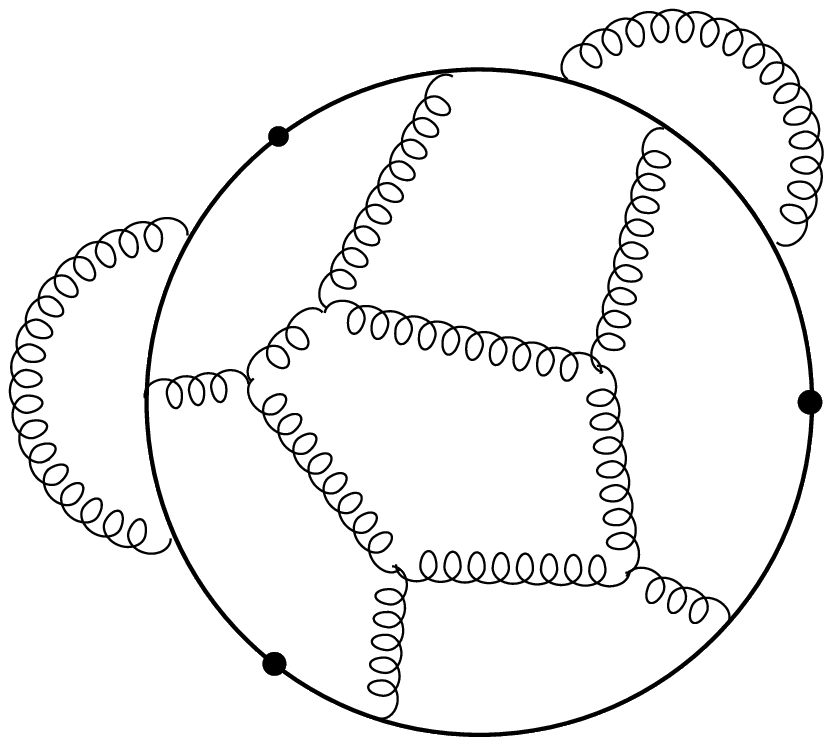} \hskip2cm 
\includegraphics[height=3.3cm]{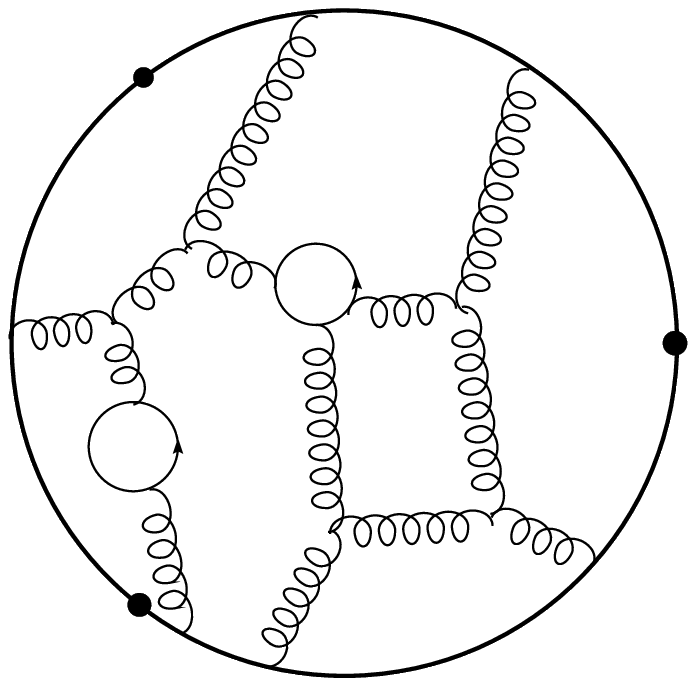}
\caption{Feynman diagrams allowed in the quenched approximation (left) or 
leading order in the topological expansion (right).}
\label{fig:quenchtop}     
\end{figure}

A better approximation to QCD is the quenched approximation
familiar in lattice gauge theory. This is a small $n_f$ expansion
at fixed $N_c$, i.e.~excluding quark loops but allowing non-planar
diagrams (Fig.~\ref{fig:quenchtop}).

An alternative is the topological expansion, which allows any number of 
internal quark loops, but restricts to planar diagrams at leading order.
Provided the sources remain attached to the same quark line,
this corresponds to taking large $N_c$ at fixed $n_f/N_c$. 
This means that quarks and gluons are treated democratically and
the order of approximation is determined solely by the topology of the
diagrams (Fig.~\ref{fig:quenchtop}).

\begin{figure}
\centering
\includegraphics[height=3.3cm]{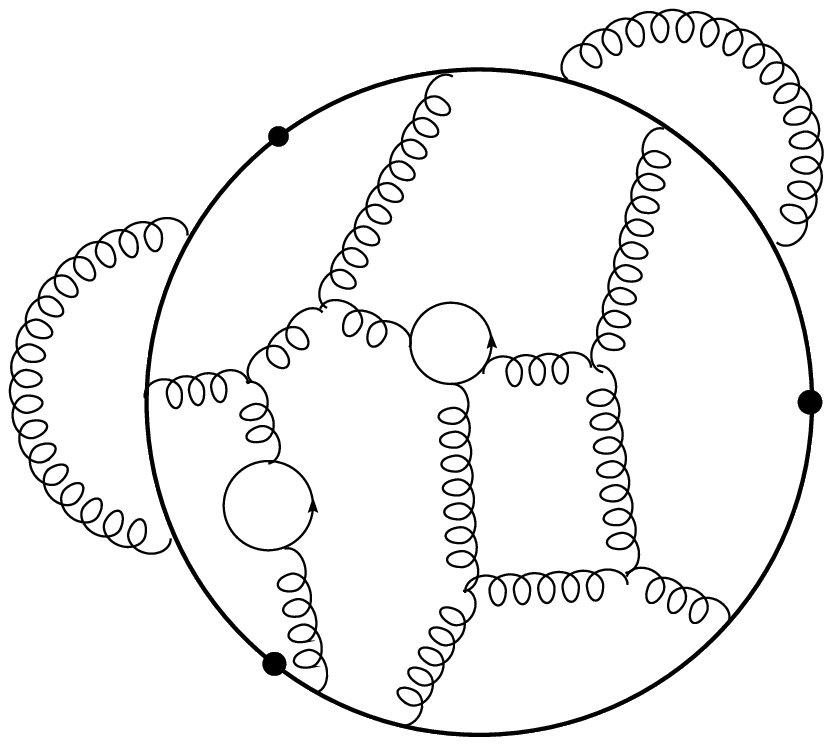} \hskip1cm 
\includegraphics[height=3cm]{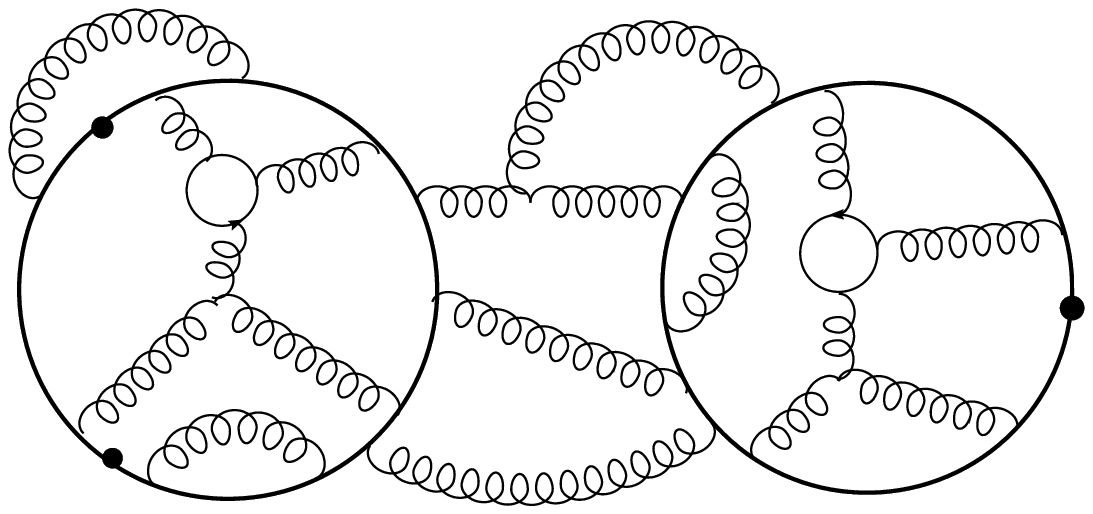}
\caption{Feynman diagrams allowed (left) and forbidden (right) by the 
OZI rule.}
\label{fig:OZI}     
\end{figure}

Finally, the OZI approximation is a still closer match to full QCD 
with dynamical quarks than either the leading order quenched or 
topological expansions. Non-planar diagrams and quark loops are retained, 
but diagrams in which the external sources are connected to different 
quark loops are still excluded (Fig.~\ref{fig:OZI}). 
This means that amplitudes which 
involve purely gluonic intermediate states are suppressed. This
is the field-theoretic basis for the original empirical OZI rule.

In each of these large-$N_c$ expansions, except the topological expansion
where $n_f/N_c$ is fixed, the $U(1)_A$ anomaly does not contribute at 
leading order. More precisely, the anomalous contribution 
$\langle 0|T~Q~\phi_5^b|0\rangle$ in the chiral Ward identity 
(\ref{eq:bf}) is suppressed by $O(1/N_c)$ relative to the current term 
$\langle 0|T~J_{\m 5}^0 ~\phi_5^b|0\rangle$.
This means that the flavour singlet current is conserved, Goldstone's
theorem applies, and conventional PCAC methods can be used to understand
the dynamics of the Green functions with a full set of $(n_f^2 - 1)$ 
massless bosons in the chiral limit. Taking this as a starting point,
we can then learn about the spectral decomposition of the actual QCD
Green functions as we relax from the leading-order limits.
In particular, this leads us to the famous Witten-Veneziano mass formula
for the $\eta'$ meson \cite{Witten:1979vv,Veneziano:1979ec}. 

The behaviour of the topological susceptibility at large $N_c$ is central
to this analysis. It is clear from looking at planar diagrams that 
at leading order in $1/N_c$, $\chi(0)$ in QCD coincides with the
topological susceptibility $\chi(0)|_{\rm YM}$ in the corresponding
pure Yang-Mills theory. Referring now to the explicit expression 
(\ref{eq:bv}) for $\chi(0)$, large-$N_c$ counting rules give $A = O(1)$ 
while $\langle\bar q q\rangle = O(N_c)$. It follows that {\it for
non-zero quark masses},
\begin{equation}
\chi(0) ~=~ - A ~+~ O(n_f/N_c)
\label{eq:baa}
\end{equation}
where $A = \C_{QQ}^{-1}$ is identified as $-\chi(0)_{\rm YM} + O(1/N_c)$.
On the other hand, if we consider the limit of $\chi(0)$ for $m_q\rta 0$
at finite $N_c$, then we have
\begin{equation}
\chi(0)|_{m_q\rta 0} ~=~ 0
\label{eq:bbb}
\end{equation}
The 't Hooft large-$N_c$ limit is therefore not smooth in QCD; the
$N_c \rta \infty$ and $m_q \rta 0$ limits do not commute 
\cite{Witten:1979vv,Veneziano:1979ec,DiVecchia:1980ve}.
This is remedied in the topological expansion, where quark loops are
retained and the $O(n_f/N_c)$ contribution in eq.(\ref{eq:baa}) allows
the smooth chiral limit $\chi(0) \rta 0$ even for large $N_c$.

\section{`$U(1)_A$ without instantons'}

The $U(1)_A$ problem has a long history, pre-dating QCD itself, and
has been an important stimulus to new theoretical ideas involving 
anomalies and gluon topology. 

At its simplest, the original `$U(1)_A$ problem' in current 
algebra is relatively straightforwardly resolved
by the existence of the anomalous contributions to the chiral Ward
identities (anomalous commutators in current algebra) and the 
consequent absence of a ninth light Nambu-Goldstone boson in 
$n_f=3$ QCD.\footnote{The existence of a light flavour-singlet 
Nambu-Goldstone boson would produce a rapid off-shell variation in the
$\eta \rta 3\pi$ decay amplitude, in contradiction with the
experimental data \cite{Weinberg:1975}.}  
However, a full resolution requires a much more detailed understanding 
of the dynamics of the pseudoscalar sector and the role of topological 
fluctuations in the anomalous Green functions.

In this section, we review the analysis of the $U(1)_A$ problem
presented by Veneziano in his seminal 1974 paper, `$U(1)_A$ without
instantons' \cite{Veneziano:1979ec}.\footnote{For reviews of the
instanton approach to the resolution of the $U(1)_A$ problem, 
see e.g.~refs.\cite{'tHooft:1976fv,Crewther:1978kq,
Christos:1984tu,'tHooft:1986nc}.}
As well as deriving the eponymous mass 
formula relating the $\eta'$ mass to the topological susceptibility, 
the essential problem resolved in ref.\cite{Veneziano:1979ec} is how to
describe the dynamics of the Green functions of the pseudoscalar operators
in QCD in terms of a spectral decomposition compatible with the 
$n_f$, $N_c$, $\theta$ and quark mass dependence imposed by the
anomalous Ward identities.

First, recall that in the absence of the anomaly, there will be 
light pseudoscalar mesons $\eta^\a$ coupling derivatively 
to the currents with decay constants $f^{a\a}$, i.e.~
$\langle 0|J_{\m5}^a|\eta^\a\rangle = i p_\m f^{a\a}$.
(We use the notation $\eta^\a$ to denote the physical mesons $\pi^0,\eta$
and $\eta'$, while the $SU(3)$ index $a = 3,8,0$.)
The mass matrix $\m^2_{\a\b}$ satisfies the Dashen, Gell-Mann--Oakes--Renner
(DGMOR) relation \cite{Gell-Mann:1968rz,Dashen:1969eg}
\begin{equation}
f^{a\a} \m^2_{\a\b} f^{T\b b} ~=~ - (M\Phi)_{ab} ~~~~~~~(\rm no~anomaly)
\label{eq:ccca}
\end{equation}
The consequences of the anomaly are determined by the interaction of the 
pseudoscalar fields $\phi_5^a$ with the topological charge density $Q$
and the subsequent mixing. This gives rise to an additional contribution
to the masses.  Moreover, we can no longer
identify the flavour singlet decay constant by the coupling 
to $J_{\m5}^0$ since this is not RG invariant. 
Instead, the physical decay constants $f^{a\a}$ are defined in terms 
of the couplings of the $\eta^\a$ to the pseudoscalar fields through the 
relation $f^{a\a}\la 0| \phi_5^b |\eta^\a\ra = d_{abc}\la\phi^c\ra$. 
This coincides with the usual definition except in the
flavour singlet case. 

The most transparent way to describe how all this works is to use an 
effective action $\C[Q,\phi_5^a]$ constructed to satisfy the
anomalous chiral Ward identities. It is important to emphasise from the
outset that this is an effective action
in the sense of section \ref{sec 2.1}, i.e.~the generating functional for
vertices which are 1PI with respect to the set of fields $Q,\phi_5^a$
only. The choice of fields is designed to capture the degrees of freedom 
essential for the dynamics.\footnote{Note especially
the frequently misunderstood point that the choice of fields in $\C$ 
is not required to be in any sense a complete set, nor does the 
restriction to a given set of fields constitute an approximation. 
Before imposing dynamical simplifications, the identities derived 
from $\C$ are {\it exact} - increasing the set of basis fields simply 
changes the definitions of the 1PI vertices.
The effective action considered here is therefore different from
the non-linear chiral Lagrangians incorporating 
the large-$N_c$ approach to the pseudoscalar mesons constructed by a
number of groups. See, for example, 
refs.~\cite{Rosenzweig:1979ay,DiVecchia:1980ve,DiVecchia:1980sq,
Kawarabayashi:1980dp,Herrera-Siklody:1996pm,Leutwyler:1997yr,Kaiser:2000gs}.}
A different choice (or linear combination)
redefines the physical meaning of the vertices, so it is important
that the final choice of fields in $\C$ results in vertices which are
most directly related to physical couplings.

The simplest effective action consistent with the anomalous Ward
identities and the renormalisation group is 
\begin{eqnarray}
&\C[Q,\phi_5^a] ~&=~ \int dx~\biggl({1\over2A}Q^2 ~+~ 
Q\bigl(\sqrt{2n_f}\d_{0a} - B_a \pl^2  \bigr)\Phi_{ab}^{-1}\phi_5^b 
~~~~~~~~~~~~~~~~~~~~~~~
\nonumber\\
&{}&~~~~~~~~~~~~~~~~~~~~~~~~~~~~
+~{1\over2} \phi_5^a \Phi^{-1}_{ac}\bigl((M\Phi)_{cd} - C_{cd}\pl^2 \bigr)
\Phi^{-1}_{db}\phi_5^b \biggr)
\label{eq:cca}
\end{eqnarray}
The constants $C_{ab}$ and $B_a$ are related to 
$\C_{V_{\m5}^a V_{\n5}^b}$ and $\C_{V_{\m5}^a Q}$ respectively.
The inclusion of the term with $B_a$ is unusual, but is required for
consistency with the RGEs derived from (\ref{eq:bzz}) beyond zero
momentum.

\vskip0.1cm
This form of $\C[Q,\phi_5^a]$ encodes three key dynamical assumptions:
\vskip0.1cm

\noindent $\bullet$~~Pole dominance. We assume that the Green functions  
are dominated by the contribution of single-particle poles associated
with the pseudoscalar mesons {\it including} the flavour singlet.
This extends the usual PCAC assumption to the singlet sector.

\noindent $\bullet$~~Smoothness. We assume that pole-free dynamical
quantities such as the decay constants and couplings (1PI vertices) 
are only weakly momentum-dependent in the range from $p=0$ to their
on-shell values. This allows us to impose relations derived from the
zero-momentum Ward identities, provided this is compatible with
the renormalisation group.

\noindent $\bullet$~~Topology. There must exist topologically non-trivial
fluctuations which can give a non-vanishing value to $\chi(0)|_{\rm YM}$
in pure gluodynamics. This is required to give the non-vanishing
coefficient in the all-important ${1\over 2A}Q^2$ term in $\C[Q,\phi_5^a]$.
Notice that we do not require a kinetic term for $Q$, which would be
associated with a (presumed heavy) pseudoscalar glueball.

\vskip0.1cm
The second derivatives of $\C[Q,\phi_5^a]$ are
\begin{eqnarray}
&\left(\matrix{\C_{QQ} &\C_{Q\phi_5^b}\cr 
&{}\cr
\C_{\phi_5^a Q} &\C_{\phi_5^a
\phi_5^b} \cr}\right) = 
\left(\matrix{A^{-1} 
&(\sqrt{2n_f}\d_{0d} + B_d p^2) \Phi_{db}^{-1} \cr
&{}\cr
\Phi_{ac}^{-1}(\sqrt{2n_f}\d_{c0} + B_c p^2)  
& \Phi^{-1}_{ac}\bigl((M\Phi)_{cd} + C_{cd}p^2\bigr)\Phi^{-1}_{db}\cr}
\right)
\nonumber\\
&{}
\nonumber\\
\label{eq:ccb}
\end{eqnarray}
The corresponding Green functions (composite operator propagators)
are given by inversion:
\begin{equation}
\left(\matrix{W_{\o\o} & W_{\o S_5^b}\cr
W_{S_5^a \o} & W_{S_5^a S_5^b} \cr}\right) ~~=~~ -
\left(\matrix{\C_{QQ} &\C_{Q\phi_5^b}\cr 
\C_{\phi_5^a Q} &\C_{\phi_5^a \phi_5^b} \cr}\right)^{-1}  
\label{eq:ccc}
\end{equation}
and we find, to leading order in $p^2$,
\begin{eqnarray}
&W_{\o\o} ~&=~ -A~ \tilde\D^{-1} 
\nonumber\\
&W_{\o S_5^b} ~&=~ W_{S_5^b \o} ~\simeq~ \sqrt{2n_f} A \D_{0d}^{-1}\Phi_{db}
\nonumber\\
&W_{S_5^a S_5^b} ~&=~ - \Phi_{ac}~ \D_{cd}^{-1}~ \Phi_{db}  
\label{eq:ccd}
\end{eqnarray}
where
\begin{equation}
\tilde \D ~=~ 1 - \Bigl(2n_f A \d_{a0}\d_{0b}
+ \sqrt{2n_f} A(\d_{a0}B_b + B_a \d_{0b})p^2\Bigr)
\bigr(M\Phi + C p^2\bigr)^{-1}_{ab}
\label{eq:cce}
\end{equation}
and
\begin{equation}
\D_{ab} ~=~ \Bigl(C_{ab} 
- \sqrt{2n_f} A (\d_{a0}B_b + B_a \d_{0b})\Bigr)p^2
+ (M\Phi)_{ab} - 2n_f A~\d_{a0}\d_{0b}~~~
\label{eq:ccf}
\end{equation}

In this form, however, the propagator matrix is not diagonal and
the operators are not normalised so as to couple with unit decay constants
to the physical states. It is therefore convenient to make a change of
variables in $\C$ so that it is written in terms of operators which are
more closely identified with the physical states. We do this is in two 
stages, since the intermediate stage allows us to make direct contact with 
the discussion in ref.\cite{Veneziano:1979ec} and will play an important 
role in some of the phenomenological applications considered later.

\vskip0.1cm
First, we define rescaled fields $\hat\eta^\a$ whose kinetic terms,
before mixing with $Q$, are canonically normalised. That is, we set
\begin{equation}
\hat\eta^\a ~=~  \hat f^{T\a a} \Phi_{ab}^{-1} \phi_5^b
\label{eq:ccg}
\end{equation}
with the `decay constants' $\hat f^{a\a}$ defined such that
${d\over dp^2}\C_{\hat\eta^\a \hat\eta^\b}|_{p=0} = \d_{\a\b}$.
This implies
\begin{equation}
(\hat f \hat f^T)_{ab}  ~=~ C_{ab} ~=~ 
{d\over dp^2} W_{S_D^a S_D^b}\Big|_{p=0} 
\label{eq:cch}
\end{equation}
where $D^a = \sqrt{2n_f}\d_{a0} Q + M_{ab}\phi_5^b$ 
is the divergence of the current $J_{\m5}^a$. In the chiral limit,
this reduces in the flavour singlet sector to 
\begin{equation}
(\hat f \hat f^T)_{00} ~=~ {d\over dp^2} \chi(p^2)\Big|_{p=0} ~=~ \chi'(0)
\label{eq:cci}
\end{equation}
a result which plays a vital role in understanding the `proton spin'
problem. Notice however that the $\hat f^{a\a}$ are {\it not} RG invariant:
in fact, $\DD \hat f^{a\a} = \c \d_{a0} \hat f^{a\a}$.
The effective action $\C[Q,\hat\eta^\a]$ is:
\begin{eqnarray}
&\C[Q,\hat\eta^a] ~&=~ \int dx~\biggl({1\over2A}Q^2 ~+~ 
Q \bigl(\sqrt{2n_f} \d_{0a} - B_a \pl^2 \bigr) 
(\hat f^{-1})^{a\a} \hat\eta^\a ~~~~~~~~~~~
\nonumber\\
&{}&~~~~~~~~~~~~~~~~~~~~~~~~~~~~
+~{1\over2} \hat\eta^\a\bigl(-\pl^2 
+ \hat f^{-1T}M\Phi \hat f^{-1} \bigr)_{\a\b}\hat\eta^\b \biggr)
\label{eq:ccj}
\end{eqnarray}
In this form, the $\hat\eta^\a$ are the canonically normalised fields
corresponding to the `would-be Nambu-Goldstone bosons' in the absence 
of the anomaly, before they acquire an additional anomaly-induced mass.
In the framework of the large-$N_c$ or OZI approximations, 
they would correspond to true Nambu-Goldstone bosons. The singlet 
$\hat\eta^0$ is what we have therefore referred to in our previous
papers as the `OZI boson' $\eta'_{OZI}$. 
As we see later, the naive current algebra relations hold when expressed 
in terms of the $\hat \eta^\a$ and $\hat f^{a\a}$, though these do
{\it not} correspond to physical states or decay constants.

\vskip0.1cm
The physical particle masses are identified with the poles in the 
two-point Green functions (\ref{eq:ccd}).  
We immediately see that due to mixing with the topological charge density 
$Q$, the physical pseudoscalar meson mass $m^2_{\a\b}$ is shifted from 
its original value. From the pole in eq.(\ref{eq:ccf}), we immediately find
\begin{equation}
f^{a\a} m_{\a\b}^2 f^{T\b b} = - (M\Phi)_{ab}
+ 2n_f A \d_{a0} \d_{b0}
\label{eq:cck}
\end{equation}
where we identify the physical, RG-invariant decay constants as
\begin{equation}
(f f^T)_{ab} ~=~ (\hat f \hat f^T)_{ab} 
- \sqrt{2n_f}A(\d_{a0} B_b + B_a \d_{0b})
\label{eq:cciii}
\end{equation}
Eq.(\ref{eq:cck}) is the key result. It generalises the original DGMOR 
relations (\ref{eq:ccca}) to the flavour-singlet sector with the anomaly 
properly incorporated and the renormalisation group constraints satisfied.
It represents a generalisation of the Witten-Veneziano mass formula which 
makes no direct reference to large-$N_c$ arguments but depends only on the 
three dynamical assumptions stated above \cite{Shore:1999tw}.

With this clarification of the distinction between the physical decay
constants $f^{a\a}$ and the RG non-invariant $\hat f^{a\a}$, we can 
rewrite eq.(\ref{eq:cce}) for the topological susceptibility 
$\chi(p^2) = W_{\o\o}(p^2)$ as
\begin{equation}
\chi(p^2) ~=~ -A \Bigl[1 - 
{\rm tr}\bigl((\hat f \hat f^T - f f^T)p^2 + 2n_f A {\bf 1}_{00}\bigr)
\bigl(\hat f \hat f^T p^2 + M\Phi\bigr)^{-1} \Bigr]^{-1}
\label{eq:ccl}
\end{equation}
It is clear that in the zero-momentum limit, this expression successfully 
reproduces eq.(\ref{eq:bv}) for $\chi(0)$. 
For one flavour, the formula simplifies to
\begin{equation}
\chi(p^2) ~=~ -A \bigl(\hat f \hat f^T p^2 + M\Phi\bigr)
\Bigl[f f^T p^2 + M\Phi + 2n_f A \Bigr]^{-1}~~~~~(n_f = 1)
\label{eq:ccm}
\end{equation}
showing clearly the pole at the shifted mass $m^2$ of eq.(\ref{eq:cck}). 
The occurrence of both $\hat f^{a\a}$ and $f^{a\a}$ in these expressions 
allows them to satisfy the RGE (\ref{eq:by}) for the topological 
susceptibility, which requires $\DD \chi(p^2) = O(p^2)$.

\vskip0.1cm
The second stage is to make a change of variable which diagonalises
the propagator matrix, so as to give the most direct possible relation 
between the operators and the physical states.
Choosing
\begin{eqnarray}
&G ~&=~Q - W_{\o S_5^a} W_{S_5^a S_5^b}^{-1} \phi_5^b~~   
~\simeq~ Q + \sqrt{2n_f} A \Phi_{0b}^{-1} \phi_5^b
\nonumber\\
&\eta^\a ~&=~ f^{T\a a} \Phi_{ab}^{-1} \phi_5^b
\label{eq:ccn}
\end{eqnarray}
defines the effective action $\C[G,\eta^\a]$ as
\begin{equation}
\C[G,\eta^\a] ~=~ \int dx~ \biggl({1\over 2A} G^2 ~+~
{1\over2} \eta^\a (-\pl^2 -m^2)_{\a\b} \eta^\b ~\biggr)  
\label{eq:cco}
\end{equation}
with $m^2_{\a\b}$ given by eq.(\ref{eq:cck}). The corresponding
propagators are
\begin{eqnarray}
&\langle 0|T~G~G|0\rangle &= -  A 
\nonumber\\
&\langle 0|T~ \eta^\a ~\eta^\b|0\rangle &= 
{-1\over p^2 - m_{\eta^\a}^2}\d^{\a\b} 
\label{eq:ccp}
\end{eqnarray}
where with no loss of generality we have taken $m^2_{\a\b}$ diagonal.

Notice also that the states mix in the complementary way to the
operators. In particular, the mixing for the states corresponding 
to eq.(\ref{eq:ccn}) for the fields $G$ and $\eta^\a$ is
\begin{eqnarray}
&|G\rangle ~&=~ |Q\rangle
\nonumber\\
&|\eta^\a\rangle ~&=~ (f^{-1})^{\a a}\bigl( \Phi_{ab} |\phi_5^b\rangle  -
\sqrt{2n_f} A \d_{a0}|Q\rangle \bigr)
\label{eq:ccq}
\end{eqnarray}
In this sense, we see that we can regard the physical $\eta'$ (and, with
$SU(3)$ breaking, the $\eta$) as an admixture of quark and gluon 
components, while the unphysical state $|G\rangle$ is purely gluonic.

An immediate corollary is the following relation, which we will
use repeatedly in deriving alternative forms of the current algebra 
identities for the pseudoscalar mesons:
\begin{equation}
\Phi_{ab}{\d\over\d\phi_5^b} ~=~ \hat f^{a\a}{\d\over\d\hat\eta^\a} ~=~
f^{a\a}{\d\over\d\eta^a} + \sqrt{2n_f} A \d_{a0} {\d\over\d G}
\label{eq:ccr}
\end{equation}

\vskip0.1cm
The formulation in terms of $\C[G,\eta^\a]$ is exactly what we need to
construct a simple `$U(1)_A$ PCAC' with which to interpret the 
low-energy phenomenology of the pseudoscalar mesons. We turn to this 
in the next section. 

Here, we focus on the intermediate formulation $\C[Q,\hat\eta^\a]$ 
in order to describe Veneziano's analysis of the $U(1)_A$ problem 
in the framework of the large-$N_c$ and topological expansions.
The starting point is the anomalous Ward identity (\ref{eq:br}) for the 
topological susceptibility:
\begin{equation}
n_f^2 \int dx~\langle 0|T~Q(x)~Q(0)|0\rangle ~=~
\int dx~m^a m^b\langle 0|T~\phi_5^a(x)~\phi_5^b(0)|0\rangle 
~+~ m^a \langle \phi^a\rangle
\label{eq:ccs}
\end{equation}
The essential problem is how to understand this relation in terms
of a spectral decomposition in the context of the $1/N_c$ expansion.

Assuming that $\chi(0)_{\rm YM} = -A + O(1/N_c)$ is non-vanishing at $O(1)$,
the l.h.s.~is $O(n_f^2)$ in leading order in $1/N_c$. On the other hand,
the r.h.s.~includes the condensate term of $O(n_f N_c m)$. 
To resolve this apparent paradox, we have to go beyond leading order 
in $1/N_c$ and consider the quark loop contributions which are included 
in the topological expansion. 
Although these are formally suppressed by powers of $(n_f/N_c)$, they
contain light intermediate states which can enhance the order of the
Green function. As we have seen above, these light states are just the
`OZI bosons' $|\hat\eta^\a\rangle$ with masses $\m^2_{\a\b}$ 
of $O(n_f m)$. 
Inserting these intermediate states, we therefore find that:
\begin{equation}
\chi(p^2) ~=~
\chi(p^2)|_{\rm YM} -\langle 0|Q|\hat\eta^\a\rangle
{1\over (p^2 - \m^2)_{\a\b}}\langle \hat\eta^\b|Q|0\rangle ~+~ \ldots 
\label{eq:cct}
\end{equation}
where the coupling $\langle 0|Q|\hat\eta^\a\rangle$ is $O(\sqrt{n_f/N_c})$.

Approximating $\chi(p^2)_{\rm YM} \sim -A$ (a low-momentum smoothness
assumption) and 
$\langle 0|Q|\hat\eta^\a\rangle \sim \sqrt{2n_f}A(f^{-1})^{0\a}$,
then summing the series of intermediate state contributions, we find
\begin{equation}
\chi(p^2) ~\simeq~ 
- {A\over 1 - 2n_f A \Bigl(f(p^2 - \m^2) f^{T}\Bigr)^{-1}_{00}}
\label{eq:cp}
\end{equation}
This expression reproduces eq.(13) of ref.\cite{Veneziano:1979ec}.
Clearly, it is dominated by the physical pseudoscalar pole
with anomaly-induced mass given by eq.(\ref{eq:cck}). It does not
completely recover our more precise expression (\ref{eq:ccl}) because
of the approximation for the coupling of $Q$ to the $|\hat \eta^\a\rangle$,
which misses the subtleties related to the introduction of $B_a$ in the
effective action $\C[Q,\hat\eta^\a]$ and the distinction of $\hat f^{a\a}$
and $f^{a\a}$. These are effects of higher order in $1/N_c$ but, as we have 
seen, they are necessary to establish full RG consistency and
will prove to be important for phenomenology.

To see how a term with the $O(n_f N_c m)$ dependence of the condensate
can arise in $n_f^2 \chi(0)$, notice from eq.(\ref{eq:cck}) that the physical 
pseudoscalar mass squared $m^2_{\eta^\a}$ has two contributions, 
the first of $O(m)$ from the conventional quark mass term and 
the new, anomaly-induced contribution of $O(n_f/N_c)$. If we are in a 
regime where the anomaly contribution dominates ($m < \L_{\rm QCD}/N_c$), 
then it follows that the above expression for $\chi(0)$ indeed becomes 
of $O(n_f^{-1}N_c m)$. 

The original Witten-Veneziano mass formula for the $\eta'$ is 
the large-$N_c$ limit of eq.(\ref{eq:cck}). In the chiral limit there is 
no flavour mixing and the singlet mass is given by
\begin{equation}
m_{\eta'}^2 ~=~ {1\over (f^{0\eta'})^2} 2n_f A ~=~
-{2n_f\over f_{\pi}^2} \chi(0)_{\rm YM} + O((n_f/N_c)^2)
\label{eq:cq}
\end{equation}
This formula provided the first link between the $\eta'$ mass and
gluon topology. For an alternative recent derivation in the context of
a $n_f/N_c$ expansion, see also ref.\cite{Giusti:2001xh}.

What we learn from all this is that the Green functions in the anomalous
chiral Ward identities admit a consistent spectral decomposition in
terms of a full set of $(n_f^2 - 1)$ pseudoscalar mesons, provided
they satisfy the generalised DGMOR mass formula (\ref{eq:cck}) 
{\it including} the all-important anomaly term. The presence of these
light poles can enhance the apparent order of the Green functions,
as is familiar with Nambu-Goldstone bosons, and the anomaly-induced
$O(n_f/N_c)$ contribution to $m_{\eta^\a}^2$ is critical in ensuring
complete consistency with the Ward identities.

Similar considerations apply to the resolution of apparent paradoxes
in the $\theta$-dependence of some Green functions. For example
\cite{Veneziano:1979ec}, we can show from the anomalous Ward identities 
that the condensate satisfies
\begin{equation}
\sum_q m_q\langle\bar q q\rangle|_{\theta} ~\equiv~
m^a \langle \phi^a \rangle ~=~
\cos(\theta/n_f)~ m^a \langle \phi^a\rangle|_{\theta =0}
\label{eq:cr}
\end{equation}
This implies
\begin{eqnarray}
&{\pl^2\over\pl\theta^2}{} m^a \langle \phi^a\rangle|_{\theta =0} ~&=~
- m^a \int dx \int dy~ \langle 0|T~Q(x)~Q(y)~\phi^a(0)|0\rangle 
\nonumber\\
&{}~&=~
-{1\over n_f^2}~ m^a \langle \phi^a \rangle|_{\theta =0}
\label{eq:cs}
\end{eqnarray}
Here, the Green function is superficially of $O(n_f/N_c)$ while the
r.h.s.~is $O(N_c/n_f)$. The resolution is simply that it contains 
pseudoscalar intermediate states contributing two light poles with
$m^2 \sim O(n_f/N_c)$. So once again we see how the spectral decomposition
in terms of the full set of pseudoscalar mesons, including the flavour
singlet, ensures consistency with the anomalous Ward identities.

\section{Pseudoscalar mesons}
\label{mesons}

This theoretical analysis provides the basis for an extension of the
conventional PCAC or chiral Lagrangian description of the phenomenology
of the pseudoscalar mesons to the flavour singlet sector. 
In this section\footnote{This section is based on 
the presentation in ref.\cite{Shore:2006mm}, where we extend and update
our original work \cite{Shore:1999tw,Shore:2001cs}
to include a detailed comparison with experimental data.} 
we describe the role of the $U(1)_A$ anomaly in the radiative decays
of $\pi^0, \eta$ and $\eta'$ and derive the $U(1)_A$ Goldberger-Treiman 
relation, first proposed by Veneziano as a resolution of the 
`proton spin' problem.

\subsection{$U(1)_A$ Dashen, Gell-Mann--Oakes--Renner relations}

The extension of the DGMOR relations to the $U(1)_A$ sector follows
from the application of the three key dynamical assumptions used
above (viz.~pole dominance by the nonet of pseudoscalar mesons, 
smoothness of decay constants and couplings over the range from 
zero to on-shell momentum, and the existence of topologically 
non-trivial gluon dynamics) to the anomalous chiral Ward identities.

The fundamental $U(1)_A$ DGMOR relation
\begin{equation}
f^{a\a} m_{\a\b}^2 f^{T\b b} = -M_{ac} \Phi_{cb}
+ 2n_f A \d_{a0} \d_{b0}
\label{eq:daaa}
\end{equation} 
has been derived above in the course of the general discussion of the
$U(1)_A$ problem. In order to make this section self-contained, we give a 
brief and direct derivation here. 

Recall that the physical meson fields are given as
$\eta^\a = f^{T\a a} \Phi_{ab}^{-1} \phi_5^b$, with the decay constants
defined so that the propagator 
$W_{S_\eta^\a S_\eta^\b} = -1/(p^2 - m_{\eta}^2)_{\a\b}$.
It follows immediately that at zero momentum,
\begin{equation}
f^{a\a} m_{\a\b}^2 f^{T\b b} = \Phi_{ac} (W_{S_5 S_5})_{cd}^{-1} \Phi_{db}
\label{eq:dbbb}
\end{equation}
Using the chiral Ward identities of section 2 together with the
identification (\ref{eq:bu}) of the topological susceptibility,
we can then show
\begin{eqnarray}
&\Phi_{ac} (W_{S_5 S_5})_{cd}^{-1} \Phi_{db}
&= (\Phi M)_{ac} \bigl(M W_{S_5 S_5} M\bigr)_{cd}^{-1} (M\Phi)_{db} 
\nonumber\\
&{}&= (M\Phi)_{ac} \Bigl(-(M\Phi) + 2n_f \chi(0) {\bf 1}_{00}
\Bigr)_{cd}^{-1} (M\Phi)_{db} 
\nonumber\\
&{}&= -(M\Phi)_{ab} + 2n_f \C_{QQ}^{-1} ~\d_{a0} \d_{b0} 
\label{eq:dccc}
\end{eqnarray}
proving the result (\ref{eq:daaa}).

Expanding this out, and assuming
the mixed decay constants $f^{0\pi}, f^{8\pi}, f^{3\eta}, f^{3\eta'}$
are all negligible, we have
\begin{eqnarray}
&(f^{0\eta'})^2 m_{\eta'}^2 + (f^{0\eta})^2 m_\eta^2 ~&=~ 
- {2\over3}\bigl(m_u \la\bar u u\ra + m_d \la\bar d d\ra 
+ m_s \la\bar s s\ra \bigr)  + 6 A ~~~~~~
\label{eq:da}\\
\nonumber\\
&f^{0\eta'} f^{8\eta'} m_\eta'^2 + f^{0\eta} f^{8\eta} m_{\eta}^2 ~&=~ 
- {\sqrt2\over3} \bigl(m_u \la\bar u u\ra + m_d \la\bar d d\ra 
- 2 m_s \la\bar s s\ra \bigr)
\label{eq:db}\\
\nonumber\\
&(f^{8\eta'})^2 m_\eta'^2 + (f^{8\eta})^2 m_{\eta}^2 ~&=~ 
-{1\over3}\bigl(m_u \la\bar u u\ra + m_d \la\bar d d\ra +  
4 m_s \la\bar s s\ra \bigr)
\label{eq:dc}\\
\nonumber\\
&~~~~~~~~~~~~~~~~~~~~~~~
f_\pi^2 m_\pi^2 ~&=~ - (m_u \la \bar u u \ra  + m_d \la \bar d d \ra)
\label{eq:dd}
\end{eqnarray}
and we can add the standard DGMOR relation for the $K^+$,
\begin{equation}
f_K^2 m_K^2 ~=~ - (m_u \la \bar u u \ra  + m_s \la \bar s s \ra)~~~~
\label{eq:de}
\end{equation}

We emphasise that these formulae, as well as the radiative decay
and $U(1)_A$ Goldberger-Treiman relations derived below, do not
depend at all on the $1/N_c$ expansion. In particular, the constant
$A$ appearing in the flavour singlet formula is defined as the
non-perturbative parameter determining the topological susceptibility
$\chi(0)$ in QCD according to the exact identity (\ref{eq:bv}). 
As explained above, large-$N_c$ ideas do indeed provide a rationale for 
extending the familiar PCAC assumptions of pole dominance and smoothness 
to the flavour singlet channel, but these assumptions can be tested 
independently against experimental data. 

The most useful form of these relations for phenomenology is to assume
exact $SU(2)$ flavour symmetry and eliminate the quark masses and
condensates in favour of $f_\pi, f_K, m_{\pi}^2$ and $m_K^2$ in the
DGMOR relations for the $\eta$ and $\eta'$. This gives
\begin{eqnarray}
&(f^{0\eta'})^2 m_{\eta'}^2 + (f^{0\eta})^2 m_\eta^2 ~&=~ 
{1\over3} \bigl(f_\pi^2 m_\pi^2 + 2 f_K^2 m_K^2\bigr) + 6 A 
\label{eq:df}\\
\nonumber\\
&f^{0\eta'} f^{8\eta'} m_\eta'^2 + f^{0\eta} f^{8\eta} m_{\eta}^2 ~&=~ 
{2\sqrt2\over3}\bigl(f_\pi^2 m_\pi^2 - f_K^2 m_K^2\bigr)
\label{eq:dg}\\
\nonumber\\
&(f^{8\eta'})^2 m_\eta'^2 + (f^{8\eta})^2 m_{\eta}^2 ~&=~ 
-{1\over3}\bigl(f_\pi^2 m_\pi^2 - 4 f_K^2 m_K^2\bigr)
\label{eq:dh}\\
\nonumber
\end{eqnarray}

We can also now clarify the precise relation of these results to
the Witten-Veneziano formula for the mass of the $\eta'$ in its
non-vanishing quark mass form, viz.
\begin{equation}
m_{\eta'}^2 + m_{\eta}^2 - 2 m_K^2 ~=~ - {6\over f_\pi^2} \chi(0)|_{YM}
\label{eq:di}
\end{equation}
Of course, only the $m_{\eta'}^2$ term on the l.h.s.~is present in the 
chiral limit. Substituting in the explicit values for the masses in this 
formula gives a prediction \cite{Veneziano:1979ec} for the topological
susceptibility, $\chi(0) \simeq -(180 ~{\rm MeV})^4$, which as we see
below is remarkably close to the subsequently calculated lattice result.

If we now add the DGMOR formulae (\ref{eq:df}) and (\ref{eq:dh}), we find
\begin{equation}
(f^{0\eta'})^2 m_{\eta'}^2 + (f^{0\eta})^2 m_\eta^2 +
(f^{8\eta})^2 m_\eta^2 + (f^{8\eta'})^2 m_{\eta'}^2 - 2 f_K^2 m_K^2 
~=~ 6A
\label{eq:dj}
\end{equation}
which we repeat is valid to all orders in $1/N_c$.
To reduce this to its Witten-Veneziano approximation, we impose the 
large-$N_c$ limit to approximate the QCD topological charge parameter 
$A$ with $-\chi(0)|_{YM}$ as explained in section 2.4. We then set the
`mixed' decay constants $f^{0\eta}$ and $f^{8\eta'}$ to zero
and all the remaining decay constants $f^{0\eta'}, f^{8\eta}$ and
$f_K$ equal to $f_\pi$. With these approximations, we recover 
eq.(\ref{eq:di}).  Eventually, after we have found explicit experimental 
values for all these quantities, we will be able to demonstrate 
quantitatively how good an approximation the large-$N_c$ Witten-Veneziano 
formula is to the generalised $U(1)_A$ DGMOR relation in full QCD.

\subsection{Radiative decay formulae for $\pi^0,\eta,\eta' \rta \c\c$}
\label{radiative}

Radiative decays of the pseudoscalar mesons are of particular interest
as they are controlled by the electromagnetic $U(1)_A$ anomaly,
\begin{equation}
\pl^\m J_{\m 5}^a - M_{ab} \phi_5^b - \sqrt{2n_f} Q \d_{a0} - a_{\rm em}^a
{\a\over8\pi} F^{\m\n} \tilde F_{\m\n} ~=~ 0
\label{eq:dk}
\end{equation}
where $F_{\m\n} = \pl_\m A_\n - \pl_\n A_\m$ is the usual electromagnetic
field strength and the anomaly coefficients $a_{\rm em}^a$ are determined
by the quark charges. The generating functional 
$\C[V_{\m 5}^a, V_\m^a, Q, \phi_5^a,
\phi^a, A_\m]$ of 1PI vertices including the photon satisfies
the Ward identity
\begin{eqnarray}
&\pl_\m \C_{V_{\m5}^a} &- \sqrt{2n_f} \d_{a0} Q 
- a_{\rm em}^a {\a\over8\pi} F^{\m\n} \tilde F_{\m\n}
- d_{abc}m^b \phi_5^c 
\nonumber\\ 
&{}&+ f_{abc} V_{\m}^b \C_{V_{\m5}^c}
+ f_{abc} V_{\m5}^b \C_{V_{\m}^c}
- d_{abc} \phi_5^c \C_{\phi^b}  
+ d_{abc} \phi^c \C_{\phi_5^b} = 0
\label{eq:dl}
\end{eqnarray}

To derive the radiative decay formulae, we first differentiate
this identity with respect to the photon field $A_\m$. This gives
\begin{equation}
ip_\m \C_{V_{\m 5}^a A^\l A^\r} ~+~\Phi_{ab} \C_{\phi_5^b A^\l A^\r}
~=~ - a_{\rm em}^a {\a\over\pi}~ \e_{\m\n\l\r} k_1^\m k_2^\n
\label{eq:dm}
\end{equation}
where $k_1, k_2$ are the momenta of the two photons. Notice that the mass
term does not contribute directly to this formula. From its definition
as 1PI w.r.t.~the pseudoscalar fields, the vertex 
$\C_{V_{\m 5}^a A^\l A^\r}$ does not have a pole at $p^2 = 0$, even in the
massless limit, so we find simply
\begin{equation}
\Phi_{ab} \C_{\phi_5^b A^\l A^\r}\bigl|_{p=0}
~=~ - a_{\rm em}^a {\a\over\pi}~ \e_{\m\n\l\r} k_1^\m k_2^\n
\label{eq:dn}
\end{equation}

The radiative couplings $g_{\eta^\a \c\c}$ for the physical mesons
$\eta^\a = \pi^0,\eta,\eta'$ are defined as usual from the decay
amplitude $\langle \c\c|\eta^\a\rangle$. With the PCAC assumptions 
already discussed, they can be identified with the 1PI vertices as follows:
\begin{equation}
\langle \c\c|\eta^\a\rangle ~=~ -i g_{\eta^\a \c\c}~\e_{\m\n\l\r}
k_1^\m k_2^\n \e^\l (k_1) \e^\r (k_2) ~=~
i \C_{\eta^\a A_\l A_\r}  \e^\l (k_1) \e^\r (k_2)
\label{eq:do}
\end{equation}
Re-expressing eq.(\ref{eq:dn}) in terms of the canonically normalised
`OZI bosons' $\hat\eta^\a$, we therefore have the first form of the
decay formula,
\begin{equation}
\hat f^{a\a} g_{\hat\eta^\a \c\c} ~=~ a_{\rm em}^a {\a\over\pi}
\label{eq:dp}
\end{equation}

Then, rewriting this in terms of the physical pseudoscalar couplings
$g_{\eta^\a \c\c}$ and decay constants according to the
relation (\ref{eq:ccr}) gives the final form for 
the generalised $U(1)_A$ PCAC formula describing 
radiative pseudoscalar decays, incorporating both the electromagnetic 
and colour anomalies:
\begin{equation}
f^{a\a} g_{\eta^\a \c\c} + \sqrt{2n_f} A g_{G\c\c} \d_{a0} ~=~ 
a_{\rm em}^a {\a\over\pi}
\label{eq:dr}
\end{equation}
Expanding this formula, we have
\begin{eqnarray}
f^{0\eta'} g_{\eta'\c\c} + f^{0\eta} g_{\eta\c\c} + {\sqrt6} A g_{G\c\c} 
~=~ a_{\rm em}^0 {\a\over\pi}
\label{eq:ds}\\
\nonumber\\
f^{8\eta'}g_{\eta'\c\c} + f^{8\eta} g_{\eta\c\c} 
~=~ a_{\rm em}^8 {\a\over\pi}~~~~~~~~~~~~~~~~
\label{eq:dt}\\ 
\nonumber\\
f_\pi g_{\pi\c\c} ~=~ a_{\rm em}^3 {\a\over\pi}
~~~~~~~~~~~~~~~~~~~~~~~~~~~~~~~
\label{eq:du}
\end{eqnarray}
where $a_{\rm em}^0 = {2\sqrt2\over3\sqrt3}N_c$,   
$a_{\rm em}^8 = {1\over3\sqrt3}N_c$ and $a_{\rm em}^3 = {1\over3}N_c$.

The new element in the flavour singlet decay formula is the gluonic
coupling parameter $g_{G\c\c}$. It takes account of the fact that because
of the anomaly-induced mixing with the gluon topological density $Q$,
the physical $\eta'$ is not a true Nambu-Goldstone boson so the naive PCAC
formulae must be modified.
$g_{G\c\c}$ is {\it not} a physical coupling and must be regarded as
an extra parameter to be fitted to data, although in view of the 
identifications in eq.(\ref{eq:ccq}) it may reasonably be thought of as the
coupling of the photons to the gluonic component of the $\eta'$.

The renormalisation group properties of these relations are readily 
derived from the RGE (\ref{eq:bzz}) for $\C$. In the `OZI boson' form,
the unphysical coupling $g_{\hat\eta^\a \c\c}$ satisfies the complementary
RGE to the decay constant $\hat f^{a\a}$ so the combination is RG invariant:
\begin{equation}
\DD \hat f^{a\a} = \c \d_{a0} \hat f^{a\a} ~~~~~~~~~~~~~
\DD \bigl(\hat f^{a\a} g_{\hat\eta^\a \c\c}\bigr) = 0
\label{eq:duuu}
\end{equation}
In contrast, {\it all} the decay constants and couplings in the relation
(\ref{eq:dr}) can be shown to be separately RG invariant, including
the gluonic coupling $g_{G\c\c}$ \cite{Shore:1991pn,Shore:2001cs}.

\subsection{The renormalisation group, OZI and $1/N_c$~: a conjecture}
\label{sec:conj}

Although these $U(1)_A$ PCAC relations have
been derived purely on the basis of the pole dominance and smoothness
assumptions, we will nevertheless find it useful in practical applications 
to exploit their OZI or large-$N_c$ behaviour, in conjunction with 
the renormalisation group.

The basic idea is that violations of the OZI rule, or equivalently
anomalous large-$N_c$ behaviour, are generally related to the existence of 
the $U(1)_A$ anomaly. Moreover, we can identify the quantities which 
will be particularly sensitive to the anomaly as
those which have RGEs involving the anomalous dimension $\c$. 
We therefore conjecture that the dependence of Green functions and
1PI vertices on $\c$ will be an important guide 
in identifying propagators and couplings which are likely to show violations 
of the OZI rule and those for which the OZI (or large-$N_c$) limit 
should be a good approximation \cite{Shore:1991dv,Shore:1991pn}. 

As an example, the large-$N_c$ order of the quantities in the flavour
singlet decay relation (\ref{eq:ds}) is as follows: 
~$f^{a\a} = O(\sqrt{N_c})$
for all the decay constants, $g_{\eta^\a\c\c} = O(\sqrt{N_c})$,
$g_{G\c\c} = O(1)$, $a_{\rm em}^a = O(N_c)$ and the topological 
susceptibility parameter $A = O(1)$. 
The renormalisation group behaviour is especially simple, with both
the meson and gluonic couplings $g_{\eta^\a \c\c}$ and $g_{G\c\c}$ 
as well as the decay constants being RG invariant.  
Putting this together, we find that all the terms in the decay formula 
are of $O(N_c)$ except the anomalous contribution $A g_{G\c\c}$ 
which is $O(1)$. Since it is RG invariant and independent of the
anomalous dimension $\c$, we conjecture that it is a quantity for
which the OZI (or large-$N_c$) approximation should be reliable
so we expect it to be numerically small compared with the other
contributions. In the next section, we test this against experiment.

As we shall see later, this conjecture has far-reaching implications
for a range of predictions related to the anomaly, particularly
in the interpretation of the $U(1)_A$ Goldberger-Treiman relation
and associated ideas on the first moment sum rules for $g_1^p$
and $g_1^\c$ in deep-inelastic scattering.

\subsection{Phenomenology}

After all this theoretical development, we finally turn to experiment
and use the data on the radiative decays $\eta,\eta' \rta \c\c$ to 
deduce values for the pseudoscalar meson decay constants $f^{0\eta'}$, 
$f^{0\eta}$, $f^{8\eta'}$ and $f^{8\eta}$ from the set of decay formulae 
(\ref{eq:ds}), (\ref{eq:dt})
and $U(1)_A$ DGMOR relations (\ref{eq:df})-(\ref{eq:dh}). 
We will also find the value of the unphysical coupling parameter $g_{G\c\c}$ 
and test the realisation of the $1/N_c$ expansion in real QCD.

The two-photon decay widths are given by
\begin{equation}
\C\bigl(\eta'(\eta)\rta\c\c\bigr) ~=~ {m_{\eta'(\eta)}^3 \over64\pi~}
|g_{\eta'(\eta)\c\c}|^2
\label{eq:dv}
\end{equation}
The current experimental data, quoted in the Particle Data Group
tables \cite{PDG}, are
\begin{equation}
\C(\eta'\rta\c\c) ~=~ 4.28 \pm 0.19 ~{\rm KeV}
\label{eq:dw}
\end{equation}
which is dominated by the 1998 L3 data \cite{L3} on the two-photon
formation of the $\eta'$ in $e^+ e^- \rta e^+ e^- \pi^+ \pi^- \c$,
and
\begin{equation}
\C(\eta\rta\c\c) ~=~ 0.510 \pm 0.026 ~{\rm KeV}
\label{eq:dx}
\end{equation}
which arises principally from the 1988 Crystal Ball \cite{Crystal} and
1990 ASP \cite{ASP} results on $e^+ e^- \rta e^+ e^- \eta$.
From this data, we deduce the following results for the couplings 
$g_{\eta'\c\c}$ and $g_{\eta\c\c}$:
\begin{equation}
g_{\eta'\c\c} ~=~ 0.031 \pm 0.001 ~{\rm GeV}^{-1}
\label{eq:dy}
\end{equation}
and
\begin{equation}
g_{\eta\c\c} ~=~ 0.025 \pm 0.001 ~{\rm GeV}^{-1}
\label{eq:dz}
\end{equation}
which may be compared with $g_{\pi\c\c} = 0.024 \pm 0.001 ~{\rm GeV}$.

We also require the pseudoscalar meson masses:
\begin{eqnarray}
m_{\eta'} ~=~ 957.78 \pm 0.14 ~{\rm MeV} ~~~~~~~~ 
m_{\eta}  ~=~ 547.30 \pm 0.12 ~{\rm MeV} \nonumber\\
m_K       ~=~ 493.68 \pm 0.02 ~{\rm MeV} ~~~~~~~~
m_\pi     ~=~ 139.57 \pm 0.00 ~{\rm MeV}
\label{eq:daa}
\end{eqnarray}
and the decay constants $f_\pi$ and $f_K$. These are defined in the
standard way, so we take the following values (in our normalisations) 
from the PDG \cite{PDG}:
\begin{equation}
f_K ~=~ 113.00 \pm 1.03 ~{\rm MeV} ~~~~~~~~
f_\pi ~=~ 92.42 \pm 0.26 ~{\rm MeV}
\label{eq:dbb}
\end{equation}
giving $f_K/f_\pi = 1.223 \pm 0.012$.

\begin{figure}
\centering
\includegraphics[height=3.3cm]{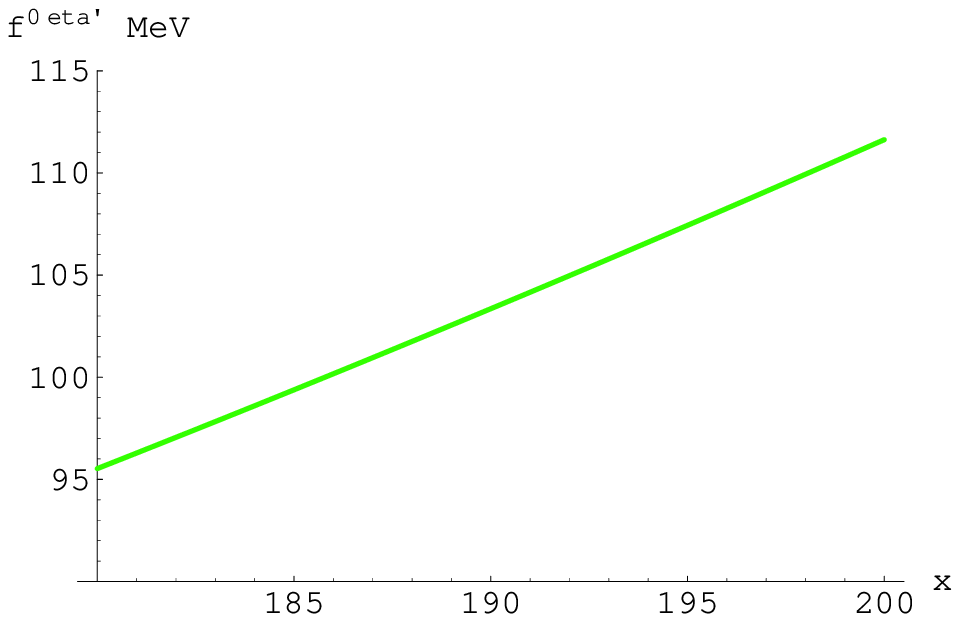} \hskip0.7cm 
\includegraphics[height=3.3cm]{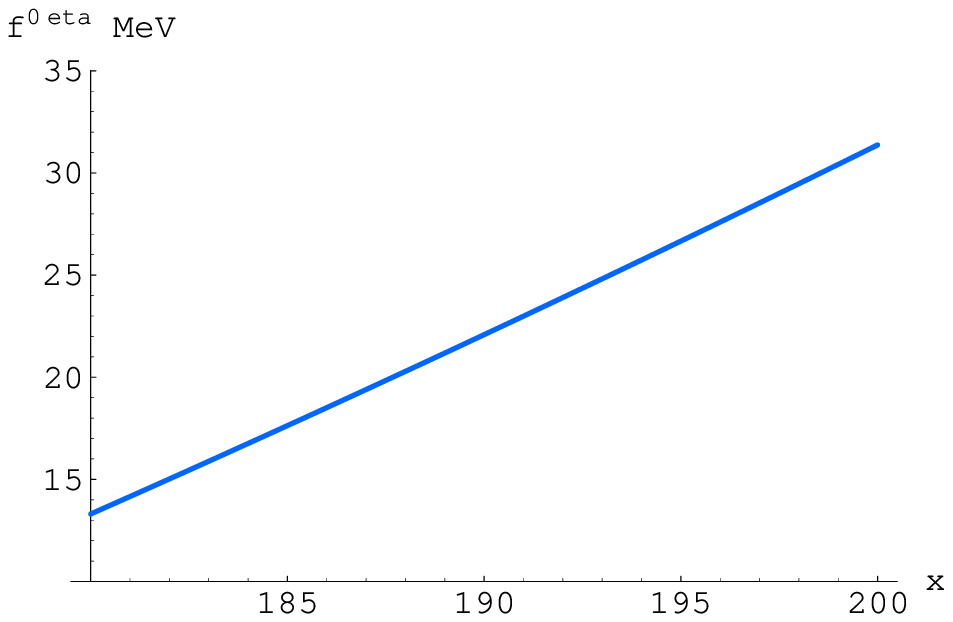}
\caption{The decay constants $f^{0\eta'}$ and $f^{0\eta}$ as functions of 
the non-perturbative parameter $A = (x~{\rm MeV})^4$ which determines the
topological susceptibility in QCD.}
\label{fig:feta}     
\end{figure}

\begin{figure}
\centering
\includegraphics[height=4.5cm]{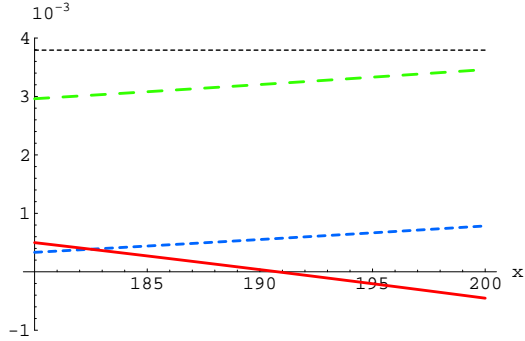}
\caption{This shows the relative sizes of the contributions to the flavour 
singlet radiative decay formula (\ref{eq:ds}) expressed as functions of the
topological susceptibility parameter $A = (x~{\rm MeV})^4$.  The dotted 
(black) line denotes ${2\sqrt2\over\sqrt3} {\a_{\rm em}\over\pi}$. The
dominant contribution comes from the term $f^{0\eta'} g_{\eta'\c\c}$, denoted
by the long-dashed (green) line, while the short-dashed (blue) line denotes
$f^{0\eta} g_{\eta\c\c}$. The contribution from the gluonic coupling,
${\sqrt6} A g_{G\c\c}$, is shown by the solid (red) line.}
\label{fig:twogamma}     
\end{figure}

The octet decay constants $f^{8\eta}$ and $f^{8\eta'}$ are obtained from
eqs.(\ref{eq:dh}) and (\ref{eq:dt}). This leaves three remaining equations
which determine the singlet decay constants $f^{0\eta'}, f^{0\eta}$ 
and the gluonic coupling $g_{G\c\c}$ in terms of the QCD topological
susceptibility parameter $A$. This dependence is plotted in 
Figs.~\ref{fig:feta} and \ref{fig:twogamma}.

To make a definite prediction, we need a theoretical input value for the
topological susceptibility. In time, lattice calculations in full QCD with
dynamical fermions should be able to determine the parameter $A$. For the
moment, however, only the topological susceptibility in pure Yang-Mills
theory is known accurately. The most recent value \cite{DelDebbio:2004ns}
is 
\begin{equation}
\chi(0)|_{YM} ~=~ -(191 \pm 5 ~{\rm MeV})^4 ~=~ 
- (1.33 \pm 0.14)\times 10^{-3} ~{\rm GeV}^4
\label{eq:dcc}
\end{equation}
This supersedes the original value $\chi(0)|_{YM} \simeq -(180~{\rm MeV})^4$
obtained some time ago \cite{DiGiacomo:1990ij}. 
Similar estimates are also obtained using QCD spectral sum rule 
methods \cite{Narison:1990cz}. At this point, therefore, we have to make
an approximation and so we assume that the $O(1/N_c)$ corrections in the 
identification
\begin{equation}
A ~=~ \chi(0)_{\rm YM} + O(1/N_c)
\label{eq:ddd}
\end{equation}
are numerically small. With this provisional input for $A$, we can
then determine the full set of decay constants:
\begin{eqnarray}
{}\nonumber\\
f^{0\eta'} ~=~ 104.2 \pm 4.0 ~{\rm MeV} ~~~~~~~~~~
f^{0\eta}  ~=~ 22.8 \pm 5.7 ~{\rm MeV} \nonumber\\
f^{8\eta'} ~=~-36.1 \pm 1.2 ~{\rm MeV} ~~~~~~~~
f^{8\eta}  ~=~ 98.4 \pm 1.4 ~{\rm MeV} 
\label{eq:dee}
\end{eqnarray}
and
\begin{equation}
g_{G\c\c} ~=~ - 0.001 \pm 0.072 ~{\rm GeV}^{-4}
\label{eq:dff}
\end{equation}
It is striking how close both the diagonal decay constants $f^{0\eta'}$
and $f^{8\eta}$ are to $f_{\pi}$. Predictably, the off-diagonal ones
$f^{0\eta'}$ and $f^{8\eta'}$ are strongly suppressed.

It is also useful to quote these results in the two-angle parametrisation
normally used in phenomenology. Defining,
\begin{equation}
\left(\matrix{f^{0\eta'} &f^{0\eta}\cr f^{8\eta'} &f^{8\eta}}\right)
~~=~~\left(\matrix{f_0 \cos\theta_0 &-f_0 \sin\theta_0\cr
f_8 \sin\theta_8 &f_8 \cos\theta_8}\right)
\label{eq:dgg}
\end{equation}
we find
\begin{eqnarray}
f_0 ~=~ 106.6 \pm 4.2 ~{\rm MeV} ~~~~~~~~
f_8 ~=~ 104.8 \pm 1.3 ~{\rm MeV} \nonumber\\
\theta_0 ~=~ -12.3 \pm 3.0 ~{\rm deg} ~~~~~~~~
\theta_8  ~=~ -20.1 \pm 0.7 ~{\rm deg}
\label{eq:dhh}
\end{eqnarray}
that is 
\begin{equation}
{f_0\over f_\pi} ~=~ 1.15 \pm 0.05 ~~~~~~~~
{f_8\over f_\pi} ~=~ 1.13 \pm 0.02
\label{eq:dii}
\end{equation}

Given these results, we can now investigate how closely our expectations 
based on OZI or $1/N_c$ reasoning are actually realised by the experimental 
data. With the input value (\ref{eq:dcc}) for $A$, the numerical magnitudes 
and $1/N_c$ orders of the terms in the flavour singlet decay relation are
as follows (see Fig.~\ref{fig:twogamma}):
\begin{eqnarray}
f^{0\eta'} g_{\eta'\c\c} ~[N_c;~3.23]
+ f^{0\eta} g_{\eta\c\c} ~[N_c;~0.57]
+ {\sqrt6} A g_{G\c\c} ~[1;~-0.005 \pm 0.23] 
\nonumber\\
=~~a_{\rm em}^0 {\a_{\rm em}\over\pi} ~[N_c;~3.79]
~~~~~~~~~~~~~~~~~~~~~~~~~~~~~~~~~~~~~~~~~~~~~~~~~~~~~~~~~~~~~~~~
\label{eq:djj}
\end{eqnarray}
The important point is that the gluonic contribution $g_{G\c\c}$, which is
suppressed by a power of $1/N_c$ compared to the others, is also 
experimentally small.
The near-vanishing for the chosen value of $A$ is presumably a coincidence,
but we see from Fig.~\ref{fig:twogamma} that across a reasonable
range of values of the topological susceptibility it is still contributing
no more than around $10\%$, in line with our expectations for a RG-invariant,
OZI-suppressed quantity. 

It is also interesting to see how the $1/N_c$ approximation is realised
in the $U(1)_A$ DGMOR generalisation (\ref{eq:dj}) of the Witten-Veneziano 
formula (\ref{eq:di}). Here we find
\begin{eqnarray}
(f^{0\eta'})^2 m_{\eta'}^2 ~[N_c;~9.96] 
~+~ (f^{0\eta})^2 m_\eta^2 ~[N_c;~0.15] 
~+~(f^{8\eta'})^2 m_\eta'^2 ~[N_c;~1.19]
\nonumber\\
~+~ (f^{8\eta})^2 m_{\eta}^2 ~[N_c;~2.90]
- 2 f_K^2 m_K^2 [N_c;-6.22]
~=~ 6A ~[1;~7.98] 
\nonumber\\
\label{eq:dkk}
\end{eqnarray}
This confirms the picture that the anomaly-induced contribution
of $O(1/N_c)$ to $m_{\eta'}^2$, which gives a sub-leading $O(1)$  
effect in $(f^{0\eta'})^2 m_{\eta'}^2$, is in fact numerically dominant 
and matched by the $O(1)$ topological susceptibility term $6A$. 
Away from the chiral limit, the conventional non-anomalous terms are
all of $O(N_c)$ and balance as expected. 
The surprising numerical accuracy of the Witten-Veneziano formula 
(\ref{eq:br}) is seen to be in part due to a 
cancellation between the underestimates of $f^{8\eta'}$ (taken to be 0) 
and $f_K$ (set equal to $f_\pi$).
This emphasises, however, that great care must be taken in using the
formal order in the $1/N_c$ expansion as a guide to the numerical importance
of a physical quantity, especially in the $U(1)_A$ channel.

Nevertheless, the fact that the RG-invariant, OZI-suppressed coupling
$g_{G\c\c}$ is experimentally small is a very encouraging result.
It increases our confidence that we are able to identify quantities 
where the OZI, or leading $1/N_c$, approximation is likely to be 
numerically good. It also shows that $g_{G\c\c}$ gives a contribution 
to the decay formula which is entirely consistent with its picturesque 
interpretation as the coupling of the photons to the anomaly-induced 
gluonic component of the $\eta'$.
{\it A posteriori}, the fact that its contribution is at most $10\%$
explains the general success of previous theoretically inconsistent
phenomenological parametrisations of $\eta'$ decays in which the
naive current algebra formulae omitting the gluonic term are used.

However, while the flavour singlet decay formula is well-defined and 
theoretically consistent, it is necessarily non-predictive. To be genuinely
useful, we would need to find another process in which the same coupling
enters. The problem here is that, unlike the decay constants which are
universal, the coupling $g_{G\c\c}$ is process-specific just like
$g_{\eta'\c\c}$ or $g_{\eta\c\c}$. There are of course many other 
processes to which our methods may be applied such as $\eta'(\eta)\rta V\c$,
where $V$ is a flavour singlet vector meson $\rho,\omega,\phi$, or $\eta'
(\eta)\rta \pi^+ \pi^- \c$. The required flavour singlet formulae may readily
be written down, generalising the naive PCAC formulae. However, each will 
introduce its own gluonic coupling, such as $g_{GV\c}$. Although strict
predictivity is lost, our experience with the two-photon decays suggests that
these extra couplings will give relatively small, at most $O(10-20\%$), 
contributions if like $g_{G\c\c}$ they can be identified as RG invariant 
and $1/N_c$ suppressed. This observation restores at least a reasonable 
degree of predictivity to the use of PCAC methods in the $U(1)_A$ sector.

\subsection{$U(1)_A$ Goldberger-Treiman relation}

A further classic application of PCAC is to the pseudoscalar couplings of
the nucleon. For the pion, the relation between the axial-vector form 
factor of the nucleon and the pion-nucleon coupling $g_{\pi NN}$ is the
famous Goldberger-Treiman relation. Here, we present its generalisation 
to the flavour singlet sector, which involves the anomaly and gluon 
topology. This $U(1)_A$ Goldberger-Treiman relation was first proposed
by Veneziano \cite{Veneziano:1989ei} in an investigation of 
the `proton-spin' problem and further developed in 
refs.\cite{Shore:1990zu,Shore:1991dv,Narison:1998aq,Shore:2006mm}.

The axial-vector form factors are defined from
\begin{equation}
\langle N|J_{\mu 5}^a|N\rangle ~=~
2m_N \Bigl( G_A^a(p^2) s_\mu  +  G_P^a(p^2) p.s p_\mu \Bigr)
\label{eq:dll}
\end{equation}
where $s_\mu = \bar u \c_\mu \c_5 u /2m_N$ is the covariant spin vector.
In the absence of a massless pseudoscalar, only the form factors $G_A^a(0)$
contribute at zero momentum.  

Expressing the matrix element in terms of the 1PI vertices derived from
the generating functional $\C[V_{\m5}^a, V_\m^a, Q, \phi_5^a,\phi^a]$,
including spectator fields $N, \bar N$ for the nucleon, we have 
\begin{equation}
\langle N|J_{\mu 5}^a|N\rangle ~=~ \bar u\Bigl(
\C_{V_5^{\m a} \bar N N} + W_{V_5^{\m a} \o} \C_{Q \bar N N}
+ W_{V_5^{\m a} S_5^b} \C_{\phi_5^b \bar N N}\Bigr) u
\label{eq:dmmm}
\end{equation}
Note that this expansion relies on the specific definition(\ref{eq:bh})
of $\C$ as a partial Legendre transform. 

We also need the following relation, valid for all momenta, 
which is derived directly from
the fundamental anomalous chiral Ward identity (\ref{eq:bi}) for $\C$:
\begin{equation}
\pl_\m \C_{V_{\m5}^a \bar N N} ~=~ - \Phi_{ab} \C_{\phi_5^b \bar N N}
\label{eq:dnnn}
\end{equation}
Now, taking the divergence of eq.(\ref{eq:dmmm}), using this Ward identity
and then\footnote{The $p\rta 0$ limit is delicate, as is the case for
the derivation of the conventional Goldberger-Treiman relation, and should
be taken in this order. Literally at $p=0$, both sides vanish since
$\bar u \c_5 u = 0$.}
taking the zero-momentum limit, 
noting that the propagators vanish at zero momentum since there
is no massless pseudoscalar, gives
\begin{equation}
2m_N G_A^a(0)~ \bar u \c_5 u ~=~ 
i \bar u ~ \Phi_{ab} \C_{\phi_5^b \bar N N}\big|_{p=0} u
\label{eq:dooo}
\end{equation}

The meson-nucleon couplings are related to the 1PI vertices by
\begin{equation}
\langle N|\eta^\a N\rangle ~=~ g_{\eta^\a NN}~ \bar u \c_5 u
~=~ i \bar u \C_{\eta^\a \bar N N} u
\label{eq:dppp}
\end{equation} 
Re-expressing eq.(\ref{eq:dooo}) in
terms of the canonically normalised `OZI boson' field $\hat \eta^\a$,
we therefore derive
\begin{equation}
2m_N G_A^a(0) ~=~ \hat f^{a\a} g_{\hat\eta^\a NN}
\label{eq:dqqq}
\end{equation}
This relation will be useful to us when we consider the `proton spin' 
problem. 
 
All that now remains to cast this into its final form is to
make the familiar change of variables from $Q,\hat\eta^\a$ to 
$G,\eta^\a$, where $\eta^\a$ are interpreted as the physical mesons.
We therefore find the generalised $U(1)_A$ Goldberger-Treiman relation:
\begin{equation}
2m_N G_A^a(0) ~=~ f^{a\a} g_{\eta^\a NN} ~+~ \sqrt{2n_f} A g_{GNN}\d_{a0}
\label{eq:dmm}
\end{equation}
For the individual components, this is
\begin{eqnarray}
&2m_N G_A^3 ~&=~ f_\pi g_{\pi NN} 
\label{eq:dnn} \\
&2m_N G_A^8 ~&=~ f^{8\eta'} g_{\eta' NN} ~+~ f^{8\eta} g_{\eta NN}
\label{eq:doo} \\
&2m_N G_A^0 ~&=~ f^{0\eta'} g_{\eta' NN} ~+~ f^{0\eta} g_{\eta NN} 
~+~ \sqrt{6} A g_{GNN}~~~~~~
\label{eq:dpp} 
\end{eqnarray}

The renormalisation group properties of these relations are described
in great detail in ref.\cite{Shore:1991dv}. 
It is clear that the flavour singlet 
axial coupling $G_A^0$ satisfies a homogeneous RGE and scales
with the anomalous dimension $\c$ corresponding to the multiplicative
renormalisation of $J_{\m5}^0$. In the form (\ref{eq:dqqq}), RG 
consistency is simply achieved by 
\begin{equation}
\DD \hat f^{a\a} ~=~ \c \d_{a0} \hat f^{a\a}~~~~~~~~~~~~~~
\DD g_{\hat\eta^\a NN} ~=~ 0
\label{eq:drrr}
\end{equation}
All the scale dependence is in the decay constant $\hat f^{0\a}$ while
the the coupling $g_{\hat\eta^\a NN}$ of the `OZI boson' to the nucleon 
is RG invariant (in contrast to $g_{\hat\eta^\a \c\c}$).
In the final form (\ref{eq:dmm}) involving the physical decay constants,
a careful analysis shows that apart from $G_A^0(0)$ the only 
other non RG-invariant quantity is the gluonic coupling $g_{GNN}$, 
which is required to satisfy the following non-homogeneous RGE to ensure 
the self-consistency of eq.(\ref{eq:dpp}):
\begin{equation}
\DD g_{GNN} ~=~ \c \Bigl(g_{GNN} + {1\over\sqrt{2n_f}}{1\over A} 
f^{0\a} g_{\eta^\a NN}\Bigr)
\label{eq:dsss}
\end{equation}

The large-$N_c$ behaviour in the flavour singlet relation is as follows:
$G_A^0 = O(N_c)$, ~$f^{0\eta}, f^{0\eta'} = O(\sqrt{N_c})$, ~$A = O(1)$,
~$g_{\eta NN}, g_{\eta' NN} = O(\sqrt{N_c})$, ~$g_{GNN} = O(1)$. 
So the final term $A g_{GNN}$ is $O(1)$, suppressed by a power of $1/N_c$ 
compared to all the others, which are $O(N_c)$.

We see that, like $g_{G\c\c}$, the gluonic coupling $g_{GNN}$
is suppressed at large $N_c$ relative to the corresponding
meson couplings. However, unlike $g_{G\c\c}$ which is RG invariant,
$g_{GNN}$ has a complicated RG non-invariance and depends on the
anomaly-induced anomalous dimension $\c$. The conjecture in section
\ref{sec:conj} then suggests that while the OZI or large-$N_c$
approximation should be a good guide to the value of $g_{G\c\c}$,
we may expect significant OZI violations for $g_{GNN}$. We would 
therefore not be surprised to find that $g_{GNN}$ makes a sizeable
numerical contribution to the $U(1)_A$ Goldberger-Treiman relation.

We now try to test these expectations against the experimental data.
We first introduce a notation that has become standard in the 
literature on deep-inelastic scattering. There, the axial couplings 
are written as
\begin{equation}
G_A^3 ~=~ {1\over2}~ a^3 ~~~~~~~~
G_A^8 ~=~ {1\over 2\sqrt{3}}~ a^8 ~~~~~~~~
G_A^0 ~=~ {1\over\sqrt{6}}~ a^0 
\label{eq:dqq}
\end{equation}
where the $a^a$ have a simple interpretation in terms of parton
distribution functions.

Experimentally,
\begin{equation} 
a^3 ~=~ 1.267 \pm 0.004 ~~~~~~~~
a^8 ~=~ 0.585 \pm 0.025
\label{eq:dss}
\end{equation}
from low-energy data on nucleon and hyperon beta decay. 
The latest result\footnote{This supersedes the result 
$a^0|_{Q^2=4{\rm GeV}^2} ~=~ 0.237{}^{+0.024}_{-0.029}$ quoted by
COMPASS in 2005 \cite{Mallot,Ageev:2005gh}, which we used as input
into our analysis of the phenomenology of the $U(1)_A$ GT relation
in ref.\cite{Shore:2006mm}. The fits presented here are updated
from those of ref.\cite{Shore:2006mm} to take account of this.
For a further discussion of the experimental situation, see section 5.}
for $a^0$ quoted by the COMPASS 
collaboration \cite{COMPASS} from deep-inelastic scattering data is
\begin{equation}
a^0|_{Q^2 \rta\infty} ~=~ 0.33 \pm 0.06
\label{eq:dtt}
\end{equation}
with a similar result from HERMES \cite{HERMES}.

The OZI expectation is that $a^0 = a^8$. In the context of DIS, this
is a prediction of the simple quark model, where it is known as the
Ellis-Jaffe sum rule \cite{Ellis:1973kp}). We return to this in 
the context of the `proton spin' problem in section 5
but for now we concentrate on the low-energy phenomenology of the
pseudoscalar meson-nucleon couplings.

The original Goldberger-Treiman relation (\ref{eq:dnn})
gives the following value for the pion-nucleon coupling,
\begin{equation}
g_{\pi NN} ~=~ 12.86 \pm 0.06
\label{eq:duu}
\end{equation}
consistent to within about $5\%$ with the experimental value $13.65 (13.80)
\pm 0.12$ (depending on the dataset used \cite{Bugg:2004cm}).
In an ideal world where $g_{\eta NN}$ and $g_{\eta' NN}$ were both known,
we would now verify the octet formula (\ref{eq:doo}) then determine
the gluonic coupling $g_{GNN}$ from the singlet Goldberger-Treiman
relation (\ref{eq:dpp}). However, the experimental situation with the 
$\eta$ and $\eta'$-nucleon couplings is far less clear. 
One would hope to determine these couplings from the near threshold 
production of the $\eta$ and $\eta'$ in nucleon-nucleon collisions, i.e. 
$pp\rightarrow pp\eta$ and $pp\rightarrow pp\eta'$, measured for example
at COSY-II \cite{Moskal:2004cm,Bass:2001ix,Moskal:2004nw}. 
However, the $\eta$ production is 
dominated by the $N(1535) S_{11}$ nucleon resonance which decays
to $N\eta$, and as a result very little is known about $g_{\eta NN}$ itself.
The detailed production mechanism of the $\eta'$ is not well understood.
However, since there is no known baryonic resonance decaying into $N\eta'$, 
we may simply assume that the reaction $pp\rightarrow pp\eta'$ is 
driven by the direct coupling supplemented by heavy-meson exchange. This 
allows an upper bound to be placed on $g_{\eta' NN}$ and on this basis 
ref.\cite{Moskal:1998pc} quotes $g_{\eta' NN}< 2.5$.  This is supported by
an analysis \cite{Nakayama:2005ts} of very recent data from CLAS 
\cite{Dugger:2005du} on the photoproduction reaction 
$\c p \rightarrow p \eta'$. Describing the cross-section data with a model 
comprising the direct coupling together with $t$-channel meson exchange 
and $s$ and $u$-channel resonances, it is found that equally good fits can 
be obtained for several values of $g_{\eta' NN}$ covering the whole region 
$0 < g_{\eta' NN} < 2.5$.  

\begin{figure}
\centering
\includegraphics[height=3.3cm]{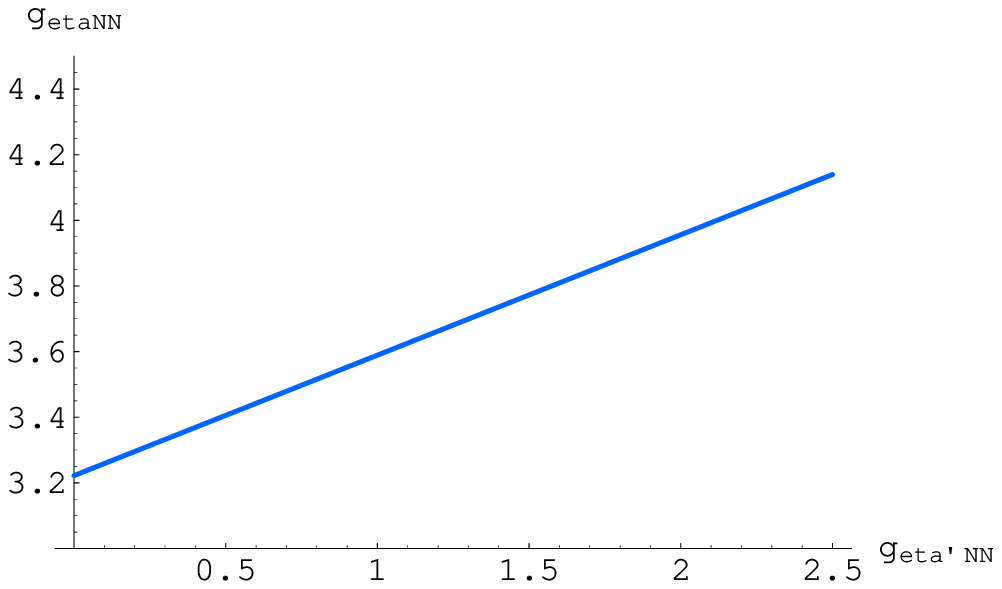} \hskip0.7cm 
\includegraphics[height=3.3cm]{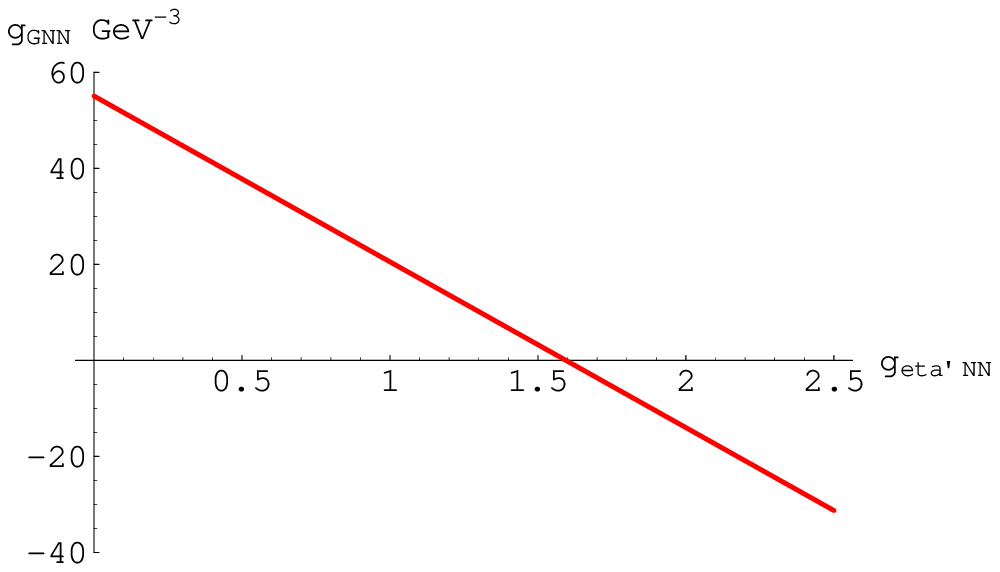}
\caption{These figures show the dimensionless $\eta$-nucleon coupling 
$g_{\eta NN}$ and the gluonic coupling $g_{GNN}$ in units of 
${\rm GeV}^{-3}$ expressed as functions of the experimentally uncertain 
$\eta'$-nucleon coupling $g_{\eta' NN}$, as determined from the flavour octet 
and singlet Goldberger-Treiman relations (\ref{eq:doo}) and (\ref{eq:dpp}).}
\label{fig:NNcouplings}     
\end{figure}

\begin{figure}
\centering
\includegraphics[height=4.5cm]{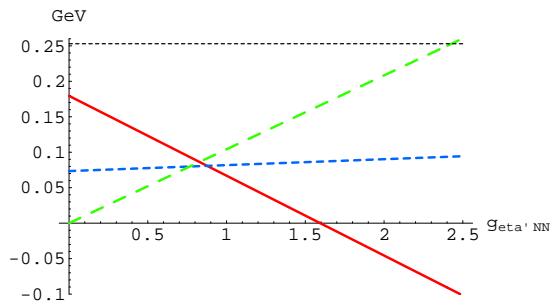} 
\caption{This shows the relative sizes of the contributions to the
$U(1)_A$ Goldberger-Treiman relation from the individual terms in 
eq.(\ref{eq:dpp}), expressed as functions of the coupling $g_{\eta' NN}$. 
The dotted (black) line denotes $2m_N G_A^0$. The long-dashed (green) line 
is $f^{0\eta'}g_{\eta' NN}$ and the short-dashed (blue) line is 
$f^{0\eta}g_{\eta NN}$. The solid (red) line shows the contribution of
the novel gluonic coupling, $\sqrt{6}Ag_{GNN}$, where $A$ determines the
QCD topological susceptibility.}
\label{fig:GTformula}     
\end{figure}

In view of this experimental uncertainty, we shall use the octet and singlet
Goldberger-Treiman relations to plot the predictions for $g_{\eta NN}$
and $g_{GNN}$ as a function of the ill-determined $\eta'$-nucleon coupling
in the experimentally allowed range $0 < g_{\eta' NN} < 2.5$. The results
(again taking the value (\ref{eq:dcc}) for $A$) are given in 
Fig.~\ref{fig:NNcouplings}. In Fig.~\ref{fig:GTformula} we have 
shown the relative magnitudes of the various contributions to the
flavour-singlet formula. 

What we learn from this is that for values of $g_{\eta' NN}$ approaching
the upper end of the experimentally allowed range, 
the contribution of the OZI-suppressed gluonic coupling $g_{GNN}$ is
quite large. The variation of $f^{0\eta'} g_{\eta' NN}$ over the
allowed range is compensated almost entirely by the variation of
$\sqrt{6} g_{GNN}$, with the $f^{0\eta} g_{\eta NN}$ contribution 
remaining relatively constant. 

For example, if experimentally we found $g_{\eta' NN} \simeq 2.5$, 
which corresponds to the cross-sections for $pp\rta pp\eta'$ and 
$\c p \rta p\eta'$ being almost entirely determined by the direct coupling, 
then we would have $g_{\eta NN} \simeq 4.14$ and $g_{GNN} \simeq -31.2~
{\rm GeV}^{-3}$. In terms of the contributions to the $U(1)_A$ 
Goldberger-Treiman relation, this would give (in GeV)
\begin{eqnarray}
2m_N G_A^0 [N_c;~0.25] ~=~
f^{0\eta'} g_{\eta' NN} [N_c;~0.26] ~+~ 
f^{0\eta} g_{\eta NN} [N_c;~0.09] ~~~~~~~
\nonumber\\
~+~ \sqrt{6} A g_{GNN} [O(1);~-0.10]~~~~
\label{eq:dvv}
\end{eqnarray}
The anomalously small value of $G_A^0$ compared to its OZI value
(the OZI approximation is $2m_N G_A^0\big|_{\rm OZI}
= \sqrt{2}~2m_N G_A^8 = 0.45$) is then due to the 
partial cancellation of the sum of the meson-nucleon coupling terms 
by the gluonic coupling $g_{GNN}$. Although formally $O(1/N_c)$ 
suppressed, numerically it gives a major contribution to the 
large OZI violation in $G_A^0$. This would give some support to
our conjecture and provide further evidence that we are able to predict 
the location of large OZI violations using the renormalisation group 
as a guide.

Of course, it may be that experimentally we eventually find a value 
for $g_{\eta' NN} \simeq 1.5$, in the region where $g_{GNN}$ contributes
only around $10\%$ or less. Although surprising, this would open the
possibility that all gluonic couplings of type $g_{GXX}$ are close to zero,
which could be interpreted as implying that the gluonic component of the
$\eta'$ wave function is simply small. 
Clearly, a reliable determination of $g_{\eta' NN}$, or equivalently
$g_{\eta NN}$, would shed considerable light on the $U(1)_A$ dynamics
of QCD.

\section{Topological charge screening and the `proton spin'}

So far, we have focused on the implications of the $U(1)_A$ anomaly for
low-energy QCD phenomenology. However, the anomaly also plays a vital role
in the interpretation of high-energy processes, in particular polarised
deep-inelastic scattering.

In this section, we discuss one of the most intensively studied topics
in QCD of the last two decades - the famous, but misleadingly named,
`proton spin' problem. We review the interpretation initially proposed
by Veneziano \cite{Veneziano:1989ei} and developed by us in a series of 
papers exploring the relation with the $U(1)_A$ GT relation and 
gluon topology \cite{Shore:1990zu,Shore:1991dv,Shore:1994zh}. In subsequent
work with Narison, we were able to quantify our prediction by using QCD
spectral sum rules to compute the slope $\chi'(0)$ of the topological
susceptibility \cite{Narison:1994hv,Narison:1998aq}. 
Remarkably, the most recent experimental data from the
COMPASS \cite{COMPASS} and HERMES \cite{HERMES} collaborations, 
released in September 2006, now confirms our original 1994 numerical 
prediction \cite{Narison:1994hv}.

\subsection{The $g_1^p$ and angular momentum sum rules}

The `proton spin' problem concerns the sum rule for the first moment of the
polarised proton structure function $g_1^p$. This is measured in polarised
DIS experiments through the inclusive processes $\m p \rta \m X$ 
(EMC, SMC, COMPASS at CERN) or $e p \rta e X$ (SLAC, HERMES at DESY) 
together with similar experiments on a deuteron target. 
The polarisation asymmetry of the cross-section is expressed as
\begin{equation}
x{d\D\s\over dx dy}~=~{Y_P\over2}{16\pi^2\a^2\over s} g_1^p(x,Q^2)~~+~~
O\Bigl({M^2 x^2\over Q^2}\Bigr)
\label{eq:ea}
\end{equation}
with conventional notation: $Q^2 = -q^2$ and 
$x = {Q^2/2p_2.q}$ are the Bjorken variables, where $p_2$, $q$ are
the momenta of the target proton and incident virtual photon respectively,
$y = {Q^2/xs}$ and $Y_p = {(2-y)/y}$.

According to standard theory, $g_1^p$ is determined by the proton
matrix element of two electromagnetic currents carrying a large spacelike
momentum. The sum rule for the first moment of $g_1^p$ is derived from 
the twist 2, spin 1 terms in the operator product expansion for the 
currents:
\begin{equation}
J^\l(q) J^\r(-q) \SIMQ
2 \e^{\l\r\n\m} {q_\n\over Q^2} ~\Bigl[\D C_1^{NS}(\a_s) 
\Bigl(J_{\m5}^3 + {1\over\sqrt3} J_{\m5 }^8\Bigr)
+ {2\sqrt{2}\over\sqrt{3}} \D C_1^{S}(\a_s) J_{\m5}^0 \Bigr]
\label{eq:eb}
\end{equation}
where $\D C_1^{NS}$ and $\D C_1^S$ are Wilson coefficients and 
$J_{\m5}^a$ ($a=3,8,0$) are the renormalised axial currents, with the
normalisations defined in section 2. It is the occurrence of the axial
currents in this OPE that provides the link between the $U(1)_A$ anomaly
and polarised DIS.
The sum rule is therefore:
\begin{equation}
\C^p_1(Q^2) \equiv
\int_0^1 dx~ g_1^p(x,Q^2) 
= {1\over12} \D C_1^{NS} \Bigl( a^3
+ {1\over3} a^8 \Bigr) + {1\over9} \D C_1^{\rm S} a^0(Q^2)  
\label{eq:ec}
\end{equation}
where the axial charges $a^3$, $a^8$ and $a^0(Q^2)$ are defined in 
terms of the forward proton matrix elements as in eq.(\ref{eq:dqq}). 
Here, we have explicitly shown the $Q^2$ scale dependence
associated with the RG non-invariance of $a^0(Q^2)$.

Since the flavour non-singlet axial charges are known from low-energy data, 
a measurement of the first moment of $g_1^p$ amounts to a determination
of the flavour singlet $a^0(Q^2)$. At the time of the original EMC
experiment in 1988 \cite{EMC} 
the theoretical expectation based on the quark model
was that $a^0 = a^8$. The resulting sum rule for $g_1^p$ is known as the
Ellis-Jaffe sum rule \cite{Ellis:1973kp}. 
The great surprise of the EMC measurement was the
discovery that in fact $a^0$ is significantly suppressed relative to $a^8$,
and indeed the earliest results suggested it could even be zero.
However, the reason the result sent shockwaves through both the 
theoretical and experimental communities (to date, the EMC paper
has over 1300 citations) was the interpretation that this implies that 
the quarks contribute only a fraction of the total spin of the proton.

In fact, this interpretation relies on the simple valence quark model
of the proton and is {\it not} true in QCD, where the axial charge 
decouples from the real angular momentum sum rule for the proton.
Rather, as we shall show, the suppression of $a^0(Q^2)$ reflects
the dynamics of gluon topology and appears to be largely
independent of the structure of the proton itself. Precisely, it is
a manifestation of {\it topological charge screening} in the QCD vacuum.

\vskip0.2cm
The angular momentum sum rule is derived by taking the forward
matrix element of the conserved angular momentum current $M^{\m\n\l}$,
defined in terms of the energy-momentum tensor as
\begin{equation}
M^{\m\n\l} ~=~ x^{[\n}T^{\l]\m} + \pl_\r X^{\r\m\n\l}
\label{eq:eea}
\end{equation}
The inclusion of the arbitrary tensor $X^{\r\m\n\l}$ just reflects the usual
freedom in QFT of defining conserved currents. This gives us some flexibility
in attempting to write $M^{\m\n\l}$ as a sum of local operators, suggesting 
interpretations of the total angular momentum as a sum of `components' of 
the proton spin. In fact, however, it is not possible to write 
$M^{\m\n\l}$ as a sum of operators corresponding to quark and gluon spin
and angular momentum in a gauge-invariant way. The best decomposition
is \cite{Jaffe:1989jz,Shore:1999be,Shore:2000ca} 
\begin{equation}
M^{\m\n\l} ~=~ O_1^{\m\n\l} + O_2^{\m[\l}x^{\n]} + O_3^{\m[\l}x^{\n]}
+ \ldots
\label{eq:eeb}
\end{equation}
where the dots denote terms whose forward matrix elements vanish.
Here,
\begin{eqnarray}
&O_1^{\m\n\l} &= {1\over2}\e^{\m\n\l\s}\bar q \c_\s \c_5 q = 
{1\over2}\e^{\m\n\l\s}\sqrt{2n_f}J_{\s 5}^0 
\nonumber\\
&O_2^{\m\l} &= i \bar q \c^\m \hD{}^\l q
\nonumber\\
&O_3^{\m\l} &= F^{\m\r}F_\r{}^\l
\label{eq:eec}
\end{eqnarray}
At first sight, $O_1^{\m\n\l}$ looks as if it could be associated with
`quark spin', since for {\it free} Dirac fermions the spin operator
coincides with the axial vector current. $O_2^{\m[\l}x^{\m]}$ would
correspond to `quark orbital angular momentum', leaving 
$O_3^{\m[\l}x^{\n]}$ as `gluon total angular momentum'. Any further 
decomposition of the gluon angular momentum is necessarily not gauge 
invariant. 

The forward matrix elements of these operators may be expressed in terms
of form factors and, as we showed in ref.\cite{Shore:1999be}, this exhibits an
illuminating cancellation. After some analysis, we find:
\begin{eqnarray}
&\langle p,s|O_1^{\m\n\l}|p,s\rangle ~&=~ a^0 m_N \e^{\m\n\l\s}s_\s
\nonumber\\
&\langle p,s|O_2^{\m[\l}x^{\n]}|p,s\rangle ~&=~ 
J_q {1\over2m_N}p_\r p^{\{\m} \e^{[\l\}\n]\r\s} s_\s -   
a^0 m_N \e^{\m\n\l\s}s_\s
\nonumber\\
&\langle p,s|O_3^{\m[\l}x^{\n]}|p,s\rangle ~&=~ 
J_g {1\over2m_N}p_\r p^{\{\m} \e^{[\l\}\n]\r\s} s_\s
\label{eq:eed}
\end{eqnarray}
The angular momentum sum rule for the proton is then just
\begin{equation}
{1\over2}~=~ J_q + J_g
\label{eq:eee}
\end{equation}
where the Lorentz and gauge-invariant form factors $J_q$ and $J_g$
may reasonably be thought of as representing quark and gluon total 
angular momentum. However, even this interpretation is not at all rigorous, 
not least because $J_q$ and $J_g$ mix under renormalisation and scale as
\begin{equation}
{d\over d\ln Q^2} \left(\matrix{J_q\cr J_g\cr}\right) ~=~
{\a_s\over4\pi} \left(\matrix{-{8\over3}C_F & {2\over3}n_f\cr
{8\over3}C_F & -{2\over3}n_f\cr}\right)~
\left(\matrix{J_q\cr J_g\cr}\right)
\label{eq:eef}
\end{equation}
Only the total angular momentum is Lorentz, gauge and scale 
invariant.\footnote{For a careful discussion of the parton interpretation
of longitudinal and transverse angular momentum sum rules, see 
ref.\cite{Bakker:2004ib}. This confirms our assertion that the 
axial charge $a^0$ is not to be identified with quark helicities 
in the parton model.}

The crucial observation, however, is that the axial charge $a^0$
explicitly {\it cancels} from the angular momentum sum rule. 
$a^0$ is an important form factor, which relates the first moment of
$g_1^p$ to gluon topology via the $U(1)_A$ anomaly, but it is {\it not}
part of the angular momentum sum rule for the proton.

Just as $a^0$ can be measured in polarised inclusive DIS, the form
factors $J_q$ and $J_g$ can be extracted from measurements
of unpolarised generalised parton distributions (GPDs) in processes
such as deeply-virtual Compton scattering $\c^* p \rta \c p$.
These can also in principle be calculated in lattice QCD. 
The required identifications with GPDs are given in ref.\cite{Shore:1999be}.

\subsection{QCD parton model}

Before describing our resolution of the `proton spin' problem, we briefly
review the parton model interpretation of the first moment sum rule
for $g_1^p$.

In the simplest form of the parton model, the proton structure at large
$Q^2$ is described by parton distributions corresponding to free valence
quarks only. The polarised structure function is given by
\begin{equation}
g_1^p(x) = {1\over2} \sum_{i=1}^{n_f} ~e_i^2 ~\D q_i(x)
\label{eq:ed}
\end{equation}
where $\D q_i(x)$ is defined as the difference of the distributions of 
quarks (and antiquarks) with helicities parallel and antiparallel to 
the nucleon spin. It is convenient to work with the conventionally-defined
flavour non-singlet and singlet combinations $\D q^{NS}$ and $\D q^S$ 
(often also written as $\D \Sigma$).

In this model, the first moment of the singlet quark distribution
$\D q^S = \int_0^1 dx~\D q^S(x)$ can be identified as the sum of
the helicities of the quarks. Interpreting the structure function
data {\it in this model} then leads to the conclusion that the quarks 
carry only a small fraction of the spin of the proton.
There is indeed a real contradiction between the experimental data
and the {\it free valence quark} parton model.

However, this simple model leaves out many important features of QCD, the 
most important being  gluons, RG scale dependence and the chiral $U_A(1)$ 
anomaly. When these effects are included, in the QCD parton model, the naive 
identification of $\D q^S$ with spin no longer holds and the experimental 
results for $g_1^p$ can be accommodated, though not predicted.

In the QCD parton model, the polarised structure function is written 
in terms of both quark and gluon distributions as follows:
\begin{eqnarray}
g_1^p(x,Q^2) = ~~~~~~~~~~~~~~~~~~~~~~~~~~~~~~~~~~~~~~~~~~~~~~~~~~~~~~~~~
~~~~~~~~~~~~~~~~~~~~~~~~~~~\nonumber\\
\int_{x}^{1} {du\over u}~ {1\over9} 
\Bigl[\D C^{NS}\Bigl({x\over u}\Bigr)\D q^{NS}(u,t) 
+ \D C^{S}\Bigl({x\over u}\Bigr)\D q^S(u,t) +
\D C^{g}\Bigl({x\over u}\Bigr)\D g(u,t) \Bigr]
\nonumber\\
{}
\label{eq:ehhh}
\end{eqnarray}
where $\D C^S, \D C^g$ and $\D C^{NS}$ are perturbatively calculable 
functions related to the Wilson coefficients and the quark and gluon
distributions have {\it a priori} a $t=\ln Q^2/\L^2$ dependence
determined by the RG evolution, or DGLAP, equations.
The first moment sum rule is therefore
\begin{equation}
\C_1^p(Q^2) = {1\over9}\Bigl[\D C_1^{NS}\D q^{NS}
+ \D C_1^S \D q^S + \D C_1^g \D g \Bigr]
\label{eq:eg}
\end{equation}
Comparing with eq.(\ref{eq:ec}), we see that the axial charge $a^0(Q^2)$ 
is identified with a linear combination of the first moments of the singlet 
quark and gluon distributions. It is often, though not always, the case 
that the moments of parton distributions can be identified in one-to-one
correspondence with the matrix elements of local operators. The polarised
first moments are special in that two parton distributions correspond
to the same local operator. 

The RG evolution equations for the first moments of the parton distributions
are derived from the matrix of anomalous dimensions for the lowest spin,
twist 2 operators. This introduces an inevitable renormalisation scheme 
ambiguity in the definitions of $\D q$ and $\D g$,
and their physical interpretation is correspondingly nuanced. The choice
closest to our own analysis is the `AB' scheme \cite{Ball:1995td}
where the parton distributions have the folowing RG evolution:
\begin{eqnarray}
&{d\over d\ln Q^2} \D q^{NS} = 0 ~~~~~~~~~~
{d\over d\ln Q^2} \D q^S = 0 
\nonumber\\
&{} \nonumber\\
&{d\over d\ln Q^2} {\a_s\over2\pi}\D g(Q^2) = 
\c \Bigl({\a_s\over2\pi}\D g(Q^2) 
-{1\over 3} \D q^S\Bigr)  
\label{eq:eh}
\end{eqnarray}
which requires $\D C_1^g = {3\a_s\over2\p}\D C_1^S$.
It is then possible to make the following identifications with the axial
charges:
\begin{eqnarray}
&a^3 &= \D u - \D d 
\nonumber\\
&a^8 &= \D u + \D d - 2 \D s 
\nonumber\\
&a^0(Q^2) &= \D u + \D d + \D s  -  {3\a_s\over2\pi} \D g(Q^2)
\label{eq:ei}
\end{eqnarray}
with $\D q^S = \D u + \D d + \D s$.
Notice that in the AB scheme, all the scale dependence of the axial charge 
$a^0(Q^2)$ is assigned to the gluon distribution $\D g(Q^2)$.

This was the identification originally introduced for the first moments
by Altarelli and Ross \cite{Altarelli:1988nr}, and resolves the `proton spin' 
problem in the context of the QCD parton model. 
In this scheme, the Ellis-Jaffe sum rule follows from the assumption
that in the proton both $\D s$ and $\D g(Q^2)$ are zero, which
is the natural assumption in the free valence quark model. 
This is equivalent to the OZI approximation $a^0(Q^2) = a^8$. 
However, in the full QCD parton model, there is no
reason why $\D g(Q^2)$, or even $\D s$, should be zero in the proton.
Indeed, given the different scale dependence of $a^0(Q^2)$ and $a^8$,
it would be unnatural to expect this to hold in QCD itself.
 
An interesting conjecture \cite{Altarelli:1988nr} 
is that the observed suppression 
in $a^0(Q^2)$ is due overwhelmingly to the gluon distribution $\D g(Q^2)$ 
alone. Although by no means a necessary consequence of QCD,
this is a reasonable expectation given that it is the 
anomaly (which is due to the gluons and is responsible for OZI
violations) which is responsible for the scale dependence in $a^0(Q^2)$ 
and $\D g(Q^2)$ whereas the $\D q$ are scale invariant.
This would be in the same spirit as our conjecture on OZI violations
in low-energy phenomenology in section \ref{sec:conj}.
To test this, however, we need to find a way to measure $\D g(Q^2)$
itself, rather than the combination $a^0(Q^2)$. The most
direct option is to extract $\D g(x,Q^2)$ from processes such as 
open charm production, $\c^* g \rta c \bar c$, which is currently being 
intensively studied by the COMPASS \cite{Procureur:2006sg}, 
STAR \cite{STAR} and PHENIX \cite{PHENIX} collaborations.

\subsection{Topological charge screening}
\label{topscreen}

We now describe a less conventional approach to deep-inelastic scattering
based entirely on field-theoretic concepts. In particular, the role of
parton distributions is taken over by the 1PI vertices of composite
operators introduced above. (For a review, see ref.\cite{Shore:1998dn}).

Once again, the starting point is the use of the OPE to express the moments
of a generic structure function $F(x,Q^2)$ as 
\begin{equation}
\int_0^1 dx~x^{n-1} F(x,Q^2) ~=~ \sum_A C_A^n(Q^2)~ 
\langle p|\OO_A^n(0)|p\rangle
\label{eq:ej}
\end{equation}
where $\OO_A^n$ denotes the set of lowest twist, spin $n$ operators and 
$C_A^n(Q^2)$ are the corresponding Wilson coefficients. The next step is
to introduce a new set of composite operators $\tilde \OO_B$, chosen to 
encompass the physically relevant degrees of freedom, and write the matrix
element as a product of two-point Green functions and 1PI vertices as
follows:
\begin{equation}
\int_0^1 dx~x^{n-1} F(x,Q^2) ~=~ \sum_A \sum_B C_A^n(Q^2)~ 
\langle 0|T~\OO_A^n~\tilde \OO_B|0\rangle~\C_{\tilde \OO_B pp}
\label{eq:ek}
\end{equation}
This decomposition splits the structure function into three parts --
first, the Wilson coefficients $C_A^n(Q^2)$ which can be calculated in
perturbative QCD; second, non-perturbative but {\it target independent}
Green functions which encode the dynamics of the QCD vacuum; third,
non-perturbative vertex functions which characterise the target by its
couplings to the chosen operators $\tilde \OO_B$.\footnote{We emphasise 
again that this decomposition of the matrix elements into products 
of Green functions and 1PI vertices is {\it exact}, independent of the 
choice of the set of operators $\tilde{\OO}_B$. In particular, it is not 
necessary for $\tilde{\OO}_B$ to be in any sense a complete set.  
If a different choice is made, the vertices 
$\C_{\tilde{\OO}_B pp}$ themselves change, becoming 1PI with respect to a 
different set of composite fields. In practice, the set of operators
$\tilde{\OO}_B$ should be as small as possible while still 
capturing the essential degrees of freedom. A good choice can also
result in vertices $\C_{\tilde{\OO}_B pp}$ which are both RG invariant 
and closely related to low-energy physical couplings.} 

Now specialise to the first moment sum rule for $g_1^p$. For simplicity, we
first present the analysis for the chiral limit, where there is no flavour
mixing. Using the anomaly (\ref{eq:bd}), we can express the flavour singlet
contribution to the sum rule as
\begin{equation}
\C_1^p(Q^2)_{singlet} ~\equiv~
\int_0^1 dx~g_1^p(x,Q^2)_{singlet} ~=~
{2\over3} {1\over 2m_N}~ \D C_1^S(\a_s)~ \langle p|Q|p\rangle
\label{eq:el}
\end{equation}
The obvious choice for the operators $\tilde\OO_B$ in this case
are the flavour singlet pseudoscalars and it is natural to choose the 
`OZI boson' field 
$\hat\eta^0 = \hat f^{00}{1\over \langle\bar q q\rangle}\phi_5^0$, 
which is normalised so that
$d/dp^2~ \C_{\hat\eta^0 \hat\eta^0}\big|_{p=0} = 1$.
As we have seen in eq.(\ref{eq:drrr}), the corresponding 1PI vertex is 
then RG invariant. Writing the 1PI vertices in terms of nucleon couplings 
as in eq.(\ref{eq:dppp}), we find
\begin{equation}
\C_1^p(Q^2)_{singlet} ~=~
{2\over3} {1\over 2m_N} \D C_1^S(\a_s)~ 
\Bigl(\langle 0|T~Q~Q|0\rangle~g_{QNN} ~+~
\langle 0|T~Q~\hat\eta^0|0\rangle~g_{\hat\eta^0 NN}\Bigr)
\label{eq:elll}
\end{equation}
\begin{figure}
\centering
\includegraphics[height=2.8cm]{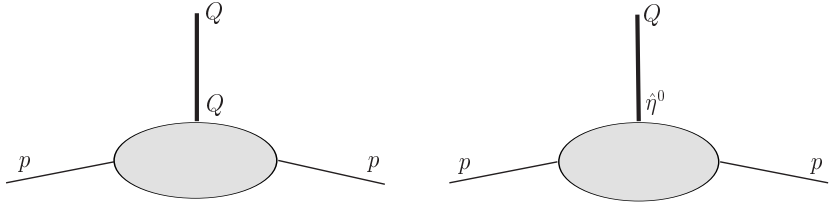} 
\caption{Illustration of the decomposition of the matrix element 
$\langle p |Q|p\rangle$ into two-point Green functions and 1PI vertices.
The Green function in the first diagram is $\chi(0)$; in the second it is
$\sqrt{\chi'(0)}$.}
\label{fig:semifignew}     
\end{figure}

Recalling that the matrix of two-point Green functions is given by
the inversion formula
\begin{equation}
\left(\matrix{W_{\o\o} & W_{\o S_{\hat\eta^0}}\cr
W_{S_{\hat\eta^0} \o} & W_{S_{\hat\eta^0} S_{\hat\eta^0}} \cr}\right) 
~=~ -
\left(\matrix{\C_{QQ} &\C_{Q \hat\eta^0}\cr 
\C_{\hat\eta^0 Q} &\C_{\hat\eta^0 \hat\eta^0} \cr}\right)^{-1}  
\label{eq:em}
\end{equation}
and using the normalisation condition for $\hat\eta^0$, we can easily
show that at zero momentum,
\begin{equation}
W_{\o S_{\hat\eta^0}}^2 ~=~ {d\over dp^2} W_{\o\o}\big|_{p=0} 
\label{eq:en}
\end{equation}

Finally, therefore, we can represent the first moment of $g_1^p$ in the
following, physically intuitive form:
\begin{equation}
\C_1^p(Q^2)_{singlet} ~=~
{2\over3} {1\over 2m_N}~ \D C_1^S(\a_s)~ 
\Bigl(\chi(0)~g_{QNN} ~+~
\sqrt{\chi'(0)}~g_{\hat\eta^0 NN}\Bigr)  
\label{eq:eo}
\end{equation}
This shows that the first moment is determined by the gluon topological
susceptibility in the QCD vacuum as well as the couplings of the proton
to the pseudoscalar operators $Q$ and $\hat\eta^0$. In the chiral limit,
$\chi(0) = 0$ so the first term vanishes. The entire flavour singlet
contribution is therefore simply
\begin{equation}
\C_1^p(Q^2)_{singlet} ~=~
{2\over3} {1\over 2m_N}~ \D C_1^S(\a_s)~ 
\sqrt{\chi'(0)}~g_{\hat\eta^0 NN}  
\label{eq:eooo}
\end{equation}
The 1PI vertex $g_{\hat\eta^0 NN}$ is RG invariant, and we see from 
eq.(\ref{eq:by}) that {\it in the chiral limit} the slope of the 
topological susceptibility scales with the anomalous dimension $\c$,
viz.
\begin{equation}
{d\over d\ln Q^2} \sqrt{\chi'(0)} ~=~ \c ~\sqrt{\chi'(0)}
\label{eq:ep}
\end{equation}
ensuring consistency with the RGE for the flavour singlet axial charge.

The formulae (\ref{eq:eo}) and (\ref{eq:eooo}) are our key result. They 
show how the first moment of $g_1^p$ can be factorised into couplings
$g_{QNN}$ and $g_{\hat\eta^0 NN}$ which carry information on the proton
structure, and Green functions which characterise the QCD vacuum. In the
case of $g_1^p$, the Green functions reduce simply to the topological
susceptibility $\chi(0)$ and its slope $\chi'(0)$. We now argue that
the experimentally observed suppression in the first moment of $g_1^p$
is due {\it not} to a suppression in the couplings, but to the
vanishing of the topological susceptibility $\chi(0)$ and an 
anomalously small value for its slope $\chi'(0)$. This is what we
refer to as {\it topological charge screening} in the QCD vacuum.

The justification follows our now familiar conjecture on the relation 
between OZI violations and RG scale dependence. We expect the source of 
OZI violations to be in those quantities which are sensitive to the anomaly,
as identified by their scaling dependence on the anomalous dimension
$\c$, in this case $\chi'(0)$. In contrast, it should be a good 
approximation to use the OZI value for the RG-invariant vertex 
$g_{\hat\eta^0 NN}$, that is $g_{\hat\eta^0 NN} \simeq \sqrt{2}
g_{\hat\eta^8 NN}$. The corresponding OZI value for $\sqrt{\chi'(0)}$ would
be $f_\pi/\sqrt{6}$. 
This gives our key formula for the flavour singlet
axial charge:
\begin{equation}
{a^0(Q^2)\over a^8} ~\simeq~ {\sqrt{6}\over f_\pi}~\sqrt{\chi'(0)}
\label{eq:eq}
\end{equation}

The corresponding prediction for the first moment of $g_1^p$ is 
\begin{equation}
\C_1^p(Q^2)_{singlet} ~=~ {1\over 9} \D C_1^S(\a_s)~a^8~
{\sqrt{6}\over f_\pi}~\sqrt{\chi'(0)}
\label{eq:er}
\end{equation}

The final step is to compute the slope of the topological susceptibility.
In time, lattice gauge theory should provide an accurate measurement of
$\chi'(0)$. However, this is a particlarly difficult correlator for 
lattice methods since it requires a simulation of QCD with light dynamical
fermions and algorithms that implement topologically non-trivial
configurations in a sufficiently fast and stable way. Instead, we have
estimated the value of $\chi'(0)$ using the QCD spectral sum rule
method. Full details and discussion of this computation can be found in
refs.\cite{Narison:1994hv,Narison:1998aq}. The result is:
\begin{equation}
\sqrt{\chi'(0)} ~=~ 26.4 \pm 4.1 ~{\rm MeV}
\label{eq:es}
\end{equation}

This gives our final prediction for the flavour singlet axial charge
and the complete first moment of $g_1^p$:
\begin{eqnarray}
&a^0\big|_{Q^2=10 {\rm GeV}^2} ~&=~ 0.33 \pm 0.05
\label{eq:et}\\
&\C_1^p\big|_{Q^2=10 {\rm GeV}^2} ~&=~ 0.144 \pm 0.009
\label{eq:eu}
\end{eqnarray}
Topological charge screening therefore gives a suppression factor of
approximately 0.56 in $a^0$ compared to its OZI value $a^8 = 0.585$.

\vskip0.2cm
In the decade since we made this prediction, the experimental measurement
has been somewhat lower than this value, in the range $a^0 \simeq 0.20-0.25$.
This would have suggested there is also a significant OZI violation in the
nucleon coupling $g_{\hat\eta^0 NN}$ itself, implicating the proton structure
in the anomalous suppression of $\C_1^p$. Very recently, however, the
COMPASS and HERMES collaborations have published new results on the
deuteron structure function which spectacularly confirm our picture that
topological charge screening in the QCD vacuum
is the dominant suppression mechanism.
\begin{figure}
\centering
\includegraphics[height=5.0cm]{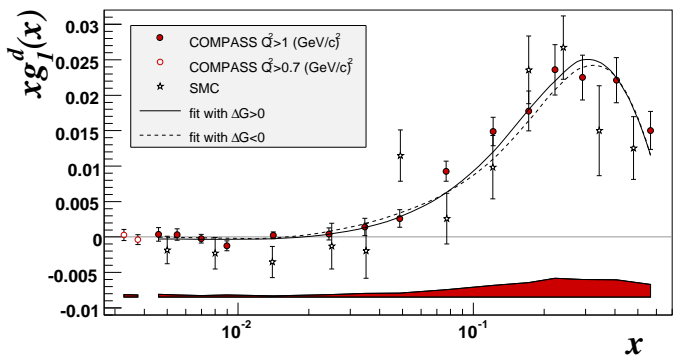} 
\caption{COMPASS and SMC data for the deuteron structure function $g_1^d(x)$.
Statistical error bars are shown with the data points. The shaded band
shows the systematic error.}
\label{fig:compassg1d}     
\end{figure}

This new data is shown in Fig.~\ref{fig:compassg1d}. This is based on data
collected by COMPASS at CERN in the years 2002-2004 and has only recently 
been published. The accuracy compared to earlier SMC data at small $x$ is 
significantly improved and the dip in $x g_1^d$ around $x\sim 10^{-2}$ 
suggested by the SMC data is no longer present (Fig.~\ref{fig:compassg1d}). 
This explains the significantly higher value
for $a^0$ found by COMPASS compared to SMC. From this data, COMPASS 
quote the first moment for the proton-neutron average $g_1^N = 
(g_1^p + g_1^n)/2$ as \cite{COMPASS}
\begin{equation}
\C_1^N\big|_{Q^2=3{\rm GeV}^2} ~=~ 0.050 \pm 0.003(stat) \pm 0.002(evol)
\pm 0.005(syst)
\label{eq:ev}
\end{equation}
Extracting the flavour singlet axial charge from the analogue of
eq.(\ref{eq:ec}) for $\C_1^N$ then gives
\begin{equation}
a^0\big|_{Q^2=3{\rm GeV}^2} ~=~ 0.35 \pm 0.03(stat) \pm 0.05(syst)
\label{eq:ew}
\end{equation}
or evolving to the $Q^2\rta\infty$ limit,
\begin{equation}
a^0\big|_{Q^2\rta\infty} ~=~ 0.33 \pm 0.03(stat) \pm 0.05(syst)
\label{eq:ex}
\end{equation}
Similar results are found by HERMES, who quote \cite{HERMES}
\begin{equation}
a^0\big|_{Q^2=5{\rm GeV}^2} ~=~ 0.330 \pm 0.011(th) \pm 0.025(exp)
\pm 0.028(evol)
\label{eq:ey}
\end{equation}
The agreement with our prediction (\ref{eq:et}) is striking. 

\vskip0.2cm
To close this section, we briefly comment on the extension of our analysis
beyond the chiral limit. In this case, the operator $\sqrt{2n_f}Q$ in
eq.(\ref{eq:el}) is replaced by the full divergence of the flavour singlet
axial current, viz. $D^0 = \sqrt{2n_f}Q + d_{0bc}m^b \phi_5^c$.
Separating the matrix element $\langle p|D^0|p\rangle$ into Green functions
and 1PI vertices, we find from the zero-momentum Ward identities that 
$\langle 0|T~D^0~Q|0\rangle = 0$ so the contribution from $g_{QNN}$ still
vanishes. The other Green function is $\langle 0|T~D^0~\hat\eta^\a|0\rangle
= -\hat f^{0\a}$, so the first moment sum rule becomes
\begin{equation}
\C_1^p(Q^2)_{singlet} ~=~
{1\over9} {1\over 2m_N}~ \D C_1^S(\a_s)~\sqrt{6}~ 
\hat f^{0\a}g_{\hat\eta^\a NN}  
\label{eq:eaa}
\end{equation}
It is clear that this is simply an alternative derivation of the $U(1)$
GT relation (\ref{eq:dqqq}) for $a^0$. We could equally use the 
alternative form (\ref{eq:dmm}) to write
\begin{equation}
\C_1^p(Q^2)_{singlet} ~=~
{1\over9} {1\over 2m_N}~ \D C_1^S(\a_s)~\sqrt{6}~ 
\Bigl(f^{0\a}g_{\eta^\a NN} + \sqrt{6}A g_{GNN}\Bigr)
\label{eq:ebb}
\end{equation}
Recalling the RGE (\ref{eq:dsss}) for $g_{GNN}$, we see that this bears a 
remarkable similarity to the expression for $a^0$ in terms of parton 
distributions in the AB scheme, eq.(\ref{eq:ei}). This was first pointed
out in ref.\cite{Shore:1990zu,Shore:1991dv}. 

Manipulating the zero-momentum Ward identities in a similar way to that
explained above in the chiral limit now shows that we can express the
decay constants $\hat f^{a\a}$ in terms of vacuum Green functions
as follows (see eq.(\ref{eq:cch}):
\begin{equation}
(\hat f \hat f^T)_{ab} ~=~ 
{d\over dp^2} \langle 0|T~D^a ~D^b|0\rangle\big|_{p=0} 
\label{eq:ecc}
\end{equation}
However, for non-zero quark masses there is flavour
mixing amongst the `OZI bosons' $\hat\eta^\a$ and we cannot extract
the decay constants simply by taking a square root, as was the case in
writing $\hat f^{00} = \sqrt{\chi'(0)}$ in the chiral limit.
Nevertheless, in ref.\cite{Narison:1998aq} we estimated
the decay constants and form factors in the approximation where we
use eq.(\ref{eq:ecc}) with the full divergence $D^a$ but neglect 
flavour mixing. Assuming OZI for the couplings, this gives the estimate
\begin{equation}
{a^0(Q^2)\over a^8} ~\simeq~ \sqrt{6}~ {\hat f^{00} \over \hat f^{88}}
\label{eq:edd}
\end{equation}
where we take
\begin{equation}
\hat f^{00} \simeq 
\sqrt{{{}^{}\over{}^{}}}{d\over dp^2} \langle 0|T~D^0 ~D^0|0\rangle\big|_{p=0}
~~~~~~~~~~~~~~
\hat f^{88} \simeq 
\sqrt{{{}^{}\over{}^{}}}{d\over dp^2} \langle 0|T~D^8 ~D^8|0\rangle\big|_{p=0}
\label{eq:eeeee}
\end{equation}
Evaluating the Green functions using QCD spectral sum rules gives
\begin{eqnarray}
&a^0\big|_{Q^2=10 {\rm GeV}^2} ~&=~ 0.31 \pm 0.02
\label{eq:eff}\\
&\C_1^p\big|_{Q^2=10 {\rm GeV}^2} ~&=~ 0.141 \pm 0.005
\label{eq:egg}
\end{eqnarray}
As we have seen in the last section, flavour mixing can be non-negligible
in the phenomenology of the pseudoscalar mesons, so we should be a little
cautious in over-estimating the accuracy of these estimates. (The quoted 
errors do not include this systematic effect.) Nevertheless,
the fact that they are consistent with those obtained in the chiral limit  
reinforces our confidence that the flavour singlet axial charge is 
relatively insensitive to the quark masses and that eqs.(\ref{eq:et}) 
and (\ref{eq:eu}) indeed provide an accurate estimate of the first moment 
of $g_1^p$.

The observation that the `proton spin' sum rule could be explained in terms
of an extension of the Goldberger-Treiman relation to the flavour singlet 
sector was made in Veneziano's original paper \cite{Veneziano:1989ei}. 
This pointed out for the first time that the suppression in $a^0$ was an 
OZI-breaking effect. Since the Goldberger-Treiman relation connects the 
pseudovector form factors with the pseudoscalar channel,
where it is known that there are large OZI violations for the flavour singlet,
it becomes natural to expect similar large OZI violations also in $a^0$.
This is the fundamental intuition which we have developed into a 
quantitative resolution of the `proton spin' problem.

\subsection{Semi-inclusive polarised DIS}

\begin{figure}
\centering
\includegraphics[height=4.1cm]{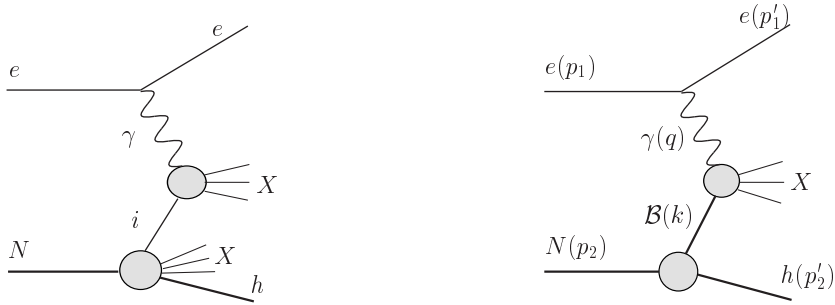} 
\caption{Semi-inclusive DIS $eN\rta ehX$ in the target fragmentation region. 
In the equivalent current fragmentation process, the detected hadron $h$ is 
emitted from the hard collision with $\c$. 
The right-hand figure shows a simple Reggeon exchange model valid for
$z\sim 1$, where $h$ carries a large target energy fraction.  }
\label{fig:semiincl}     
\end{figure}

While the agreement between our prediction for the first moment of $g_1^p$
and experiment is now impressive, it would still be interesting to find other
experimental tests of topological charge screening. A key consequence of this
mechanism is that the OZI violation observed in $a^0$ is not a property
specifically of the proton, but is {\it target independent}. This leads us to 
look for ways to make measurements of the polarised structure functions
of other hadronic targets besides the proton and neutron. We now show how
this can effectively be done by studying semi-inclusive DIS 
$eN \rta ehX$ in the target fragmentation region (Fig.~\ref{fig:semiincl}).

The differential cross-section in the target fragmentation region can be
written analogously to eq.(\ref{eq:ea}) in terms of fracture functions:
\begin{equation}
x{d \D \s^{target}\over dx dy dz dt} = {Y_P\over2}{4\pi\a^2\over s}
\D M_1^{hN}(x,z, t, Q^2)
\label{eq:ehh}
\end{equation}
where $x = Q^2/2p_2.q$, $x_{\BB} = Q^2/2k.q$,
$z=p_2^{\prime}.q/p_2.q$ so that $1-z = x/x_{\BB}$, 
and the invariant momentum transfer $t = K^2 = -k^2$, where 
$k$ is the momentum of the struck parton. For $K^2 \ll Q^2$, $z \simeq
E_h/E_N$ (in the photon-nucleon CM frame) is the energy fraction of the
target nucleon carried by the detected hadron $h$. 
 
$\D M_1^{hN}$ is the fracture function \cite{Trentadue:1993ka}
equivalent of the inclusive structure function $g_1^N$, so in the 
same way as in eq.(\ref{eq:ed}) we have
\begin{equation}
\D M_1^{hN}(x,z, t, Q^2) = {1\over2}\sum_i e_i^2
\D M_i^{hN}(x,z, t, Q^2) 
\label{eq:eii}
\end{equation}
Here, $\D M_i^{hN}(x,z, t, Q^2)$ is an extended fracture function,
introduced by Grazzini, Trentadue and Veneziano \cite{Grazzini:1997ih},
which carries an explicit dependence on $t$. One of the advantages
of these fracture functions is that they satisfy a simple, homogeneous 
RG evolution equation analogous to the usual inclusive parton distributions.

Our proposal \cite{Shore:1997tq,deFlorian:1997th} 
(see also \cite{Shore:1997ck}) is to study 
semi-inclusive DIS in the kinematical region where the detected hadron $h$
($\pi$, $K$ or $D$) carries a large target energy fraction, i.e. $z$
approaching 1, with a small invariant momentum transfer $t$. In this region,
it is useful to think of the target fragmentation process as being 
simply modelled by a single Reggeon exchange (see Fig.~\ref{fig:semiincl}),
i.e.
\begin{equation}
\D M_1^{hN}(x,z, t, Q^2)\big|_{z\sim 1} ~\simeq~ F(t) (1-z)^{-2\a_{\BB}(t)}
g_1^{\BB}(x_{\BB},t,Q^2)
\label{eq:ejj}
\end{equation}
If we consider ratios of cross-sections, the dynamical Reggeon emission
factor $F(t)(1-z)^{-2\a_{\BB}(t)}$ will cancel and we will be able to 
isolate the ratios of $g_1^{\BB}(x_{\BB},t,Q^2)$ for different effective
targets $\BB$. Although single Reggeon exchange is of course only an 
approximation to the more fundamental QCD description in terms of fracture 
functions (see ref.\cite{Grazzini:1999vz} for a more technical discussion), 
it shows particularly clearly how observing semi-inclusive 
processes at large $z$ with particular choices of $h$ and $N$ amounts 
in effect to performing inclusive DIS on virtual hadronic targets $\BB$.
Since our predictions will depend only on the $SU(3)$ properties
of $\BB$, together with target independence, they will hold equally 
well when $\BB$ is interpreted as a Reggeon rather than a pure hadron 
state. 

The idea is therefore to make predictions for the ratios $\RR$ of 
the first moments of the polarised fracture functions 
$\int_0^{1-z} dx \D M_1^{hN}(x,z, t, Q^2)$ or equivalently
$\int_0^1 dx_{\BB} g_1^{\BB}(x_{\BB},t, Q^2)$ for various reactions.
The first moments $\C_1^{\cal B}$ are calculated as in eq.(\ref{eq:ec})
in terms of the axial charges $a^3$, $a^8$ and $a^0(Q^2)$ for a state
with the $SU(3)$ quantum numbers of ${\cal B}$. We then use  
topological charge screening to say that 
$a^0(Q^2) \simeq s(Q^2) a^0\big|_{\rm OZI}$, i.e.~the flavour singlet 
axial charge is suppressed relative to its OZI value by a universal,
target-independent, suppression factor $s(Q^2)$. From our calculation
of $\sqrt{\chi'(0)}$ and the experimental results for $g_1^p$, we have
$s|_{Q^2=10{\rm GeV}^2} \simeq 0.33/0.585 = 0.56$.

Some of the more interesting predictions obtained in ref.\cite{Shore:1997tq}
are as follows. The ratio
\begin{equation}
{\cal R} \biggl({en\rta e\pi^+ X\over ep\rta e\pi^- X}\biggr)_{z\sim 1}
 ~\simeq~~ {2s-1\over2s+2} 
\label{eq:ekk}
\end{equation}
is calculated by comparing $\C_1$ for the $\D^-$ and $\D^{++}$.
It is particularly striking because the physical value of $s(Q^2)$ is
close to one half, so the ratio becomes very small.
For strange mesons, on the other hand, the ratio depends on whether
the exchanged object is in the $\bf 8$ (where
the reduced matrix elements involve the appropriate $F/D$ ratio) or 
$\bf 10$ representation, so the prediction is less conclusive, viz.
\begin{equation}
{\cal R}\biggl({en\rta eK^+ X\over ep\rta eK^0 X}\biggr)_{z\sim 1} ~\simeq~~
{2s-1-3(2s-1)F/D \over 2s-1-3(2s+1)F/D}~~~({\bf 8}) ~~~~~~{\rm or}~~~~~~
{2s-1\over 2s+1}~~~({\bf 10}) 
\label{eq:ell}
\end{equation}
which we find by comparing $\C_1$ for either the $\Sigma^-$ and $\Sigma^+$
in the ${\bf 8}$ representation or $\Sigma^{*-}$ and $\Sigma^{*+}$ in the
${\bf 10}$.
For charmed mesons, we again find
\begin{equation}
{\cal R} \biggl({en\rta eD^0 X\over ep\rta eD^- X}\biggr)_{z\sim 1} ~\simeq~~
{2s-1\over2s+2} 
\label{eq:emm}
\end{equation}
corresponding to the ratio for $\Sigma_c^0$ to $\Sigma_c^{++}$.

At the other extreme, for $z$ approaching 0, the detected hadron carries
only a small fraction of the target nucleon energy. In this limit, the 
ratio $\RR$ of the fracture function moments becomes simply the ratio
of the structure function moments for $n$ and $p$, i.e.~using the
current experimental values,
$\RR_{z\sim 0} ~\simeq~ \C_1^n / \C_1^p = -0.30$.  This is to be compared
with the corresponding OZI or Ellis-Jaffe value of $-0.12$.

The differences between the OZI, or valence quark model, 
expectations and our predictions based on topological charge
screening can therefore be quite dramatic and should give a very clear
experimental signal. In ref.\cite{deFlorian:1997th}, together with De Florian,
we analysed the potential for realising these experiments in some detail.
Since we require particle identification in the target fragmentation
region, fixed-target experiments such as COMPASS or HERMES are not
appropriate. The preferred option is a polarised $ep$ collider.

The first requirement is to measure particles at extremely small
angles ($\theta \le 1$ mrad), corresponding to $t$ less than around
1 ${\rm GeV}^2$. This has already been achieved at HERA in measurements of 
diffractive and leading proton/neutron scattering using a forward
detection system known as the Leading Proton Spectrometer (LPS). 
The technique for measuring charged particles involves placing detectors 
commonly known as `Roman Pots' inside the beam pipe itself. 

The next point is to notice that the considerations above apply equally
to $\r$ as to $\pi$ production, since the ratios ${\cal R}$ are determined
by flavour quantum numbers alone. The particle identification  
requirements will therefore be less stringent, especially as the
production of leading strange mesons from protons or neutrons is strongly
suppressed. However, we require the forward detectors to have good 
acceptance for both positive and negatively charged mesons $M = \pi,\r$
in order to measure the ratio (\ref{eq:ekk}).

The reactions with a neutron target can be measured if the polarised
proton beam is replaced by polarised ${}^3 He$. In this case, if we assume
that ${}^3He = Ap + Bn$, the cross section for the production of positive
hadrons $h^+$ measured in the LPS is given by
\begin{equation}
\s\bigl({}^3He \rta h^+\bigr) \simeq A \s\bigl(p\rta h^+\bigr)
+ B \s\bigl(n \rta p\bigr) + B\s\bigl(n\rta M^+\bigr)
\label{eq:enn}
\end{equation}
The first contribution can be obtained from measurements with the proton
beam. However, to subtract the second one, the detectors must have sufficient
particle identification at least to distinguish protons from positively 
charged mesons.

Finally, estimates of the total rates \cite{deFlorian:1997th}
suggest that around $1\%$ of the total DIS events will contain a leading 
meson in the target fragmentation region where a LPS would have 
non-vanishing acceptance ($z>0.6$) and in the dominant domain $x<0.1$. 
The relevant cross-sections are therefore sufficient to allow the 
ratios ${\cal R}$ to be measured.

The conclusion is that while our proposals undoubtedly
pose a challenge to experimentalists, they are nevertheless possible.
Given the theoretical importance of the `proton spin' problem
and the topological charge screening mechanism, there is therefore 
strong motivation to perform target fragmentation experiments at
a future polarised $ep$ collider \cite{DeRoeck:2001ir}.

\section{Polarised two-photon physics and a sum rule for $g_1^\c$}

The $U(1)_A$ anomaly plays a vital role in another sum rule arising
in polarised deep-inelastic scattering, this time for the polarised
photon structure function $g_1^\c(x_\c,Q^2;K^2)$. For real photons,
the first moment of $g_1^\c$ vanishes as a consequence of electromagnetic
current conservation \cite{Bass:1992}. For off-shell photons, we proposed
a sum rule in 1992 \cite{Narison:1992fd,Shore:1992pm}
whose dependence on the virtual momentum of the 
target photon encodes a wealth of information about the anomaly,
chiral symmetry breaking and gluon dynamics in QCD. This is of 
special current interest since, given the ultra-high luminosity 
of proposed $e^+ e^-$ colliders designed as $B$ factories, 
a detailed measurement of our sum rule is about to become possible 
for the first time.

\subsection{The first moment sum rule for $g_1^\c$}

\begin{figure}
\centering
\includegraphics[height=4.5cm]{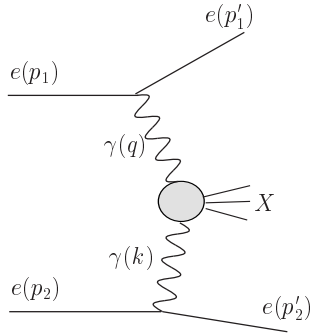} 
\caption{Kinematics for the two-photon DIS process $e^+ e^- \rta e^+ e^- X$.}
\label{fig:g1gammakin}     
\end{figure}

The polarised structure function $g_1^\c$ is measured in the process
$e^+ e^- \rta e^+ e^- X$, which at sufficiently high energy is dominated
by the two-photon interaction shown in Fig.~\ref{fig:g1gammakin}.
The deep-inelastic limit is characterised by $Q^2\rta \infty$ with
$x = Q^2/2p_2.q$ and $x_\c = Q^2/2k.q$ fixed, where $Q^2 = -q^2$,
$K^2 = -k^2$ and $s = (p_1+p_2)^2$. The target photon is assumed to be
relatively soft, $K^2 \ll Q^2$.
 
We are interested in the dependence of the photon structure function
$g_1^\c(x_\c,Q^2;K^2)$ on the invariant momentum $K^2$ of the target photon.
Experimentally, this is given by $K^2 \simeq E E'_2\theta_2^2$ where
$E'_2$ and $\theta_2$ are the energy and scattering angle of the
target electron. For the values $K^2\sim m_\r^2$ of interest in the 
sum rule, the target electron is nearly-forward and $\theta_2$ is 
very small. If it can be tagged, then the virtuality $K^2$ is simply 
determined from $\theta_2$; otherwise $K^2$ can be inferred indirectly 
from a measurement of the total hadronic energy.

The total cross-section $\s$ and the spin asymmetry $\D\s$ can be 
expressed formally in terms of `electron structure functions' as
follows \cite{Narison:1992fd}
\begin{equation}
\s~=~2\pi\a^2~{1\over s}~\int_0^\infty {dQ^2\over Q^2}\int_0^1
{dx\over x^2}~\Bigl[F_2^e ~{1\over y}\Bigl(1 -y +{y^2\over2}\Bigr)
- F_L^e {y\over2} \Bigr] 
\label{eq:fa}
\end{equation}
\begin{equation}
\D\s~=~2\pi\a^2~{1\over s}~\int_0^\infty {dQ^2\over Q^2} \int_0^1 
{dx\over x}~g_1^e \Bigl(1- {y\over2}\Bigr)~~~~~~~~~~~~~~~~~~~~~~~~~
\label{eq:fb}
\end{equation} 
where $\s = {1\over 2}(\s_{++} + \s_{+-})$ and $\D\s = {1\over 2}(\s_{++} - 
\s_{+-})$ with $+,-$ referring to the electron helicities. The parameter
$y = Q^2/x s \ll 1$ and only the leading order terms are retained below.

These electron structure functions can be expressed as convolutions of 
the photon structure functions with appropriate splitting functions.
In particular, we have
\begin{equation}
g_1^e(x,Q^2)~=~{\a\over2\pi}\int_0^\infty{dK^2\over K^2} \int_{x}^1
{dx_\c\over x_\c} \D P_{\c e}\Bigl({x\over x_\c}\Bigr)~
g_1^\c(x_\c,Q^2;K^2) 
\label{eq:fc}
\end{equation}
where $\D P_{\c e}(x) = (2 - x)$. This allows us to relate the 
$x_\c$-moments of the photon structure functions to the $x$-moments of the
cross-sections. For the first moment of $g_1^\c$, we find:
\begin{equation}
\int_0^1 dx~x {d^3\D\s\over dQ^2 dx dK^2} ~~=~~
{3\over2} \a^3 {1\over s Q^2 K^2}
\int_0^1 dx_\c~g_1^\c(x_\c,Q^2;K^2)
\label{eq:fd}
\end{equation}

The first moment sum rule follows, as for the proton, by using the OPE 
(\ref{eq:eb}) to express the product of electromagnetic currents for 
the incident photon in terms of the axial currents $J_{\m5}^a$.
The matrix elements $\langle \c^*(k)|J_{\m5}^a|\c^*(k)\rangle$ with
the target photon are then expressed in terms of the 3-current AVV 
Green function involving one axial and two electromagnetic currents.
We define form factors for this fundamental correlator as follows:
\begin{eqnarray}
&-i\langle0|J_{\m5}^a(p) J_\l(k_1)  J_\r(k_2)|0\rangle ~~&=~~ 
A_1^a ~\e_{\m\l\r\a}k_1^\a ~+~ A_2^a ~\e_{\m\l\r\a}k_2^\a 
\nonumber\\
&{}&~+A_3^a ~\e_{\m\l\a\b}k_1^\a k_2^\b k_{2\r} ~+~ 
A_4^a ~\e_{\m\r\a\b}k_1^\a k_2^\b k_{1\l} 
\nonumber\\
&{}&~+A_5^a ~\e_{\m\l\a\b}k_1^\a k_2^\b k_{1\r} ~+~
A_6^a ~\e_{\m\r\a\b}k_1^\a k_2^\b k_{2\l}
\nonumber\\
&{}&{}\label{eq:fe}
\end{eqnarray}
where the six form factors are functions of the invariant momenta, i.e.
$A_i^a = A_i^a(p^2,k_1^2,k_2^2)$. We also abbreviate $A_i^a(0,k^2,k^2)
= A_i^a(K^2)$.

\vskip0.1cm
The first moment sum rule for $g_1^\c$ is then \cite{Narison:1992fd}:
\begin{equation}
\int_0^1 dx_\c~g_1^\c(x_\c,Q^2;K^2) ~=~ 4\pi\a \sum_{a=3,8,0} \D C_1^a(Q^2)
\Bigl(A_1^a(K^2) - A_2^a(K^2)\Bigr)
\label{eq:ff}
\end{equation}
where the Wilson coefficients are related to those in eq.(\ref{eq:ec}) by
$\D C_1^3 = \D C_1^{NS}$, $\D C_1^8 = {1\over\sqrt3}\D C_1^{NS}$ and
$\D C_1^0 = {2\sqrt2\over\sqrt3}\D C_1^{S}$.\footnote{Explicitly, 
$$
\D C_1^{NS} = {1\over3}\Bigl(1- {\a_s(Q^2)\over\pi}\Bigr), ~~~~~~~ 
\D C_1^{S}  = {1\over3}\Bigl(1- {\a_s(Q^2)\over\pi}\Bigr)~
\exp \int_0^{t(Q)} dt' ~\c(\a_s(t'))
$$
at leading order, where $t(Q) = {1\over2}\ln{Q^2\over\m^2}$ and
$\c = -{3\over4}{\a_s^2\over(4\pi)^2}$ is the 
anomalous dimension corresponding to the $U(1)_A$ current renormalisation.}

\vskip0.1cm
Now, just as the sum rule for the proton structure function $g_1^p$
could be related to low-energy meson-nucleon couplings via the $U(1)_A$
Goldberger-Treiman relations, we can relate this sum rule for $g_1^\c$
to the pseudoscalar meson radiative decays using the analysis
in section \ref{radiative}. Introducing the {\it off-shell} radiative
pseudoscalar couplings for photon virtuality $K^2$, we define form
factors
\begin{equation}
F^a(K^2) ~=~ 1 - \Bigl(a^a_{\rm em}{\a\over\pi}\Bigr)^{-1}~
\hat f^{a\a} g_{\hat\eta^\a \c\c}(K^2)
\label{eq:fg}
\end{equation}
or alternatively,
\begin{eqnarray}
&F^3(K^2) ~&=~ 1 - \Bigl(a^3_{\rm em}{\a\over\pi}\Bigr)^{-1}~
f_\pi g_{\pi\c\c}(K^2)
\nonumber\\
&F^8(K^2) ~&=~ 1 - \Bigl(a^8_{\rm em} {\a\over\pi}\Bigr)^{-1}~
\Bigl(f^{8\eta} g_{\eta\c\c}(K^2) + f^{8\eta'}g_{\eta'\c\c}(K^2) \Bigr)
\nonumber\\
&F^0(K^2) ~&=~ 1 - \Bigl(a^0_{\rm em} {\a\over\pi}\Bigr)^{-1}~
\Bigl(f^{0\eta}g_{\eta\c\c}(K^2) + f^{0\eta'} g_{\eta'\c\c}(K^2) 
+ \sqrt{6}A g_{G\c\c}(K^2) \Bigr)
\nonumber\\
&{}&{}\label{eq:fh}
\end{eqnarray}
where the $a^a_{\rm em}$ are the electromagnetic $U(1)_A$ anomaly 
coefficients defined earlier. We may then rewrite the sum rule as
\begin{equation}
\int_0^1 dx_\c~g_1^\c(x_\c,Q^2;K^2) ~=~ {1\over2}{\a\over\pi}
\sum_{a=3,8,0} \D C_1^a(Q^2)~ 
a^a_{\rm em}~F^a(K^2)
\label{eq:fi}
\end{equation}

\vskip0.1cm
The dependence of the $g_1^\c$ on the invariant momentum $K^2$ of the
target photon reflects many key aspects of both perturbative and
non-perturbative QCD dynamics. For on-shell photons, $K^2=0$, we have
simply \cite{Bass:1992,Narison:1992fd}
\begin{equation}
\int_0^1 dx_\c~g_1^\c(x_\c,Q^2;K^2=0) ~=~ 0
\label{eq:fj}
\end{equation}
This is a consequence of electromagnetic current conservation. 
This follows simply by taking the divergence of eq.(\ref{eq:fe}) and
observing that in the limit $p\rta 0$, both $A_1$ and $A_2$ are 
of $O(K^2)$.\footnote{Electromagnetic current conservation in
eq.(\ref{eq:fe}) implies
$$
A_1^a = A_3^a k_2^2 + A_5^a {1\over2}(p^2 - k_1^2 - k_2^2),~~~~~~~~~
A_2^a = A_4^a k_1^2 + A_6^a {1\over2}(p^2 - k_1^2 - k_2^2)
$$
The chiral limit is special since the form factors can have massless
poles and is considered in detail in ref.\cite{Shore:1992pm}. The sum rule 
(\ref{eq:fj}) still holds.}

In the asymptotic limit where $K^2 \ll m_\r^2$, a relatively straightforward
renormalisation group analysis combined with the anomaly equation
shows that, for the flavour non-singlets,
the $A_i^a$ tend to the value ${1\over2}{\a\over\pi}a_{\rm em}^a$.
while in the flavour singlet sector, $A_i^0$ has an additional factor 
depending on the anomalous dimension $\c$. Using the explicit expressions
for the Wilson coefficients, we find
\begin{eqnarray}
\int_0^1{} dx_\c~g_1^\c(x_\c,Q^2;K^2\ll m_\r^2)
~~~~~~~~~~~~~~~~~~~~~~~~~~~~~~~~~~~~~~~~~~~~~~~~~~~~~~~~~~~
\nonumber\\
=~{1\over6} {\a\over\pi} \Bigl(1 - {\a_s(Q^2)\over\pi}\Bigr)
\biggl(a_{\rm em}^3 + {1\over\sqrt3}a_{\rm em}^8 + 
{2\sqrt2\over\sqrt3}a_{\rm em}^0 
\exp\biggl[ \int_{t(K)}^{t(Q)}{} dt'~\c(\a_s(t'))\biggr] \biggr)
\nonumber\\
{}\nonumber\\
=~{1\over3} {\a\over\pi} 
\biggl[1 ~-~ {4\over9}{1\over \ln Q^2/\L^2} ~+~
{16\over81} \biggl({1\over \ln Q^2/\L^2}- {1\over \ln K^2/\L^2}
\biggr) \biggr]~~~~~~~~~~~~~~~~~~~~~
\label{eq:fk}
\end{eqnarray}
The asymptotic limit is therefore determined by the electromagnetic
$U(1)_A$ anomaly, with logarithmic corrections  reflecting the
anomalous dimension of the flavour singlet current due to the
colour $U(1)_A$ anomaly. (See also ref.\cite{Sasaki:2006bt} for   
a NNLO analysis.)

In between these limits, the first moment of $g_1^\c$ provides a 
measure of the form factors defining the 3-current $AVV$ Green function, 
which encodes a great deal of information about the dynamics of QCD,
especially the non-perturbative realisation of chiral symmetry 
\cite{Shore:1992pm}.
Equivalently, in the form (\ref{eq:fi}), it measures the momentum dependence 
of the off-shell radiative couplings of the pseudoscalar mesons as
the form factors $F^a(K^2)$ vary from 0 to 1.

Just as for $g_1^p$, we can again isolate a dependence on the topological
susceptibility through the identification of the flavour singlet decay
constant $\hat f^{00}$ in eq.(\ref{eq:fg}) with $\sqrt{\chi'(0)}$ in the
chiral limit. This time, however, it is unlikely to be a good approximation
to set the corresponding coupling $g_{\hat\eta^0 \c\c}$ equal to its OZI value
since it is not RG invariant. A more promising approximation is to
recall from section 4 that the RG invariant gluonic
coupling $g_{G\c\c}(0)$ is OZI suppressed and likely to be small. This was
confirmed by the phenomenological analysis. If we assume this is also true
of the off-shell coupling, then we may approximate the sum rule for
$g_1^\c$ entirely in terms of the off-shell couplings of the physical
mesons $\pi^0$, $\eta$ and $\eta'$. 

In general, the momentum dependence of the form factors $(A_1^a - A_2^a)$
or $F^a$ will depend on the fermions contributing to the AVV Green
function \cite{Shore:1992pm}. In the case of leptons, or heavy quarks, 
the crossover scale as the form factors $F^a(K^2)$ rise from 0 to 1 with 
increasing $K^2$ will be given by the fermion mass. For the
light quarks, however, we expect the crossover scale to be a typical
hadronic scale $\sim m_\r$ rather than $m_{u,d,s}$. This can be justified
by a rough OPE argument and is consistent with old ideas of vector meson
dominance \cite{Shore:1992pm,Ueda:2006cp}. 
This behaviour would be an interesting manifestation of the 
spontaneous breaking of chiral symmetry. 

Once again, therefore, we see a close relation between the realisation 
of sum rules in high-energy deep-inelastic scattering and low-energy meson 
physics. All these issues are discussed at some length in our earlier papers,
but here we now turn our attention to the vital question
of whether the $g_1^\c$ sum rule can be measured in current or future
collider experiments \cite{Shore:2004cb}.

\subsection{Cross-sections and spin asymmetries at polarised B factories}

The spin-dependent cross-sections for the two-photon DIS process
$e^+ e^- \rta e^+ e^- X$ were analysed in 
refs.\cite{Narison:1992fd,Shore:2004cb}
taking account of the experimental cuts on the various kinematical
parameters. Keeping the lower cut on $Q^2$ as a free parameter, we 
found the following results for the total cross-section and spin
asymmetry:
\begin{equation}
\s ~\simeq~
0.5 \times 10^{-8}~{1\over Q^2_{\rm min}}~\log{Q^2_{\rm min}\over \L^2}
~\biggl(\log{s\over Q^2_{\rm min}}\biggr)^2
\label{eq:ffa}
\end{equation}
and 
\begin{equation}
{\D\s\over\s} ~=~
{1\over2}~{Q^2_{\rm min}\over s}~ \log{s\over 4 Q^2_{\rm min}}~
\biggl[ 1 + \log{s\over 4\L^2}\biggl(\log{Q^2_{\rm min}\over\L^2}\biggr)^{-1}
\biggr]
\label{eq:ffb}
\end{equation}
In order to measure the $g_1^\c$ sum rule, we need to find collider
parameters such that the spin asymmetry is significant in a kinematic region
where the total cross-section is still large. A useful statistical measure
of the significance of the asymmetry is that $\sqrt{L\s}\D\s/\s \gg 1$,
where $L$ is the luminosity.

When we first proposed the first moment sum rule for $g_1^\c$,
the luminosity available from the then current accelerators was inadequate 
to allow it to be studied. For example, for a polarised version of LEP 
operating at $s=10^4~{\rm GeV}^2$ with an annual integrated luminosity of 
$L=100~{\rm pb}^{-1}$, and optimising the cut at 
$Q_{\rm min}^2 = 10~{\rm GeV}^2$, we only have $\s \simeq 35~{\rm pb}$ 
and $\D\s/\s \simeq 0.01$. The corresponding annual event rate would be 
$3.5\times 10^3$ and the statistical significance 
$\sqrt{L\s} \D\s/\s \simeq 0.5$, so even a reliable 
measurement of the spin asymmetry could not be made. 

Clearly, a hugely increased luminosity is required and this has now become
available with proposals for machines with projected annual integrated 
luminosities measured in inverse attobarns. However, as noted in 
ref.\cite{Narison:1992fd}, if this increased luminosity is associated with 
increased CM energy, then the $1/s$ factor in the spin asymmetry 
(\ref{eq:ffb}) sharply reduces the possibility of extracting $g_1^\c$. 
There is also a competition as $Q^2_{\rm min}$ is varied between 
increasing spin asymmetry and decreasing total cross-section.
This is particularly evident when we analyse the potential of the ILC
\cite{ILCone,ILCtwo} for measuring the sum rule \cite{Shore:2004cb}. 
We find that even optimising the
$Q^2_{\rm min}$ cut, the spin asymmetry is still only of order
$\D\s/\s \simeq 0.002$ when $\s$ itself has fallen to around $15~{\rm pb}$. 
While, given the high luminosity, this would allow a measurement of the 
first moment of $g_1^\c$ integrated over $K^2$, a detailed study of the 
$K^2$-dependence of the sum rule requires a much greater spin asymmetry.

This leads us to consider instead the new generation of ultra-high 
luminosity $e^+ e^-$ colliders. Although these are envisaged as 
$B$ factories, these colliders operating with polarised beams would, 
as we now show, be extremely valuable for studying polarisation phenomena 
in QCD. As an example of this class, we take the proposed SuperKEKB collider. 
(The analysis for PEPII is very similar, the main difference being the 
additional ten-fold increase in luminosity in the current SuperKEKB proposals.)

\begin{figure}
\centering
\includegraphics[height=3.2cm]{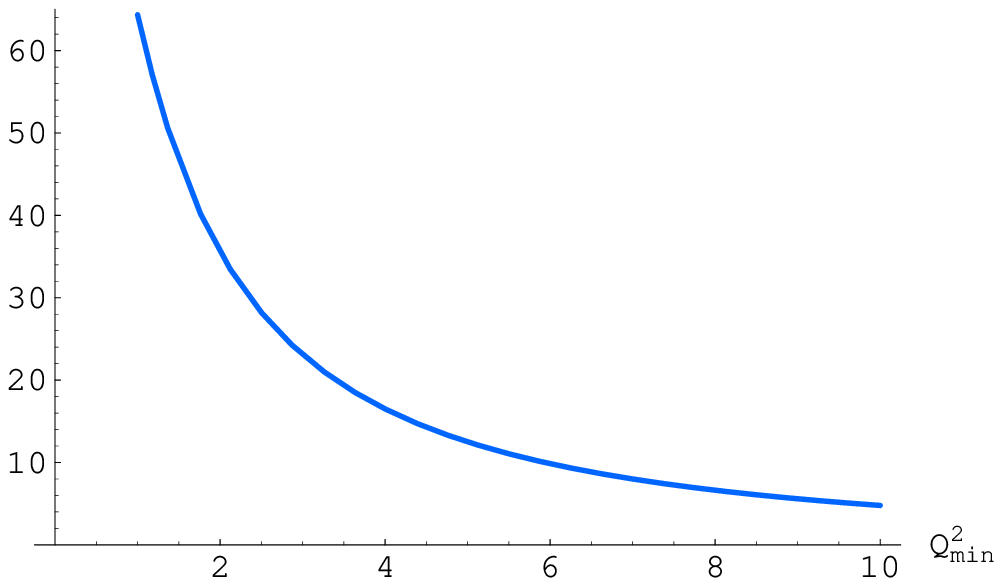} \hskip0.5cm
\includegraphics[height=3.2cm]{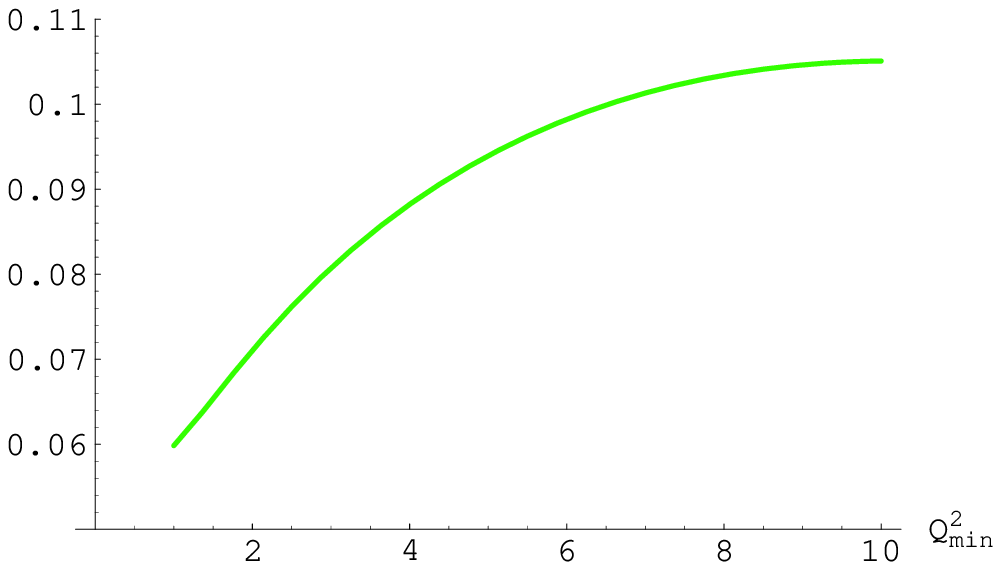}
\caption{The left-hand graph shows the total cross-section $\s$ (in pb)
at SuperKEKB as the experimental cut $Q^2_{\rm min}$ is varied from 
1 to 10 ${\rm GeV}^2$. The right-hand graph shows the spin asymmetry
$\D\s/\s$ over the same range of $Q^2_{\rm min}$.}
\label{fig:SuperKEKBfig}     
\end{figure}

SuperKEKB is an asymmetric $e^+ e^-$ collider with $s= 132~{\rm GeV}^2$,
corresponding to electron and positron beams of 8 and 3.5~GeV respectively.
The design luminosity is $5\times 10^{35}$ ${\rm cm}^{-2} s^{-1}$, which
gives an annual integrated luminosity of $5~{\rm ab}^{-1}$ \cite{SuperKEKB}. 
To see the effects of the experimental cut on $Q_{\rm min}^2$ in this case, 
we have plotted the total cross-section and the spin asymmetry in 
Fig.~\ref{fig:SuperKEKBfig}, in the range of $Q_{\rm min}^2$ from 
1 to $10~{\rm GeV}^2$. In this range $\s$ is falling like $1/Q_{\rm min}^2$ 
while $\D\s/\s$ rises to what is actually a maximum at 
$Q_{\rm min}^2=10~{\rm GeV}^2$. 

Taking $Q_{\rm min}^2 =5~{\rm GeV}^2$, we find $\s \simeq 12.5~{\rm pb}$
with spin asymmetry $\D\s/\s \simeq 0.1$. The annual event rate
is therefore $6.25 \times 10^7$, with $\sqrt{L\s} \D\s/\s \simeq 750$.
This combination of a very high event rate and the large $10\%$ spin 
asymmetry means that SuperKEKB has the potential not only to measure 
$\D\s$ but to access the full first moment sum rule for $g_1^\c$ itself. 
Recall from eq.(\ref{eq:fd}) that to measure $\int_0^1 dx~g_1^\c(x,Q^2;K^2)$ 
we need not just $\D\s$ but the fully differential cross-section w.r.t.~$K^2$ 
as well as $x$ and $Q^2$ if the interesting non-perturbative 
QCD physics is to be accessed. To measure this, we need to divide the 
data into sufficiently fine $K^2$ bins in order to plot the explicit $K^2$ 
dependence of $g_1^\c$, while still maintaining the statistical significance 
of the asymmetry.
The ultra-high luminosity of SuperKEKB ensures that the event rate is
sufficient, while its moderate CM energy means that the crucial spin asymmetry
is not overly suppressed by its $1/s$ dependence.

Our conclusion is that the new generation of ultra-high luminosity, moderate 
energy $e^+ e^-$ colliders, currently conceived as $B$ factories, could also
be uniquely sensitive to important QCD physics if run with polarised beams. 
In particular, they appear to be the only accelerators capable of accessing
the full physics content of the sum rule for the first moment of the
polarised structure function $g_1^\c(x,Q^2;K^2)$. The richness of this physics,
in particular the realisation of chiral symmetry breaking, the manifestations
of the axial $U(1)_A$ anomaly and the role of gluon topology,
provides a strong motivation for giving serious consideration to an attempt to
measure the $g_1^\c$ sum rule at these new colliders.

\vskip1cm

\noindent{\bf Acknowledgements}
\vskip0.3cm

In addition to Gabriele, I would like to thank Daniel de Florian,
Massimiliano Grazzini, Stephan Narison and Ben White for their collaboration
on the original research presented here. This paper has been prepared with
the partial support of PPARC grant PP/D507407/1.

\vfill\eject


\begin{thebibliography}{99.}

\bibitem{Veneziano:1979ec}
  G.~Veneziano,
  Nucl.\ Phys.\ \textbf{B159}, 213 (1979).

\bibitem{Shore:1999tw}
  G.~M.~Shore,
  Nucl.\ Phys.\ \textbf{B569}, 107 (2000),
  [arXiv:hep-ph/9908217].

\bibitem{Shore:2006mm}
  G.~M.~Shore,
  Nucl.\ Phys.\ \textbf{B744}, 34 (2006),
  [arXiv:hep-ph/0601051].

\bibitem{Veneziano:1989ei}
  G.~Veneziano,
  Mod.\ Phys.\ Lett.\ \textbf{A4}, 1605 (1989).

\bibitem{Narison:1992fd}
  S.~Narison, G.~M.~Shore and G.~Veneziano,
  Nucl.\ Phys.\ \textbf{B391}, 69 (1993).

\bibitem{Shore:1992pm}
  G.~M.~Shore and G.~Veneziano,
  Mod.\ Phys.\ Lett.\ \textbf{A8}, 373 (1993).

\bibitem{Shore:2004cb}
  G.~M.~Shore,
  Nucl.\ Phys.\ \textbf{B712}, 411 (2005),
  [arXiv:hep-ph/0412192].

\bibitem{Shore:1990zu}
  G.~M.~Shore and G.~Veneziano,
  Phys.\ Lett.\ \textbf{B244}, 75 (1990).

\bibitem{Shore:1991dv}
  G.~M.~Shore and G.~Veneziano,
  Nucl.\ Phys.\ \textbf{B381}, 23 (1992).

\bibitem{Narison:1994hv}
  S.~Narison, G.~M.~Shore and G.~Veneziano,
  Nucl.\ Phys.\ \textbf{B433}, 209 (1995),
  [arXiv:hep-ph/9404277].

\bibitem{Shore:1997tq}
  G.~M.~Shore and G.~Veneziano,
  Nucl.\ Phys.\ \textbf{B516}, 333 (1998),
  [arXiv:hep-ph/9709213].

\bibitem{Veneziano:Okubo}
G.~Veneziano: ``The `spin' of the proton and the OZI limit of QCD.''
In: \textit{From Symmetries to Strings: Forty Years of Rochester
Conferences}, ed. A.~Das (World Scientific, 1990), 13-26.

\bibitem{Adler:1969}
S.~L.~Adler, Phys.~Rev.~\textbf{177}, 2426 (1969).

\bibitem{BellJackiw:1969}
J.~S.~Bell and R.~Jackiw, Nuovo Cimento \textbf{60A}, 47 (1969).

\bibitem{AdlerBardeen:1969}
S.~L.~Adler and W.A.~Bardeen, Phys.~Rev.~\textbf{182}, 1517 (1969).

\bibitem{Steinberger:1949}
J.~Steinberger, Phys.~Rev.~\textbf{76}, 1180 (1949).

\bibitem{Schwinger:1951}
J.~Schwinger, Phys.~Rev.~\textbf{82}, 664 (1951).

\bibitem{Fujikawa:1979}
K.~Fujikawa, Phys.~Rev.~Lett.~\textbf{42}, 1195 (1979);
Phys.~Rev.~\textbf{D21}, 2848 (1980), 
erratum-ibid.~\textbf{D22}, 1499 (1980).

\bibitem{Shore:1998dm}
  G.~M.~Shore,
  ``$U(1)_A$ problems and gluon topology: anomalous symmetry in QCD'',
  In: \textit{Hidden Symmetries and Higgs Phenomena}, Zuoz Summer School, 
  Switzerland, 1998, pp 201-223; 
  arXiv:hep-ph/9812354.

\bibitem{Witten:1979vv}
  E.~Witten,
  Nucl.\ Phys.\ \textbf{B156}, 269 (1979).

\bibitem{DiVecchia:1980ve}
  P.~Di Vecchia and G.~Veneziano,
  Nucl.\ Phys.\ \textbf{B171}, 253 (1980).

\bibitem{Espriu:1982bw}
  D.~Espriu and R.~Tarrach,
  Z.\ Phys.\ \textbf{C16}, 77 (1982).

\bibitem{Shore:1990wp}
  G.~M.~Shore,
  Nucl.\ Phys.\ \textbf{B362}, 85 (1991).

\bibitem{Shore:1991pn}
  G.~M.~Shore and G.~Veneziano,
  Nucl.\ Phys.\ \textbf{B381}, 3 (1992).

\bibitem{Hooft:1974}
G.~'t Hooft, Nucl.~Phys.~\textbf{B72}, 461 (1972).

\bibitem{Veneziano:TE}
G.~Veneziano, Phys.~Lett.~\textbf{52B}, 220 (1974); 
Nucl.~Phys.~\textbf{B117}, 519 (1976).

\bibitem{Okubo}
S.~Okubo, Phys.~Lett.~\textbf{5}, 165 (1963).

\bibitem{Zweig}
G.~Zweig, CERN report 8419/TH412 (1964).

\bibitem{Iizuka}
J.~Iizuka, Prog.~Theor.~Phys.~Suppl.~\textbf{37-38}, 21 (1966).

\bibitem{Weinberg:1975}
S.~Weinberg, Phys.~Rev.~\textbf{D11}, 3583 (1975).

\bibitem{'tHooft:1976fv}
  G.~'t Hooft,
  Phys.\ Rev.\ \textbf{D14}, 3432 (1976);
  [Erratum-ibid.\ \textbf{D18}, 2199 (1978)].

\bibitem{Crewther:1978kq}
  R.~J.~Crewther,
  Riv.\ Nuovo Cim.\  \textbf{2N8}, 63 (1979).

\bibitem{Christos:1984tu}
  G.~A.~Christos,
  Phys.\ Rept.\  \textbf{116}, 251 (1984).

\bibitem{'tHooft:1986nc}
  G.~'t Hooft,
  Phys.\ Rept.\  \textbf{142}, 357 (1986).

\bibitem{Gell-Mann:1968rz}
  M.~Gell-Mann, R.~J.~Oakes and B.~Renner,
  Phys.~Rev.~\textbf{175}, 2195 (1968).
  
\bibitem{Dashen:1969eg}
  R.~F.~Dashen,
  Phys.~Rev.~\textbf{183}, 1245 (1969).

\bibitem{Rosenzweig:1979ay}
  C.~Rosenzweig, J.~Schechter and C.~G.~Trahern,
  Phys.\ Rev.\ \textbf{D21}, 3388 (1980).

\bibitem{DiVecchia:1980sq}
  P.~Di Vecchia, F.~Nicodemi, R.~Pettorino and G.~Veneziano,
  Nucl.\ Phys.\ \textbf{B181}, 318 (1981).

\bibitem{Kawarabayashi:1980dp}
  K.~Kawarabayashi and N.~Ohta,
  Nucl.\ Phys.\ \textbf{B175}, 477 (1980).

\bibitem{Herrera-Siklody:1996pm}
  P.~Herrera-Siklody, J.~I.~Latorre, P.~Pascual and J.~Taron,
  Nucl.\ Phys.\ \textbf{B497}, 345 (1997);~~ 
  Phys.\ Lett.\ \textbf{B419}, 326 (1998).

\bibitem{Leutwyler:1997yr}
  H.~Leutwyler,
  Nucl.\ Phys.\ Proc.\ Suppl.\ \textbf{64}, 223 (1998),
  [arXiv:hep-ph/9709408].

\bibitem{Kaiser:2000gs}
  R.~Kaiser and H.~Leutwyler,
  Eur.\ Phys.\ J.\ \textbf{C17}, 623 (2000),
  [arXiv:hep-ph/0007101].

\bibitem{Giusti:2001xh}
  L.~Giusti, G.~C.~Rossi, M.~Testa and G.~Veneziano,
  Nucl.\ Phys.\ \textbf{B628} (2002) 234
  [arXiv:hep-lat/0108009].

\bibitem{Shore:2001cs}
  G.~M.~Shore,
  Phys.\ Scripta \textbf{T99}, 84 (2002),
  [arXiv:hep-ph/0111165].

\bibitem{PDG}
  Particle Data Group, Review of Particle Properties, 
  Phys.~Lett.~\textbf{B592}, 1 (2004).

\bibitem{L3}
  M.~Acciarri {\it et al.}, L3 Collaboration,
  Phys.~Lett.~\textbf{B418}, 399 (1998).

\bibitem{Crystal}
  D.~A.~Williams {\it et al.}, Crystal Ball Collaboration,
  Phys.~Rev.~\textbf{D38}, 1365 (1988).

\bibitem{ASP}
  N.~A.~Roe {\it et al.}, ASP Collaboration,
  Phys.~Rev.~\textbf{D41}, 17 (1990).

\bibitem{DelDebbio:2004ns}
  L.~Del Debbio, L.~Giusti and C.~Pica,
  Phys.\ Rev.\ Lett.\ \textbf{94}, 032003 (2005).
  [arXiv:hep-th/0407052].

\bibitem{DiGiacomo:1990ij}
  A.~Di Giacomo,
  Nucl.\ Phys.\ Proc.\ Suppl.\ \textbf{23B}, 191 (1991).

\bibitem{Narison:1990cz}
  S.~Narison,
  Phys.\ Lett.\ \textbf{B255},101 (1991);
  Z.\ Phys.\ \textbf{C26}, 209 (1984).

\bibitem{Narison:1998aq}
  S.~Narison, G.~M.~Shore and G.~Veneziano,
  Nucl.\ Phys.\ \textbf{B546}, 235 (1999),
  [arXiv:hep-ph/9812333].

\bibitem{COMPASS}
  V.~Y.~Alexakhin {\it et al.}  [COMPASS Collaboration],
  ``The deuteron spin-dependent structure function g1(d) and its first
  moment,''
  arXiv:hep-ex/0609038.

\bibitem{HERMES}
  A.~Airapetian {\it et al.}  [HERMES Collaboration],
  ``Precise determination of the spin structure function g(1) of the proton,
  deuteron and neutron,''
  arXiv:hep-ex/0609039.

\bibitem{Mallot}
  G.~Mallot, S.~Platchkov and A.~Magnon, CERN-SPSC-2005-017; SPSC-M-733.

\bibitem{Ageev:2005gh}
  E.~S.~Ageev {\it et al.}  [COMPASS Collaboration],
  Phys.\ Lett.\ \textbf{B612}, 154 (2005),
  [arXiv:hep-ex/0501073].

\bibitem{Ellis:1973kp}
  J.~R.~Ellis and R.~L.~Jaffe,
  Phys.\ Rev.\ \textbf{D9}, 1444 (1974),
  [Erratum-ibid.\ \textbf{D10}, 1669 (1974)].

\bibitem{Bugg:2004cm}
  D.~V.~Bugg,
  Eur.\ Phys.\ J.\ \textbf{C33}, 505 (2004).

\bibitem{Moskal:2004cm}
  P.~Moskal,
  ``Hadronic interaction of eta and eta$'$ mesons with protons,''
  arXiv:hep-ph/0408162.

\bibitem{Bass:2001ix}
  S.~D.~Bass,
  Phys.\ Scripta \textbf{T99}, 96 (2002),
  [arXiv:hep-ph/0111180].

\bibitem{Moskal:2004nw}
  P.~Moskal {\it et al.},
  Int.\ J.\ Mod.\ Phys.\ \textbf{A20}, 1880 (2005),
  [arXiv:hep-ex/0411052].

\bibitem{Moskal:1998pc}
  P.~Moskal {\it et al.},
  Phys.\ Rev.\ Lett.\ \textbf{80}, 3202 (1998),
  [arXiv:nucl-ex/9803002].

\bibitem{Nakayama:2005ts}
  K.~Nakayama and H.~Haberzettl,
  ``Analyzing eta$'$ photoproduction data on the proton at energies of 
  1.5GeV -- 2.3GeV,''
  arXiv:nucl-th/0507044.

\bibitem{Dugger:2005du}
  M.~Dugger  [CLAS Collaboration],
  ``S=0 pseudoscalar meson photoproduction from the proton,''
  arXiv:nucl-ex/0512005.

\bibitem{Shore:1994zh}
  G.~M.~Shore,
  Nucl.\ Phys.\ Proc.\ Suppl.\ \textbf{39BC}, 101 (1995),
  [arXiv:hep-ph/9410383].

\bibitem{EMC}
  J.~Ashman {et al.}, Phys.~Lett.~\textbf{B206}, 364 (1988); ~
  Nucl.~Phys.~\textbf{B328}, 1 (1990).  

\bibitem{Jaffe:1989jz}
  R.~L.~Jaffe and A.~Manohar,
  Nucl.\ Phys.\ \textbf{B337}, 509 (1990).

\bibitem{Shore:1999be}
  G.~M.~Shore and B.~E.~White,
  Nucl.\ Phys.\ \textbf{B581}, 409 (2000),
  [arXiv:hep-ph/9912341].

\bibitem{Shore:2000ca}
  G.~M.~Shore,
  Nucl.\ Phys.\ Proc.\ Suppl.\ \textbf{96},171 (2001),
  [arXiv:hep-ph/0007239].

\bibitem{Bakker:2004ib}
  B.~L.~G.~Bakker, E.~Leader and T.~L.~Trueman,
  Phys.\ Rev.\ \textbf{D70}, 114001 (2004),
  [arXiv:hep-ph/0406139].

\bibitem{Ball:1995td}
  R.~D.~Ball, S.~Forte and G.~Ridolfi,
  Phys.\ Lett.\ \textbf{B378}, 255 (1996),
  [arXiv:hep-ph/9510449].

\bibitem{Altarelli:1988nr}
  G.~Altarelli and G.~G.~Ross,
  Phys.\ Lett.\ \textbf{B212}, 391 (1988).

\bibitem{Procureur:2006sg}
  S.~Procureur  [COMPASS Collaboration],
  ``New measurement of Delta(G)/G at COMPASS,''
  arXiv:hep-ex/0605043.

\bibitem{STAR}
  R.~Fatemi  [STAR Collaboration],
  ``Using jet asymmetries to access Delta(G), the gluon helicity distribution
  of the proton at STAR,''
  arXiv:nucl-ex/0606007.

\bibitem{PHENIX}
  Y.~Fukao  [PHENIX Collaboration],
  ``The overview of the spin physics at RHIC-PHENIX experiment,''
  AIP Conf.\ Proc.\ \textbf{842}, 321 (2006),
  [arXiv:hep-ex/0607033].

\bibitem{Shore:1998dn}
  G.~M.~Shore,
  ``The proton spin crisis: Another ABJ anomaly?'',
  In: \textit{From the Planck length to the Hubble radius}, Erice 1998,
  ed.~A.~Zichichi, World Scientific, Singapore, pp 79-105;~
  arXiv:hep-ph/9812355.

\bibitem{Trentadue:1993ka}
  L.~Trentadue and G.~Veneziano,
  Phys.\ Lett.\ \textbf{B323}, 201 (1994).

\bibitem{Grazzini:1997ih}
  M.~Grazzini, L.~Trentadue and G.~Veneziano,
  Nucl.\ Phys.\ \textbf{B519}, 394 (1998),
  [arXiv:hep-ph/9709452].

\bibitem{deFlorian:1997th}
  D.~de Florian, G.~M.~Shore and G.~Veneziano,
  ``Target fragmentation at polarized HERA: A test of universal topological
  charge screening in QCD,''~
  In: \textit{Proceedings of the 1997 Workshop with Polarized Protons at Hera},
   ed.~A.~de Roeck and T.~Gehrmann, Hamburg/Zeuthen 1997, 
  pp 696-703;~
  arXiv:hep-ph/9711353.

\bibitem{Shore:1997ck}
  G.~M.~Shore,
  Nucl.\ Phys.\ Proc.\ Suppl.\ \textbf{64}, 167 (1998),
  [arXiv:hep-ph/9710367].

\bibitem{Grazzini:1999vz}
  M.~Grazzini, G.~M.~Shore and B.~E.~White,
  Nucl.\ Phys.\ \textbf{B555}, 259 (1999),
  [arXiv:hep-ph/9903530].

\bibitem{DeRoeck:2001ir}
  A.~De Roeck,
  Nucl.\ Phys.\ Proc.\ Suppl.\ \textbf{105}, 40 (2002),
  [arXiv:hep-ph/0110335].

\bibitem{Bass:1992}
  S.D.~Bass, Int.~J.~Mod.~Phys.~\textbf{A7}, 6039 (1992).

\bibitem{Sasaki:2006bt}
  K.~Sasaki, T.~Ueda and T.~Uematsu,
  Phys.\ Rev.\ \textbf{D73}, 094024 (2006),
  [arXiv:hep-ph/0604130].

\bibitem{Ueda:2006cp}
  T.~Ueda, T.~Uematsu and K.~Sasaki,
  Phys.\ Lett.\ \textbf{B640}, 188 (2006),
  [arXiv:hep-ph/0606267].

\bibitem{ILCone}
  F.~Richard {\it et al.}, ``TESLA: The Superconducting electron 
  positron linear collider with an integrated X-ray laser laboratory.
  Technical Design Report, Part I''; hep-ph/0106314.

\bibitem{ILCtwo}
  M.~Woods {\it et al.}, ``Luminosity, Energy and Polarization
  Studies for the Linear Collider'', In: \textit{ Proc. 5th International 
  Workshop on Electron-Electron Interactions at TeV Energies}, 
  Santa Cruz, 2003; physics/0403037. 

\bibitem{SuperKEKB}
  A.~G.~Akeroyd {\it et al.} (SuperKEKB Physics Working Group),
  ``Physics at Super B Factory''; hep-ex/0406071.


\end{thebibliography}
\end{document}